\documentclass[ALICE,manyauthors]{cernphprep}

\usepackage[comma,square,numbers,sort&compress]{natbib}
\usepackage{hyperref}
\usepackage{lineno}
\usepackage[utf8]{inputenc}
\usepackage{color}

\newcommand{\snn}{$\sqrt{s_{\mathrm{NN}}}$}
\newcommand{\pt}{$p_{\mathrm{T}}$}
\newcounter{vers}
\newcommand{\ITS}{\rm{ITS}}
\newcommand{\SPD}{\rm{SPD}}
\newcommand{\SDD}{\rm{SDD}}
\newcommand{\SSD}{\rm{SSD}}
\newcommand{\TPC}{\rm{TPC}}

\newcommand{\deta}{$\mathrm{\Delta}\eta$}
\newcommand{\dphi}{$\mathrm{\Delta}\varphi$}

\newcommand{\sigmadeta}{$\sigma_{\mathrm{\Delta}\eta}$}
\newcommand{\sigmadphi}{$\sigma_{\mathrm{\Delta}\varphi}$}

\newcommand{\snnpp}{$\sqrt{s} = 7$~TeV}
\newcommand{\snnpPb}{$\sqrt{s_{\mathrm{NN}}} = 5.02$~TeV}
\newcommand{\snnPbPb}{$\sqrt{s_{\mathrm{NN}}} = 2.76$~TeV}
\newcommand{\ptLow}{$0.2 < p_{\mathrm{T,assoc}} < p_{\mathrm{T,trig}} < 2.0$~GeV/$c$}
\newcommand{\ptMed}{$2.0 < p_{\mathrm{T,assoc}} < 3.0 < p_{\mathrm{T,trig}} < 4.0$~GeV/$c$}
\newcommand{\ptHigh}{$3.0 < p_{\mathrm{T,assoc}} < 8.0 < p_{\mathrm{T,trig}} < 15.0$~GeV/$c$}

\begin{document}%

\begin{titlepage}
\PHyear{2015}
\PHnumber{263}      
\PHdate{17 September}  
%

\title{Multiplicity and transverse momentum evolution of charge-dependent correlations in pp, p--Pb, and Pb--Pb 
collisions at the LHC}
\ShortTitle{Charge-dependent correlations in pp, p--Pb, and Pb--Pb}   

\Collaboration{ALICE Collaboration\thanks{See Appendix~\ref{app:collab} for the list of collaboration members}}
\ShortAuthor{ALICE Collaboration} 

\begin{abstract}
We report on two-particle charge-dependent correlations in pp, p--Pb, and Pb--Pb collisions as a function of the pseudorapidity and azimuthal angle 
difference, $\mathrm{\Delta}\eta$ and $\mathrm{\Delta}\varphi$ respectively. These correlations 
are studied using the balance function that probes the charge creation time
and the development of collectivity in the produced system. 
The dependence of the balance function on the event multiplicity as well as on the trigger and associated particle transverse momentum 
($p_{\mathrm{T}}$) in pp, p-Pb, and Pb--Pb collisions at $\sqrt{s_{\mathrm{NN}}} = 7$, 5.02, and 2.76~TeV, respectively, are presented. In the low transverse momentum region, for $0.2 < p_{\mathrm{T}} < 2.0$~GeV/$c$, the balance function becomes narrower in both $\mathrm{\Delta}\eta$ and $\mathrm{\Delta}\varphi$ directions in all three systems for events with higher multiplicity.
The experimental findings 
favor models that either incorporate some collective behavior (e.g. AMPT) or different mechanisms that lead to effects that resemble collective behavior (e.g. PYTHIA8 with color reconnection). For higher values of transverse momenta the 
balance function becomes even narrower but exhibits no multiplicity dependence, indicating that the observed 
narrowing with increasing multiplicity at low $p_{\mathrm{T}}$ is a feature of bulk particle production.
\end{abstract}
\end{titlepage}
\setcounter{page}{2}

%
%
\section{Introduction}
\label{Sec:Introduction}
Angular correlations between two particles have been established as a powerful tool to study the properties of the system 
created in high energy collisions of hadrons and nuclei 
\cite{Alt:2004gx,Alt:2007hk,Adler:2002tq,Adams:2005ph,Adare:2006nr,Adams:2006yt,Alver:2008aa,Adare:2008ae,Abelev:2009af,Alver:2009id,Adare:2010ry,Aamodt:2011by,Aamodt:2011vg,Chatrchyan:2011eka,Chatrchyan:2012wg,ATLAS:2012at}. 
These measurements are usually performed in a two dimensional space as a function of $\mathrm{\Delta}\eta$ and 
$\mathrm{\Delta}\varphi$. Here \deta~and \dphi~are the differences in pseudorapidity 
$\eta = -\mathrm{ln}\Big[\tan(\theta/2)\Big]$ (where $\theta$ is the polar angle of a particle relative to the beam axis) 
and in azimuthal angle $\varphi$ of the two particles.

In heavy--ion collisions at both the Relativistic Heavy Ion Collider (RHIC) \cite{Adler:2002tq,Adams:2005ph,Adare:2006nr,Adams:2006yt,Alver:2008aa,Adare:2008ae,Abelev:2009af,Alver:2009id,Adare:2010ry} and at the Large Hadron Collider 
(LHC) \cite{Aamodt:2011by,Aamodt:2011vg,Chatrchyan:2011eka,Chatrchyan:2012wg,ATLAS:2012at}, these 
correlations exhibit characteristic structures: (a) a peak at (\deta,\dphi)~\\=~(0,0), usually referred to as the near--side 
jet peak, resulting from intra-jet correlations as well as correlation due to decay of resonances and quantum statistics 
correlations, (b) an elongated structure over $\mathrm{\Delta}\eta$ 
at \dphi~=~$\pi$ originating partially from correlations between particles from 
back--to--back jets and from collective effects such as anisotropic flow, and (c) a similar 
component at \dphi~=~0 extending to large values of \deta, usually called the near--side ridge, whose origin was 
subject of a theoretical debate \cite{Armesto:2004pt,Majumder:2006wi,Chiu:2005ad,Wong:2008yh,Dumitru:2008wn,Gavin:2008ev,Dusling:2009ar,Romatschke:2006bb,Shuryak:2007fu,Voloshin:2004th,Takahashi:2009na,Alver:2010gr,Alver:2010dn,Hama:2009vu,Bozek:2012en}. Although initially the near--side ridge was also attributed to jet--medium interactions \cite{Armesto:2004pt,Majumder:2006wi,Chiu:2005ad,Wong:2008yh}, it is now believed to be associated to the development of collective motion \cite{Romatschke:2006bb,Shuryak:2007fu,Voloshin:2004th,Takahashi:2009na,Alver:2010gr,Alver:2010dn,Hama:2009vu,Bozek:2012en} and to initial state density fluctuations, including the initial state effects within the framework of the Color Glass Condensate (CGC) \cite{Dumitru:2008wn,Gavin:2008ev,Dusling:2009ar}.

Similar structures have recently been reported in two-particle correlation analyses in smaller systems. In particular, the 
CMS Collaboration, by studying angular correlations between two particles in 
\deta~and \dphi, reported the development of an enhancement of correlations on the near--side (i.e. \dphi~=~0) in high- compared to low-multiplicity pp collisions at \snnpp~that 
persists over large values of \deta~\cite{Khachatryan:2010gv}. In the subsequent data taking periods at 
the LHC, similar ridge structures were observed on both the near-- and the away--side in high-multiplicity p--Pb collisions at \snn~= 5.02~TeV 
\cite{CMS:2012qk,Abelev:2012ola,Aad:2012gla,Chatrchyan:2013nka,Aad:2013fja,Abelev:2014mda}. The origin of these effects, 
appearing in small systems, is still debated theoretically. In particular, it was suggested in \cite{Bozek:2010pb,Bozek:2012gr,Bozek:2013uha} that in high-multiplicity collisions the small system develops collective motion during a short hydrodynamic 
expansion phase. On the other hand, in \cite{Dumitru:2010iy,Dusling:2013qoz,Dusling:2012wy} the authors suggested 
that the ridge structure can be understood within the CGC framework. 

The ALICE Collaboration also reported 
a particle mass ordering in the extracted $v_{2}$ (i.e. the second coefficient of the Fourier expansion of the azimuthal distribution of particles 
relative to the symmetry plane) values for $\pi^{\pm}$, $\mathrm{K}^{\pm}$, and p($\mathrm{\overline{p}}$) in high-multiplicity p--Pb collisions \cite{ABELEV:2013wsa}. This mass ordering becomes evident once the correlations 
observed in the lowest multiplicity class are subtracted from the ones recorded in the highest multiplicity class. 
The ordering is less pronounced, yet still present, if this subtraction procedure is not applied. Similar mass ordering 
in Pb--Pb collisions \cite{Abelev:2014pua} is usually attributed to the interplay between radial and elliptic flow induced by 
the collective motion of the system. These observations in p--Pb collisions were reproduced by models 
incorporating a hydrodynamic expansion of the system \cite{Bozek:2013ska,Werner:2013ipa}. Recently, it was suggested in 
\cite{Romatschke:2015dha} that the signatures of collective effects observed in experiments could 
be partially described by models that couple the hot QCD matter created in these small systems, described as an ensemble of non-interacting particles, to a late stage hadronic cascade model. More recently, the CMS Collaboration demonstrated 
that the effects responsible for the observed correlations in high-multiplicity p-Pb events are of multiparticle nature \cite{Khachatryan:2015waa}. This strengthens the picture of the development of collective effects even in these small 
systems.

The charge-dependent part of two-particle correlations is traditionally studied with the balance function (BF) \cite{Bass:2000az}, described in detail in Section~\ref{Sec:BalanceFunction}. Such studies have emerged as a powerful tool to probe the properties of the system created in high energy collisions. 
Particle production is governed by conservation laws, such as local charge conservation. The latter ensures that each 
charged particle is balanced by an oppositely-charged partner, created at the same location in space and time. The 
BF reflects the distribution of balancing charges in momentum 
space. It is argued to be a sensitive probe of both the time when charges are created \cite{Bass:2000az,Jeon:2001ue} and 
of the collective motion of the system \cite{Voloshin:2004th,Bozek:2004dt}. In particular, the width of the balance function is expected to be small in the case of a system consisting of particles that are created close to the end 
of its evolution and are affected by radial flow \cite{Voloshin:2004th,Bass:2000az,Jeon:2001ue,Bozek:2004dt}. On the other 
hand, a wide balance function distribution might signal the creation of balancing charges at the first stages of the system's 
evolution \cite{Voloshin:2004th,Bass:2000az,Jeon:2001ue,Bozek:2004dt} and the reduced contribution or absence of 
radial flow. 

In this article, we extend the previous measurements~\cite{Abelev:2013csa} by reporting results on the balance 
function in pp, p--Pb, and Pb--Pb collisions at $\sqrt{s_{\mathrm{NN}}} = 7$, 5.02, and 2.76~TeV, respectively. The data were recorded 
with the ALICE detector \cite{Carminati:2004fp,Alessandro:2006yt,Aamodt:2008zz}. The results are presented as a 
function of multiplicity and transverse momentum (\pt) to investigate potential scaling properties and 
similarities or differences between the three systems. The article is organized as follows: Section~\ref{Sec:ExpSetup} 
briefly describes the experimental setup, while details about the data sample and the selection criteria are introduced in Section~\ref{Sec:Analysis}. In Section~\ref{Sec:BalanceFunction}, the analysis technique and the applied corrections 
are illustrated. In Section~\ref{Sec:Systematics}, the specifics about the estimation of the systematic uncertainties are described. 
Section~\ref{Sec:Results} discusses the results followed by a detailed comparison with models 
to investigate the influence of different mechanisms (e.g. unrelated to hydrodynamic effects) on the 
balance functions. In the same section, the comparison of the results among the three systems is presented. 

\section{Experimental setup}
\label{Sec:ExpSetup}

ALICE \cite{Aamodt:2008zz} is one of the four major detectors at the LHC. It is designed to efficiently reconstruct and identify particles 
 in the high-particle density environment of central Pb--Pb collisions \cite{Aamodt:2010pb,Aamodt:2010cz}. The experiment 
consists of a number of central barrel detectors positioned inside a solenoidal magnet providing a $0.5$~T field parallel to the beam direction, and a set of forward detectors. The central 
detector systems of ALICE provide full azimuthal coverage for track reconstruction within a pseudorapidity window of $|\eta| < 0.9$. 
The experimental setup is also optimized to provide good momentum resolution (about $1\%$ at $p_{\mathrm{T}}~< 1$~GeV/$c$) and 
particle identification (PID) over a broad momentum range \cite{Abelev:2014ffa}. 

For this analysis, charged particles were reconstructed using the Time Projection Chamber (\TPC) 
\cite{Alme:2010ke} and the Inner Tracking System (\ITS) \cite{Aamodt:2008zz}. The \TPC~is the main 
tracking detector of the central barrel \cite{Alme:2010ke}, consisting of $159$ pad rows grouped into $18$ sectors that cover the full azimuth within $|\eta| < 0.9$. The inner and outer radii of the detector are 85 and 247~cm, respectively. 
The \ITS~consists of six layers of silicon detectors employing three different technologies. The two innermost 
layers, positioned at $r = 3.9$ and 7.6~cm,  are Silicon Pixel Detectors (\SPD), followed by two layers of Silicon Drift Detectors (\SDD) at $r = 15$~cm and 23.9~cm. 
Finally, the two outermost layers are double--sided Silicon Strip Detectors (\SSD) at $r = 38$~cm and 43~cm.

A set of forward detectors, the V0 scintillator arrays~\cite{Abbas:2013taa}, were used in the trigger logic and the multiplicity determination. The V0~consists of two systems, the V0A 
and the V0C, positioned on both sides of the interaction point along the beam. They cover the pseudorapidity ranges 
$2.8 < \eta < 5.1$ and $-3.7 < \eta < -1.7$ for the V0A and the V0C, respectively.

For more details on the ALICE detector setup and its performance in the LHC run 1, see \cite{Aamodt:2008zz,Abelev:2014ffa}.


\section{Analysis details}
\label{Sec:Analysis}

This analysis is based on data from pp, p--Pb, and Pb--Pb collisions. The data were recorded for pp collisions during the 2010 run at \snnpp, for p--Pb collisions during the 2013 run at \snnpPb, and for Pb--Pb collisions during the 2010 and 2011 runs at \snnPbPb. In p--Pb collisions, the nucleon--nucleon 
centre-of-mass system was shifted with respect to the ALICE laboratory system by a rapidity of -0.465 in the direction 
of the proton beam. For simplicity, the pseudorapidity in the laboratory frame is denoted, throughout this article, with $\eta$ for all systems (note that for pp and Pb--Pb collisions the laboratory and the 
centre-of-mass systems coincide).

Minimum-bias p--Pb and Pb--Pb events were triggered by the coincidence between signals from the two sides of the 
V0 detector. For the pp run, the minimum-bias trigger definition was modified to require at least one hit in the \SPD~
or either of the V0 detectors. In addition, for Pb--Pb, an online selection based on the V0 detectors 
was used  to increase the number of events with high multiplicity. An offline event selection exploiting the 
signal arrival time in V0A and V0C, with a 1~ns resolution, was used to discriminate background (e.g.~beam-gas) 
from collision events. This led to a reduction of background events in the analyzed samples to a negligible fraction 
($< 0.1 \%$) for all systems~\cite{Abelev:2014ffa}. All events retained in the analysis had a reconstructed primary vertex position along 
the beam axis ($z_{vtx}$) within 10~cm from the nominal interaction point. Finally, events with multiple reconstructed 
vertices were rejected, leading to a negligible amount of pile-up events for all systems~\cite{Abelev:2014ffa}.

After all the selection criteria, approximately $240~\times$ $10^6$, $100~\times~10^6$, and $35~\times~10^6$ events 
were analyzed for pp, p--Pb, and Pb--Pb, respectively. 

Tracks are reconstructed from a collection of space points (clusters) inside the \TPC. The tracking algorithm, based on the Kalman filter, provides the quality of the fit by calculating its $\chi^2$ value. Each space-point is reconstructed at one of the TPC padrows, where the deposited ionazation energy is also measured. The specific ionization energy loss ($\langle \mathrm{d}E/\mathrm{d}x \rangle$) is estimated by averaging this ionization over all clusters associated to the track. The procedure has an uncertainty, which we later refer to as $\sigma_{\mathrm{d}E/\mathrm{d}x}$.
    
To select primary tracks with high efficiency and to minimize the contribution from background tracks (i.e. secondary 
particles originating either from weak decays or from the interaction of particles with the detector material), all selected tracks 
were required to have at least 70 reconstructed space points out of the maximum of 159 possible in the \TPC. In addition, 
the $\chi^2$ per degree of freedom per TPC space point of the momentum fit was required to be below 
2. To further reduce the contamination from background tracks, only tracks with a distance of closest 
approach (DCA) to the primary vertex in both the $xy$-plane ($\mathrm{DCA}_{\mathrm{xy}}$) 
and the z coordinate ($\mathrm{DCA}_{\mathrm{z}}$) below a threshold value (i.e. $\mathrm{DCA}_{\mathrm{xy}} < 2.4$~cm and $\mathrm{DCA}_{\mathrm{z}} < 3.0$~cm) were analyzed. These requirements lead to a reconstruction efficiency 
of about $80\%$ for primary particles and a contamination from secondaries of about $5\%$ at $p_{\mathrm{T}} = 1$~GeV/$c$ 
\cite{Abelev:2013bla} in pp collisions. The efficiency is similar in p--Pb collisions and it is lower by about 3--5$\%$ in central Pb--Pb collisions, according to detailed Monte Carlo simulations. In addition, 
electrons originating from $\gamma$-conversion and $\pi^0$--Dalitz decays were removed based 
on the energy loss $(\mathrm{d}E/\mathrm{d}x)$ measured by the TPC. Tracks for which the measured $\mathrm{d}E/\mathrm{d}x$ lied within $3\sigma_{\mathrm{d}E/\mathrm{d}x}$ of the Bethe-Bloch parametrization of $\langle \mathrm{d}E/\mathrm{d}x \rangle$ for electrons and at least $3\sigma_{\mathrm{d}E/\mathrm{d}x}$ away from the relevant parametrizations for pions, kaons, and protons, were removed.

All particles were reconstructed within $|\eta| < 0.8$. This selection excludes possible biases from the tracking efficiency that becomes lower for 
$|\eta| > 0.8$ as compared to $|\eta| < 0.8$. The particles  selected in this analysis have a transverse momentum 
in the range $0.2 < p_{\mathrm{T}} < 15.0$~GeV/$c$.

In order to reduce the contribution from track splitting (i.e. incorrect reconstruction of a signal produced by one track 
as two tracks) and merging (i.e. two nearby tracks being reconstructed as one track) in the active 
volume of the TPC, a selection based on the closest distance of two tracks in the \TPC~volume was applied when forming 
particle pairs. This was done by excluding pairs with a minimum pseudorapidity difference of $|\mathrm{\Delta}\eta|<0.02$ 
and angular distance $|\mathrm{\Delta}\varphi^{*}|<0.02$~rad. Here $\mathrm{\Delta}\varphi^*$ is the angular distance 
between two tracks, accounting also for their curvature due to their charge, according to:

\begin{equation}
\mathrm{\Delta}\varphi^{*}=\varphi_{1}-\varphi_{2}-\alpha_1+\alpha_2,
\label{eq:hbt}
\end{equation}

\noindent where $\varphi_{1}$ and $\varphi_{2}$ are the azimuthal angles of the two tracks at the vertex, and $\alpha_{i}$ (with $i = 1,2$) is given by 

\begin{equation}
\alpha_{i} = q_i\left| \arcsin \left(\frac{0.0075B_{z}(\mathrm{T}) r(\mathrm{cm})}{p_{\mathrm{T}i}(\mathrm{GeV/}c)}\right)\right| 
\label{eq:alpha}
\end{equation}

\noindent In Eq.~\ref{eq:alpha}, $q_{1}$ and $q_{2}$ stand for the charge of each track, $B_{z}$ is the magnetic field in the $z$ direction, $r$ corresponds to the radius of the smallest distance of the tracks in the detector used ($0.8 < r < 2.5$~m with a step of $\Delta r = 0.2$~cm, for the TPC) and $p_{\mathrm{T}1}$ 
and $p_{\mathrm{T}2}$ are the transverse momentum values of the two particles forming the pair.

\subsection{\textbf{Multiplicity classes in pp, p--Pb, and Pb--Pb collisions}}
\label{Sec:Multiplicity}

The analyzed events were divided into multiplicity classes using the V0A detector. 
Since this detector does not provide any tracking information, the amplitude of the signal from each cell, which is 
proportional to the number of particles that hit a cell, was used as a proxy for multiplicity~\cite{Adam:2014qja}. The choice of the V0A 
as the default multiplicity estimator was driven by the fact that in p--Pb collisions\footnote{Note that ALICE also recorded Pb--p collisions but this sample was smaller than the one analysed and reported in this article.} this detector is located in the 
direction of the Pb--ion and thus is sensitive to its fragmentation~\cite{Adam:2014qja}. In addition, this choice allowed for reducing autocorrelation biases introduced when the multiplicity class was estimated in the same $\eta$ range as the one used to measure correlations. For consistency, the same multiplicity estimator was used for the 
other two systems. For the V0 detectors, a calibration procedure~\cite{Abelev:2014ffa,Abbas:2013taa} (i.e. gain 
equalization) was performed to account for fluctuations induced by the hardware performance, and for the different 
conditions of the LHC machine for each running period. 

For each multiplicity class, the raw transverse momentum spectrum for charged particles with 
$p_{\mathrm{T}} > 0.2$ GeV/$c$ reconstructed in $|\eta| < 0.8$ was extracted. These raw spectra were
corrected for detector acceptance and efficiency using Monte Carlo simulations with PYTHIA \cite{Sjostrand:2006za}, 
DPMJET~\cite{Ranft:1999qe}, and HIJING~\cite{Wang:1991hta} event generators for pp, p--Pb, and Pb--Pb, respectively. The ALICE detector response for 
these events was determined using a GEANT3~\cite{Brun:1994aa} simulation. In addition to the reconstruction 
efficiency, a correction related to the contamination from secondaries originating from weak decays and from the interaction 
of particles with the material of the detector was applied. This correction was estimated with both the aforementioned simulations and also using a data-driven method, based on fitting the DCA distributions with templates extracted from Monte Carlo for primary particles and secondaries originating either from weak decays or from the interaction of other particles with the detector material, as described in~\cite{Aamodt:2010dx}. The resulting 
corrected charged-particle multiplicity was calculated by integrating the corrected transverse momentum spectrum over the region with 
$p_{\mathrm{T}} > 0.2$~GeV/$c$.

Table \ref{Table:correctedMultiplicity} presents the multiplicity classes in terms of percentage of the multiplicity distribution, and the corresponding number of charged particles with \pt $> 0.2$ GeV/$c$ reconstructed at $|\eta| < 0.8$ for all three 
systems. The resulting values for $\mathrm{N}_{\mathrm{charged}}$ are subject to an overall tracking efficiency uncertainty of 
$4\%$~\cite{Abelev:2013haa}.

\begin{table}[ht]
\centering
\begin{tabular*}{140mm}{@{\extracolsep{\fill}}lcccc}
\hline
Multiplicity classes & \multicolumn{3}{c}{$\langle\mathrm{N}_{\mathrm{charged}} \rangle$ (corrected)} \\
 & pp & p--Pb & Pb--Pb \\
\hline
\hline
70--80$\%$ & $ 4.1 \pm 0.2   $ & $	11.2	\pm	0.4	$ & $ 45 \pm 2   $\\
60--70$\%$ & $ 5.0 \pm 0.2   $ & $	16.3	\pm	0.7	$ & $ 103 \pm 4 $\\
50--60$\%$ & $ 6.1 \pm 0.3   $ & $	18.5	\pm	0.7	$ & $ 204 \pm 8 $\\
40--50$\%$ & $ 7.4 \pm 0.3   $ & $	24.1	\pm	1.0	$ & $ 364 \pm 15 $\\
30--40$\%$ & $ 9.0 \pm 0.4   $ & $	29.0	\pm	1.2	$ & $ 603 \pm 24 $\\
20--30$\%$ & $ 11.0 \pm 0.4 $ & $	34.7	\pm	1.4	$ & $ 943 \pm 38 $\\ 
10--20$\%$ & $ 13.8 \pm 0.6 $ & $	41.9	\pm	1.7	$ & $ 1419 \pm 57 $\\
0--10$\%$ & $ 18.7 \pm 0.8   $ & $	56.3	\pm	2.3	$ & - \\
5--10$\%$ & - & - & $ 1918 \pm 77 $ \\
0--5$\%$ & - & - & $ 2373 \pm 95 $ \\

\hline
\end{tabular*}
\caption{Corrected mean charged particle multiplicities (for \pt $> 0.2$
  GeV/$c$, and $|\eta| < 0.8$) for event classes defined by the percentage of the V0A
  multiplicity distribution for pp, p--Pb, and Pb--Pb collisions at $\sqrt{s_{\mathrm{NN}}} = 7$, 5.02, and 2.76~TeV, respectively.}
\label{Table:correctedMultiplicity}
\end{table}


\section{Balance function}
\label{Sec:BalanceFunction}

The charge-dependent correlations are studied using the balance function~\cite{Bass:2000az} for pairs of 
charged particles with angular differences \deta~and \dphi. For each pair, the first (``trigger'') particle has a transverse 
momentum $p_{\mathrm{T,trig}}$, while the second (``associated'') charged particle has a transverse momentum 
$p_{\mathrm{T,assoc}}$.

The associated yield per trigger particle is then calculated for different charge combinations. For one charge combination (+,-), it is defined as

\begin{equation}
c_{(+,-)} = \frac{1}{N_{\mathrm{trig},+}}\frac{\mathrm{d}^2N_{\mathrm{assoc},-}}{\mathrm{d}\Delta\eta \mathrm{d}\Delta\varphi} = S_{(+,-)}/f_{(+,-)}
\label{Eq:perTriggerYield}
\end{equation}

\noindent and similarly for the other charge combinations. The signal $S_{(+,-)}= 1/N_{\mathrm{trig},+}\mathrm{d}^2N_{\mathrm{same},(+,-)}/\mathrm{d}\Delta\eta\mathrm{d}\Delta\varphi$ is constructed from the number of positive trigger particles $N_{\mathrm{trig},+}$ and the particle pair distribution \\$\mathrm{d}^2N_{\mathrm{same},(+,-)}/\mathrm{d}\Delta\eta\mathrm{d}\Delta\varphi$, formed in \deta-\dphi~with positive and negative particles from the same event. Both terms are corrected for detector inefficiencies and contamination from secondary particles on a track-by-track basis, using the corrections described in Section~\ref{Sec:Multiplicity} as an inverse weight. $S_{(+,-)}$ is computed after summing separately over all events the two components $N_{\mathrm{trig},+}$ and $\mathrm{d}^2N_{\mathrm{same},(+,-)}/\mathrm{d}\Delta\eta\mathrm{d}\Delta\varphi$.

The background distribution $f_{(+,-)}= \alpha \mathrm{d}^2N_{mixed,+-}/\mathrm{d}\Delta\eta\mathrm{d}\Delta\varphi$ corrects for particle 
pair-acceptance. It is constructed by combining a trigger particle from one event with associated particles from other events. This procedure is known as the event mixing technique. These mixed pairs are formed from events having the same multiplicity 
classes and $z_{vtx}$ within $\pm 2$~cm of each other. Each trigger particle is mixed with associated particles from at least 5 events. The 
coefficient $\alpha$ in Eq.~\ref{Eq:perTriggerYield} is used to normalize the mixed-event distribution to unity in the $\Delta\eta$ region of maximal pair acceptance. Finally, the associated yield per trigger particle is computed by calculating the weighted-average of the corresponding yields for several intervals of $V_z$. This is done to account for the different pair acceptance and efficiency as a function of $V_z$.

The balance function is then defined as the difference of the associated yields per trigger particle for unlike and 
like-sign combinations~\cite{Bass:2000az}, according to

\begin{equation} 
B(\mathrm{\Delta}\eta,\mathrm{\Delta}\varphi) = \frac{1}{2} \Big[ c_{(+,-)} + c_{(-,+)} - c_{(+,+)} - c_{(-,-)} \Big]
\label{Eq:bfDefinition}
\end{equation}

The resulting two-dimensional distributions are projected separately onto $\mathrm{\Delta} \eta$ and $\mathrm{\Delta} \varphi$ 
and the widths, $\sigma_{\mathrm{\Delta} \eta}$ and $\sigma_{\mathrm{\Delta} \varphi}$, are calculated as the 
standard deviation of the distributions. In this analysis, the projection in \deta~is done on the near-- 
($-\pi/2 < \mathrm{\Delta}\varphi < \pi/2$) and on the away--side ($\pi/2<\Delta\varphi<3\pi/2$), separately. 

Three transverse momentum intervals are used in the analysis: the 
low (\ptLow), intermediate (\ptMed), and high (\ptHigh) \pt~regions. Note that the integral of the balance function reported in this article does not reach unity but rather 0.5 due to the requirement imposed on the $p_{\mathrm{T}}$ of the ``trigger'' and the ``associated'' particles.

For \ptLow, the width in $\mathrm{\Delta}\eta$ and $\mathrm{\Delta}\varphi$ is calculated in $|\mathrm{\Delta}\eta|<1.6$ 
and $-\pi/2<\mathrm{\Delta}\varphi<\pi/2$. For higher values of transverse momentum, the balance function distributions 
are fitted with a sum of a Gaussian and a constant. The width is then calculated within $3\sigma_{\mathrm{Gauss}}$, with 
$\sigma_{\mathrm{Gauss}}$ extracted from the Gaussian of the aforementioned fit. The statistical error of the width 
is calculated using the subsample method~\cite{Politis:SubSample1,Book:SubSample2}. The values of $\sigma_{\mathrm{\Delta} \eta}$ and $\sigma_{\mathrm{\Delta} \varphi}$ are calculated for each subsample (maximum 10 subsamples were used) and the statistical uncertainty is estimated from the spread of these independent results.


\begin{table}[th]
\centering
\begin{tabular*}{140mm}{@{\extracolsep{\fill}}lcccc}
\hline
Category & \multicolumn{3}{c}{Systematic uncertainty (max. value) } \\
 & pp & p--Pb & Pb--Pb \\
\hline
\hline
Magnetic field & - & - & $ 1.5 \%$ \\
\hline
LHC periods & $ 1.1 \%$  & $< 0.1\%$ & $1.0\%$ \\ 
\hline
Tracking & $ 1.2 \%$ & $0.2 \%$ & $1.2 \%$   \\ 
\hline
V0 equalization & $ < 0.1 \%$ & - & - \\ 
\hline
Electron variation& $ < 0.1 \%$ & $0.1 \%$ & $0.2 \%$ \\
\hline
Split/merged pairs variation & $ < 0.1 \%$ & $0.2 \%$ & $0.7 \%$ \\
\hline
Efficiency and contamination correction & $0.4 \%$ & $0.4 \%$ & $1.1\%$\\
\hline
\end{tabular*}
\caption{The maximum value of the systematic uncertainties on the width of the balance 
function over all multiplicity classes for each of the sources studied for the pp, p--Pb and Pb--Pb systems.}
\label{tab:systematic}
\end{table}

\section{Systematic uncertainty}
\label{Sec:Systematics}

\begin{figure}[th!]
\centering
\includegraphics[width=\textwidth]{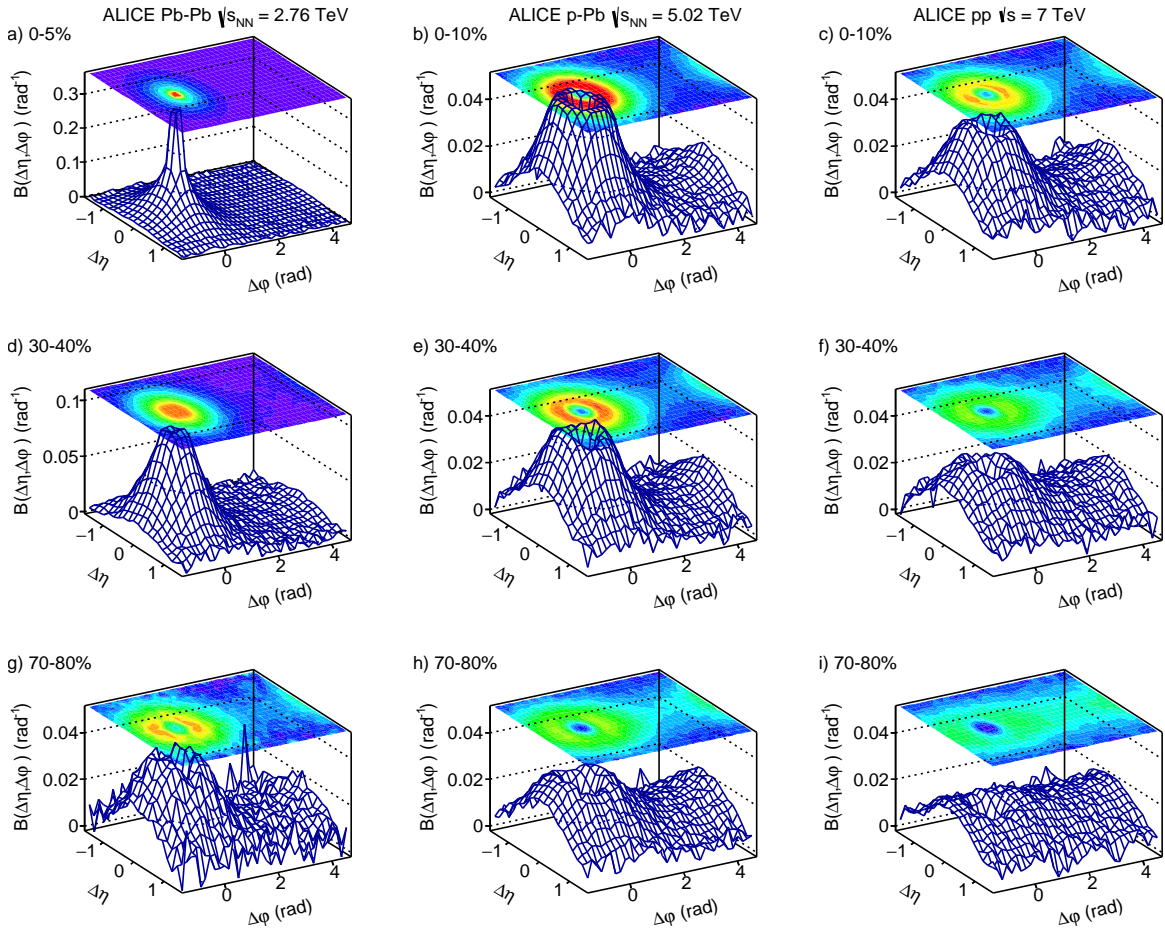}
\caption{The balance function $B(\mathrm{\Delta}\eta,\mathrm{\Delta}\varphi$) for charged particles with \ptLow~in Pb--Pb, p--Pb, and pp collisions at $\sqrt{s_{\mathrm{NN}}} = 2.76$, 5.02, and 7~TeV, respectively. From top to bottom the $0-5\%$ for Pb--Pb and $0-10\%$ for p--Pb and pp collisions, $30-40\%$, and the 
$70-80\%$ multiplicity classes are shown.}
\label{fig:bf2d}
\end{figure}

In all figures except Fig.~\ref{fig:bf2d}, the data points are plotted with their statistical and systematic uncertainties 
indicated by error bars and open boxes around each point, respectively. The systematic 
uncertainty was obtained by varying the event, track, and pair selection criteria, as will be 
explained in the following paragraphs. The contribution of each source was calculated as the spread of the 
values of each data point, extracted from 
variations of the selection criteria. If statistically significant, each contribution was added in quadrature to obtain the final 
systematic uncertainty. Following this procedure, the resulting maximum values of the systematic uncertainty over all multiplicity classes and systems 
for the balance function projections in $\mathrm{\Delta} \eta$ and $\mathrm{\Delta} \varphi$ were less than 5$\%$. 
In what follows, we report the maximum systematic uncertainties over all multiplicity classes for each system for $\sigma_{\mathrm{\Delta} \eta}$ and $\sigma_{\mathrm{\Delta} \varphi}$.

The Pb--Pb data samples were analyzed separately for two magnetic field configurations. 
The difference of $1.5\%$ in the results was taken as a systematic uncertainty. For all systems, different LHC 
periods, reflecting different machine conditions and detector configurations (e.g. non-working channels), were 
analyzed separately. The corresponding maximum systematic uncertainties over all multiplicity classes was $1.1\%$. 
Furthermore, the influence on the results of different tracking strategies was studied by repeating the analysis 
using tracks reconstructed by the combination of signals from the \TPC~and the \ITS. The relevant maximum systematic uncertainties from this 
source were $1.2\%$, $0.2\%$, and $1.2\%$ for pp, p--Pb, and Pb--Pb, respectively. Finally, the contribution coming 
from the V0 gain equalization in pp collisions was investigated by equalizing the signal per V0 ring, per channel, and per detector. The study did not reveal any systematic differences in the obtained results.

In addition, several of the track quality criteria defined by the tracking algorithm described in Section~\ref{Sec:Analysis} were varied. The uncertainty related to the 
electron rejection criterium was studied by varying the requirement on the expected Bethe--Bloch 
parameterization of the momentum dependence of $\mathrm{d}E/\mathrm{d}x$ for electrons from $3\sigma$ 
to $5\sigma$. This contribution was negligible in the pp system, while it was $0.1\%$ and $0.2\%$ for p--Pb 
and Pb--Pb, respectively. The requirement on the closest 
distance of two tracks of a pair in the TPC was varied from 
$\mathrm{\Delta}\eta = 0.01$ to $\mathrm{\Delta}\eta = 0.03$ and from $\mathrm{\Delta}\varphi^{*} = 0.01$ rad to 
$\mathrm{\Delta}\varphi^{*} = 0.03$ rad. This source was found to yield negligible systematic uncertainty for the 
pp system, while the maximum contribution for p--Pb and Pb--Pb systems were $0.2\%$ and $0.7\%$, respectively. 
The systematic uncertainty of the track-by-track correction for efficiency and contamination was estimated from 
Monte Carlo simulations. For this, the results of the analysis of a sample at the event generator level (i.e. without 
invoking either the detector geometry or the reconstruction algorithm) were compared with the results of the analysis over the 
output of the full reconstruction chain, using the corrections for detector inefficiencies and acceptance discussed in Section~\ref{Sec:Analysis}. This source resulted into a partially correlated uncertainty of around $0.4\%$ for the case of pp and p--Pb, and $1.1\%$ 
for the Pb--Pb system.

The resulting values for the systematics are summarized in Table~\ref{tab:systematic}, for all systems. The table provides the maximum 
value for every source over all multiplicity classes and transverse momentum ranges.

Finally, different multiplicity estimators were used to study the variations coming from the multiplicity class definition. There was 
no systematic uncertainty assigned for this contribution. The results obtained with the two forward detectors 
(e.g. V0A and V0C) show no significant difference. On the other hand, a slightly weaker narrowing of the balance function with increasing multiplicity is observed when 
the central barrel detector is used for both measuring the correlations and the multiplicity class definition, in the pp and p--Pb systems. These differences 
are coming from physics processes (e.g. back--to--back jets), whose contribution is reduced 
if one defines multiplicity classes using a detector located further away from mid-rapidity. This also justifies the reason why the V0A detector was chosen as the multiplicity estimator in this analysis


\begin{figure}[th!]
\centering
\includegraphics[width=0.32\textwidth]{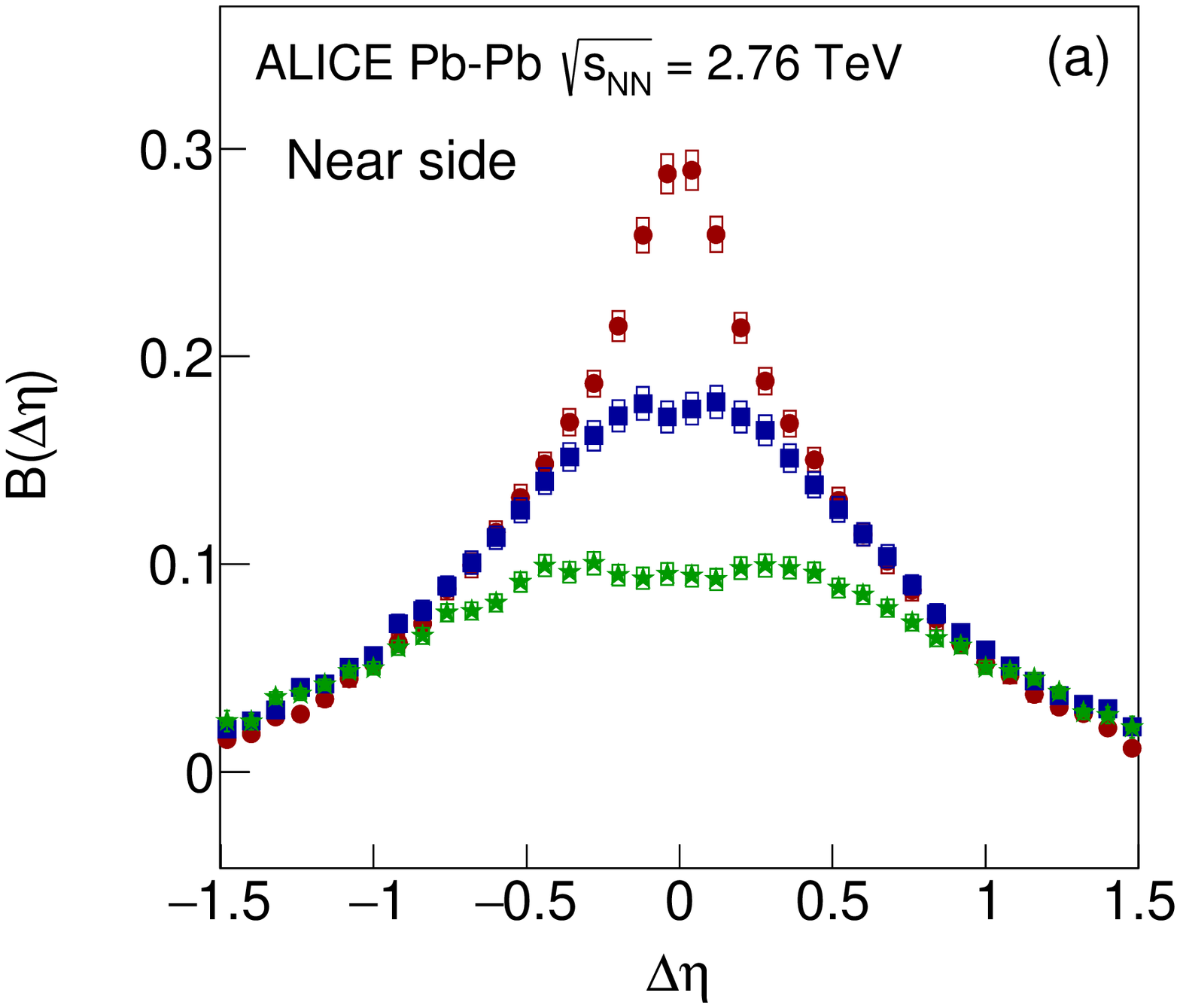}
\includegraphics[width=0.32\textwidth]{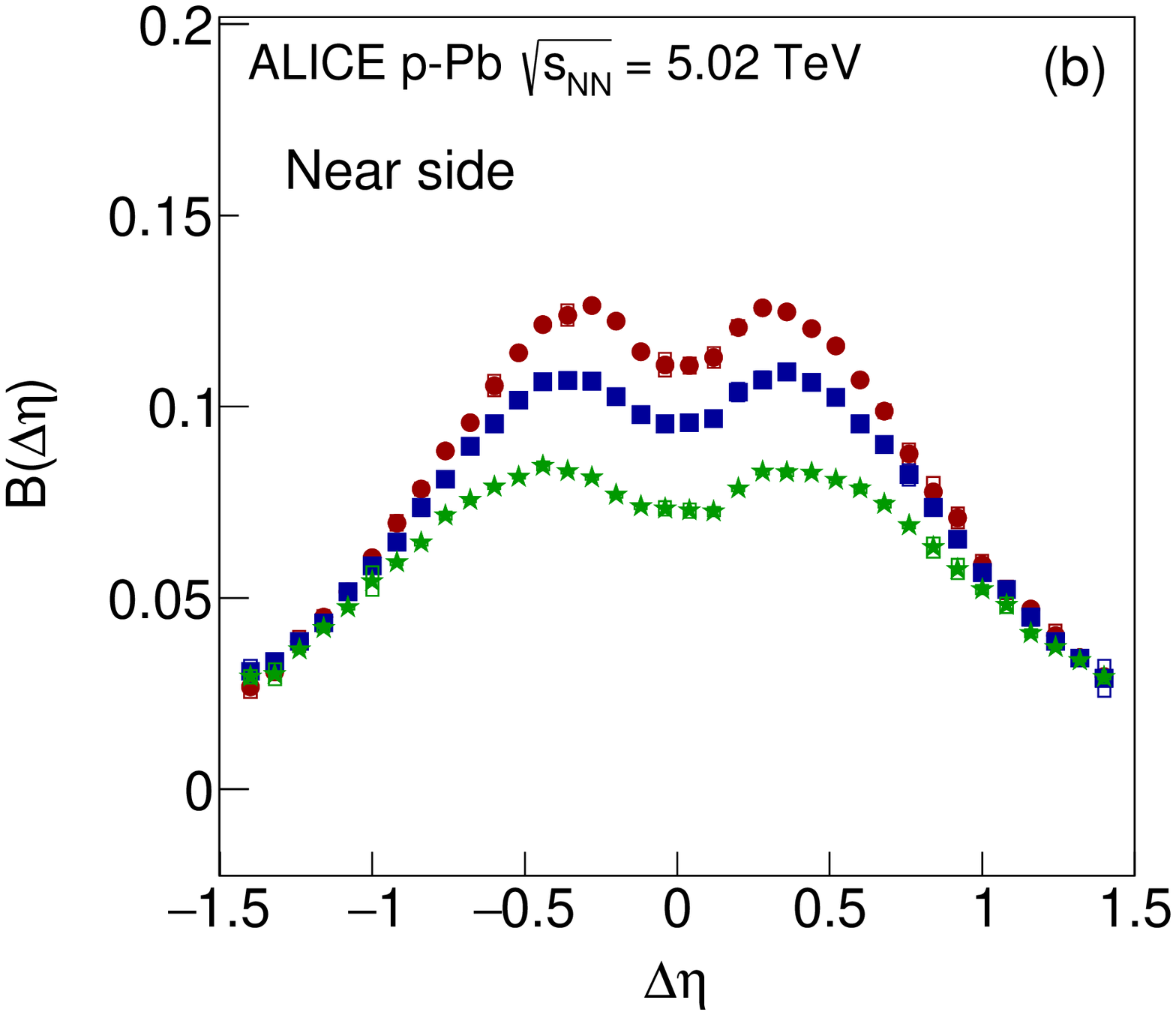}
\includegraphics[width=0.32\textwidth]{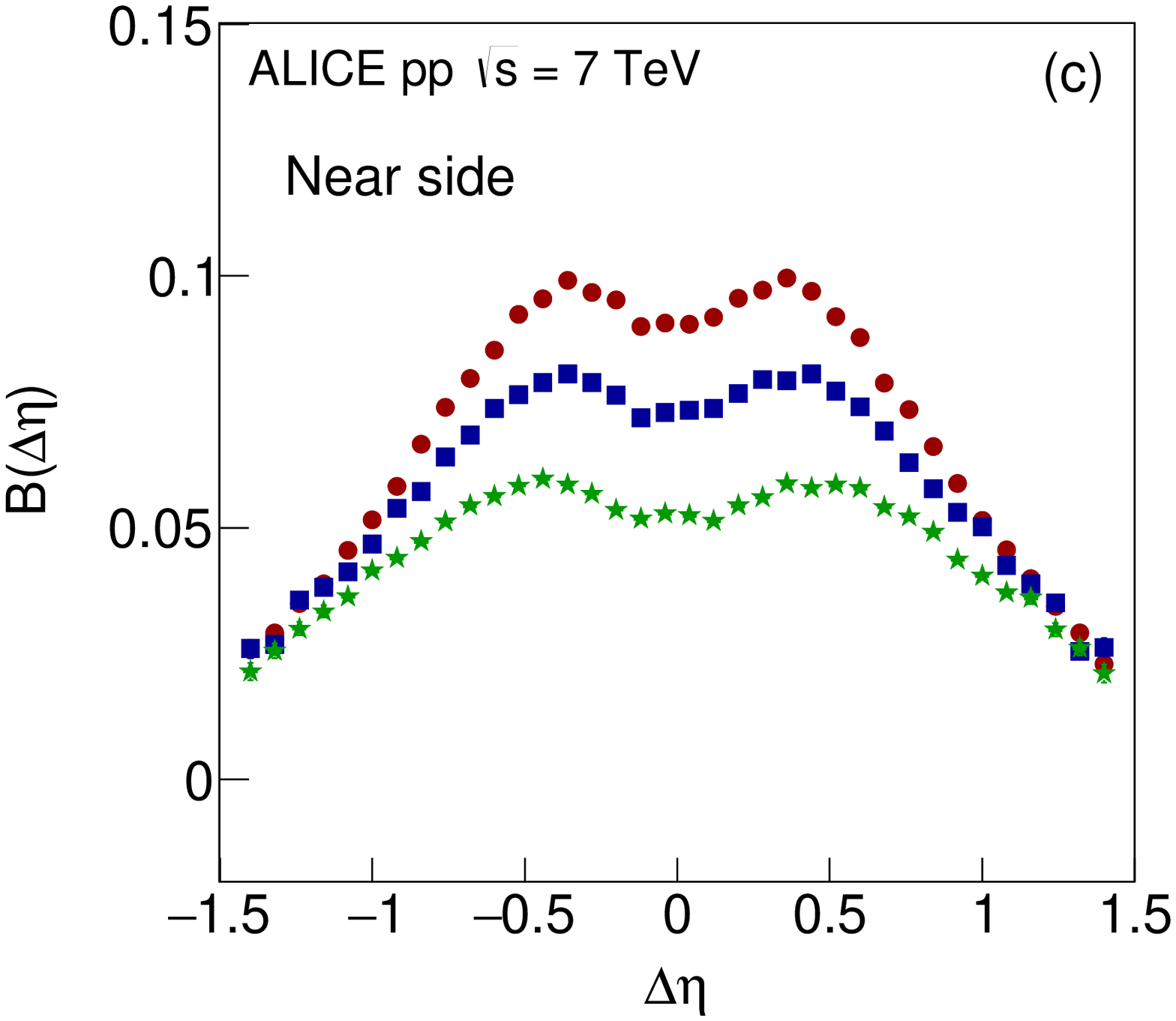}
\includegraphics[width=0.32\textwidth]{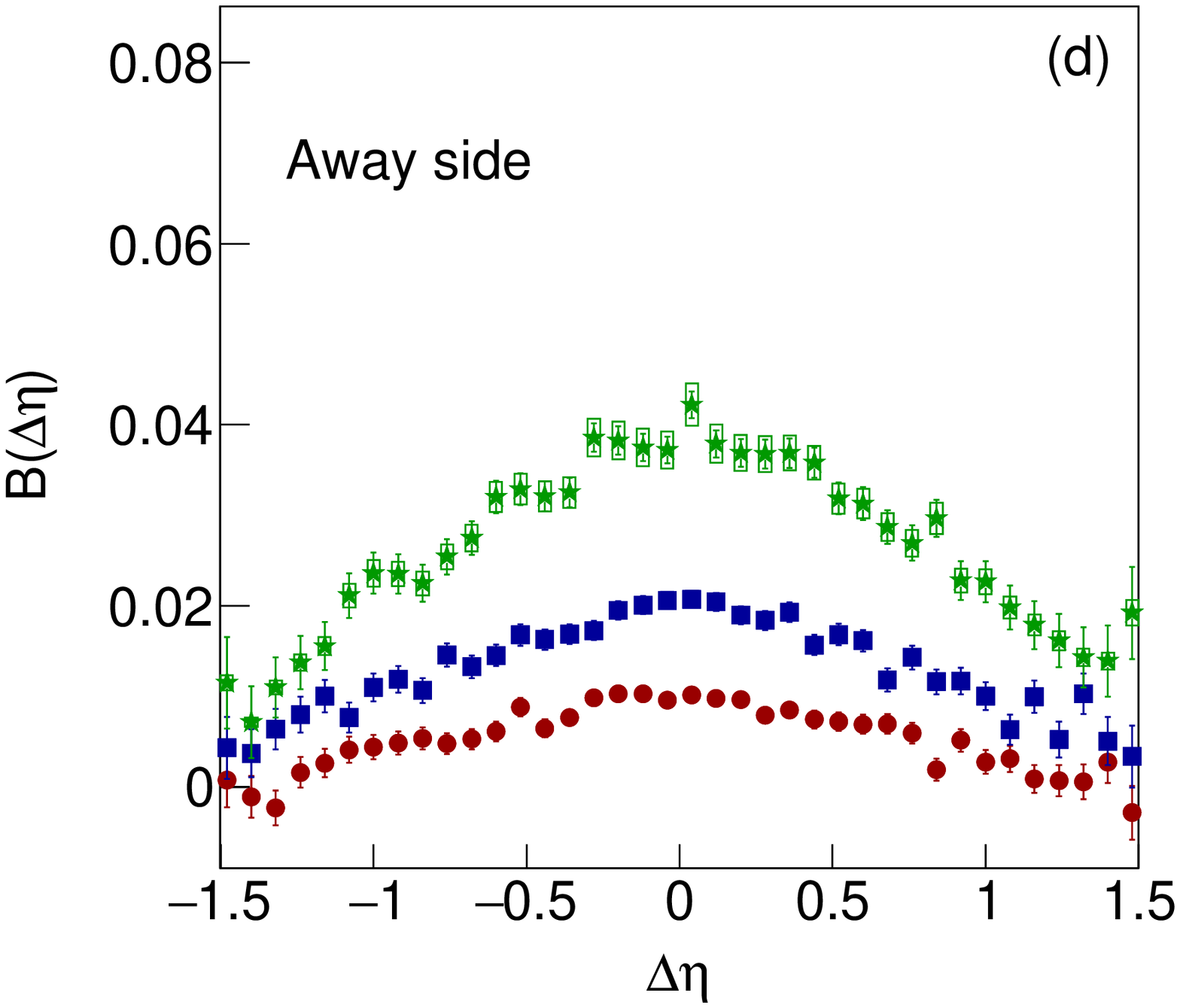}
\includegraphics[width=0.32\textwidth]{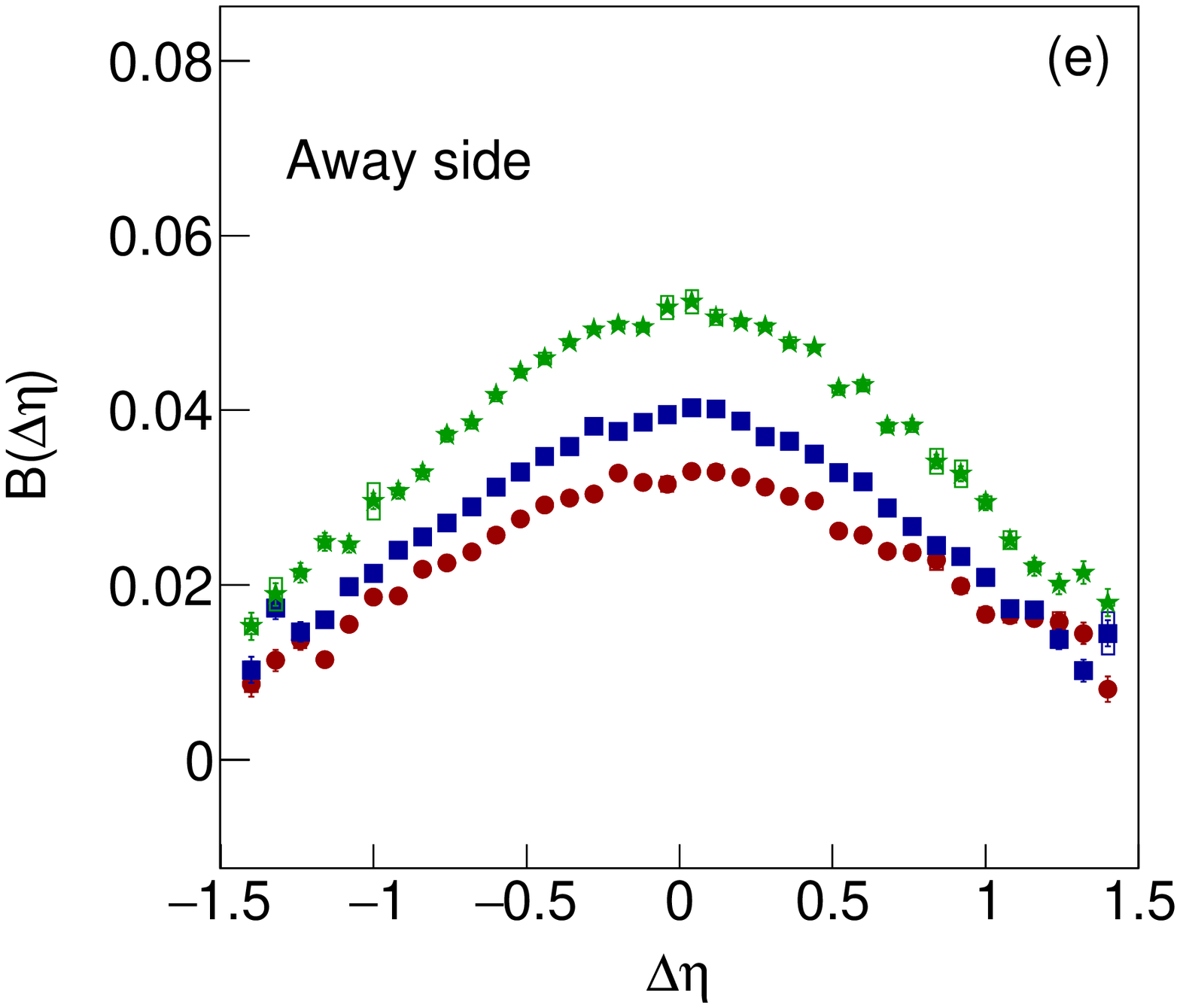}
\includegraphics[width=0.32\textwidth]{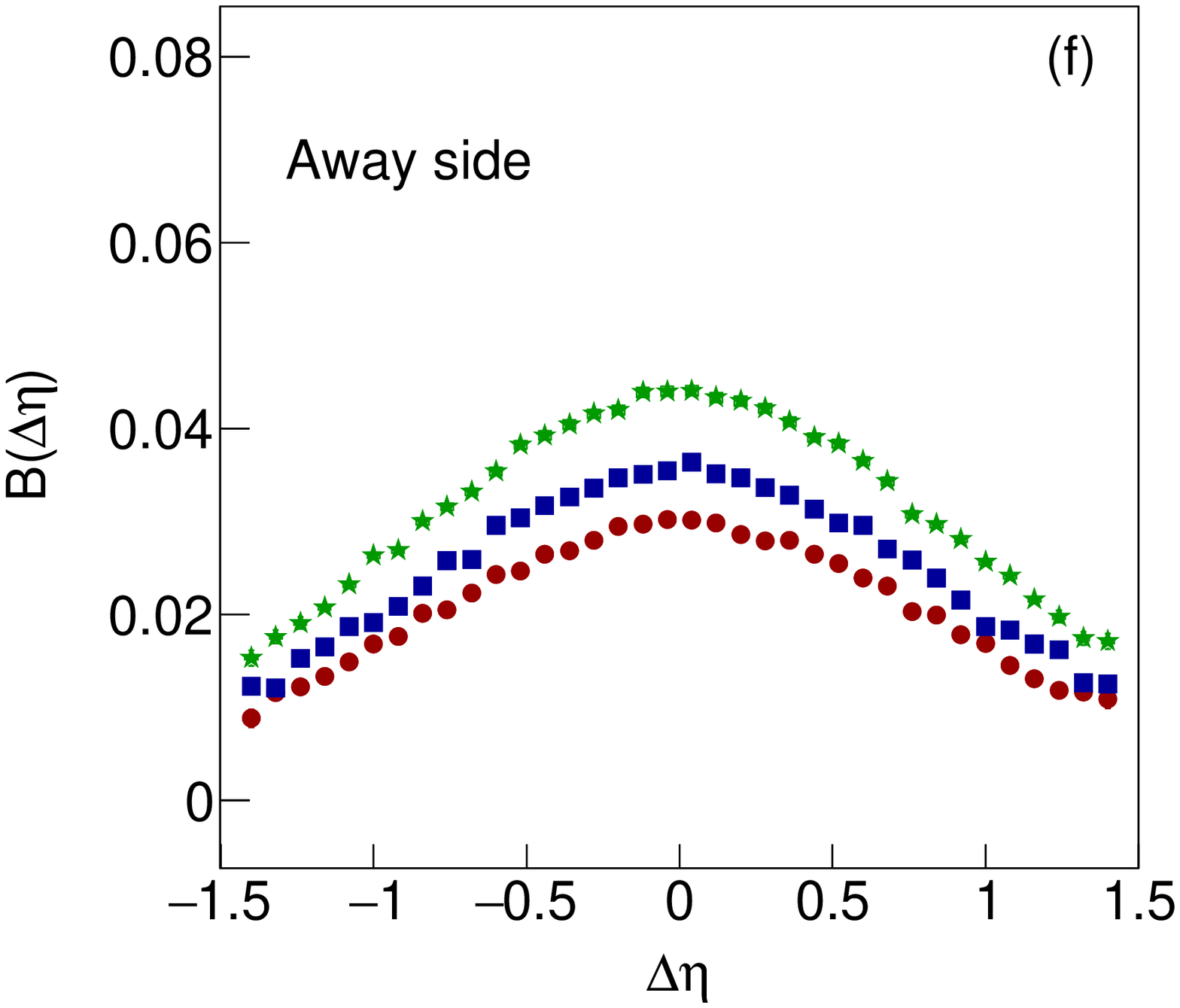}
\includegraphics[width=0.32\textwidth]{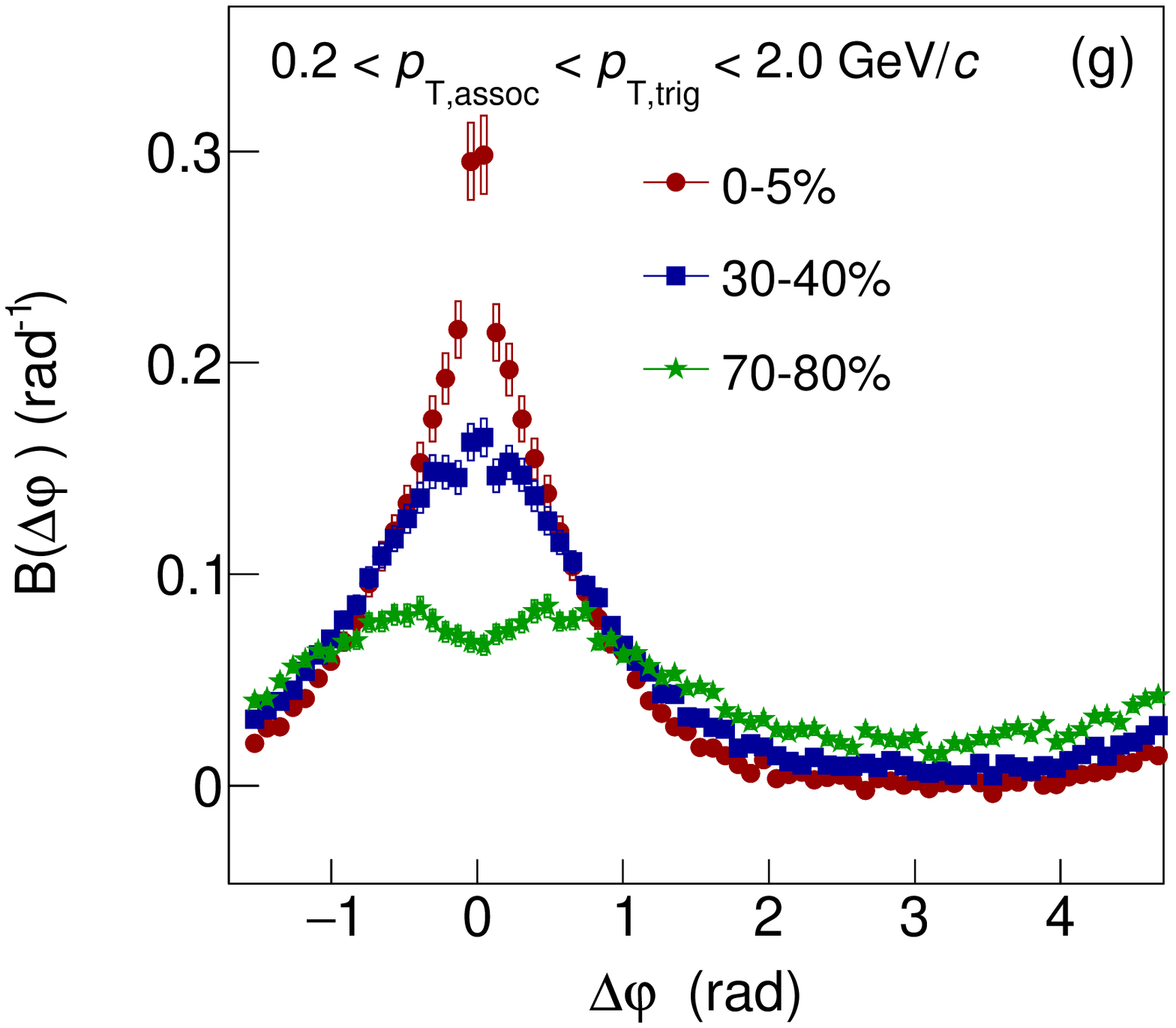}
\includegraphics[width=0.32\textwidth]{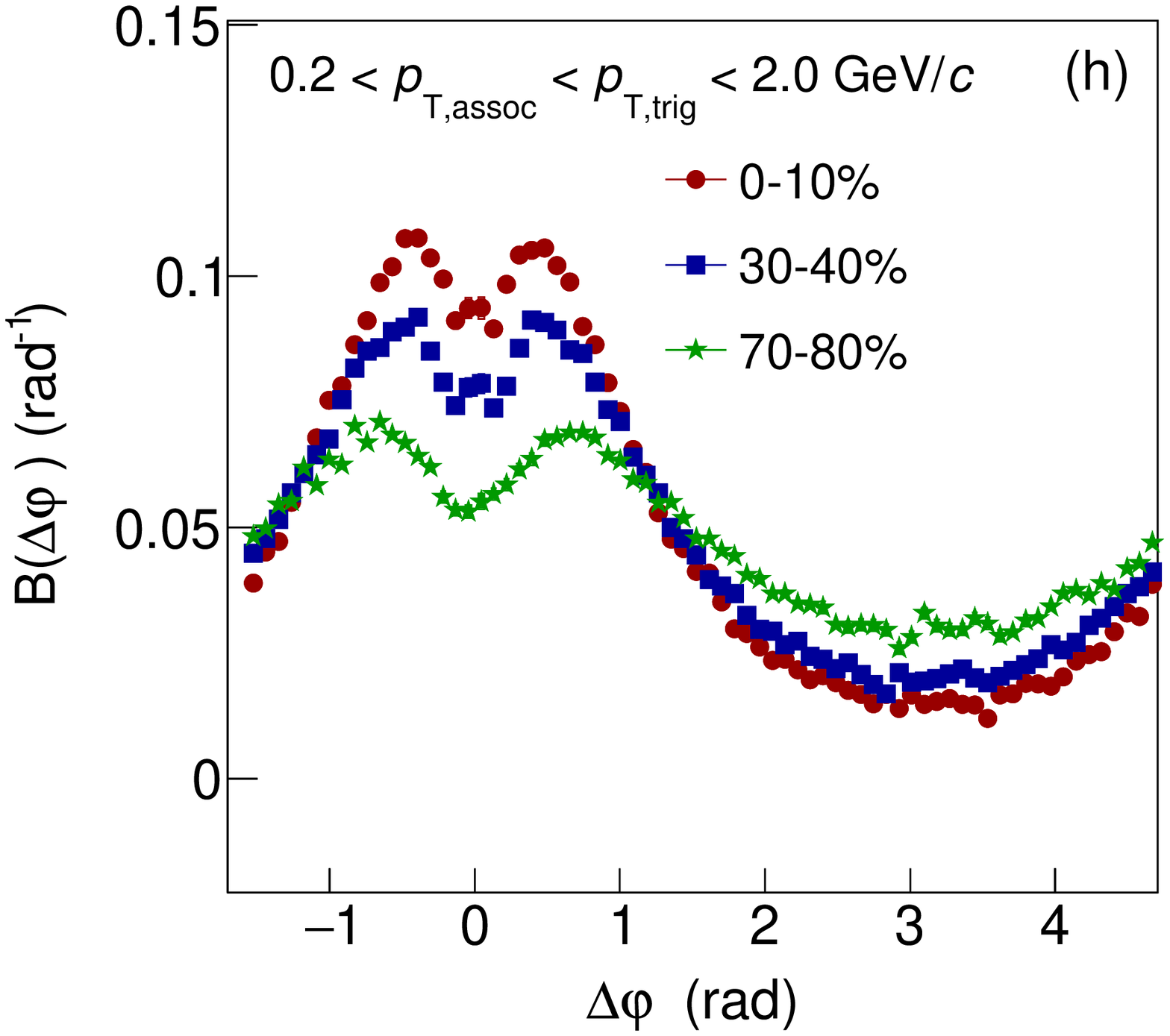}
\includegraphics[width=0.32\textwidth]{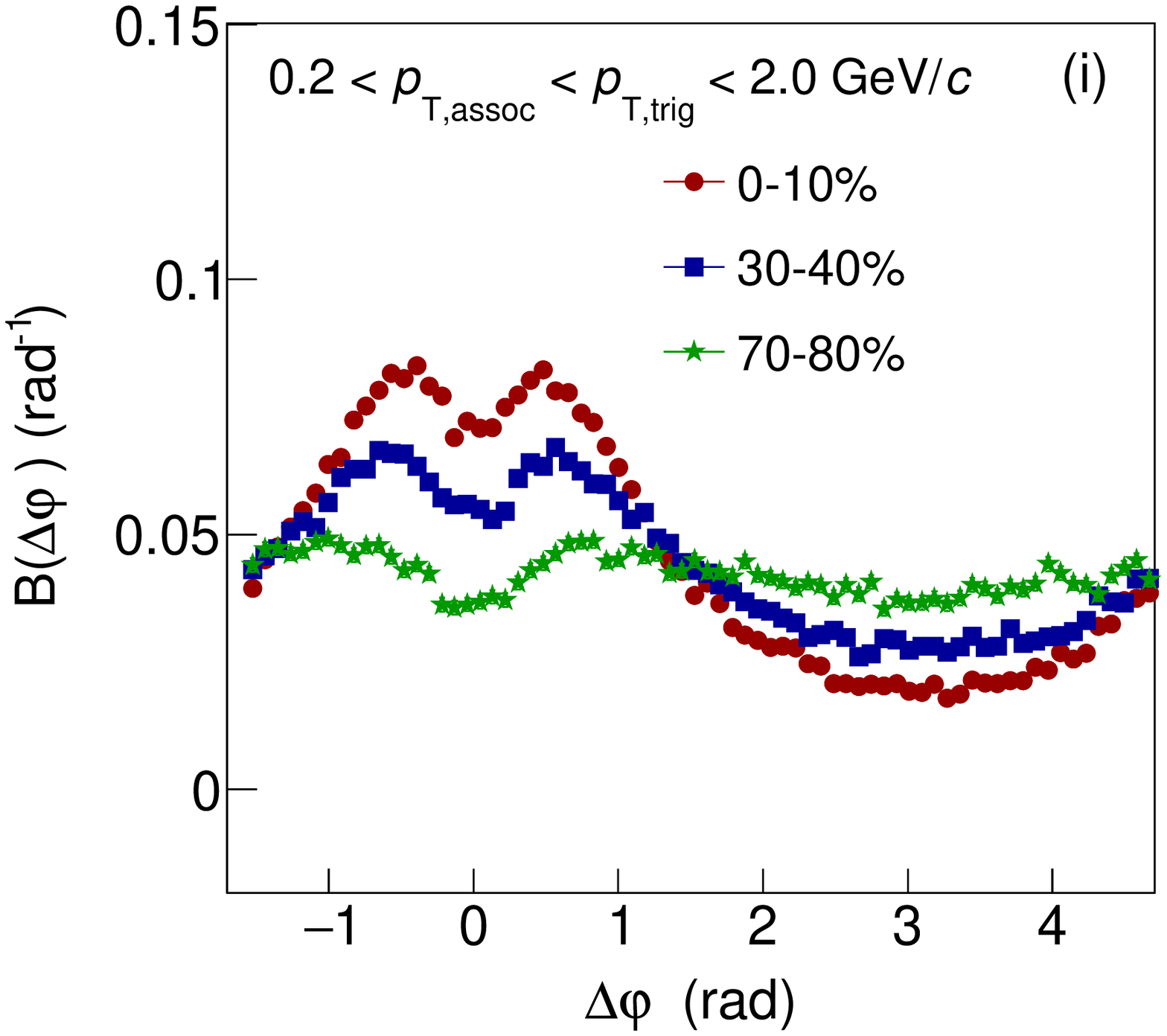}
\caption{The balance function for charged particles with \ptLow~as a function of \deta~on the near--side (upper row) and away--side (middle row) and \dphi~(lower row) in 
different multiplicity classes of Pb--Pb in panels (a), (d) and (g), p--Pb in panels (b), (e) and (h), and pp collisions in panels (c), (f) and (i) at $\sqrt{s_{\mathrm{NN}}} = 2.76$, 5.02, and 7~TeV, respectively.}
\label{fig:bfIn1dLowpT}
\end{figure}

\section{Results}
\label{Sec:Results}
\subsection{\textbf{Balance function in the low transverse momentum region}}
Figure \ref{fig:bf2d} presents the balance function for charged particles in \deta~and 
\dphi~for three multiplicity classes of Pb--Pb, p--Pb, and pp collisions at $\sqrt{s_{\mathrm{NN}}} = 2.76$, 5.02, and 7~TeV, respectively. From top 
to bottom the results for the highest (i.e. 0-5$\%$ for Pb--Pb collisions and 0-10$\%$ for p--Pb and pp collisions), intermediate (i.e. 30-40$\%$), and lowest (i.e. 70-80$\%$) multiplicity classes are shown. The trigger and associated particles are selected from the low 
transverse momentum region \ptLow. The bulk of the charge-dependent correlation yield is located on the 
near--side ($-\pi/2<\Delta\varphi<\pi/2$). In this region, the balance function becomes 
narrower with increasing multiplicity for all three collision systems. The peak values of the balance function 
also change with multiplicity, with higher values corresponding to collisions with higher multiplicity. On the away--side ($\pi/2 < \mathrm{\Delta}\varphi < 3\pi/2$), the balance function has a larger magnitude 
for lower multiplicity events. In addition, a depletion in the correlation pattern around 
$(\mathrm{\Delta}\eta,\mathrm{\Delta}\varphi)=(0,0)$ 
starts to emerge in mid-central (e.g. 30-40$\%$ multiplicity class) events in 
Pb--Pb collisions and becomes more pronounced in p--Pb and pp collisions with decreasing multiplicity. The origin of this structure will be discussed later.

\begin{figure}[th!]
\centering
\includegraphics[width=0.32\textwidth]{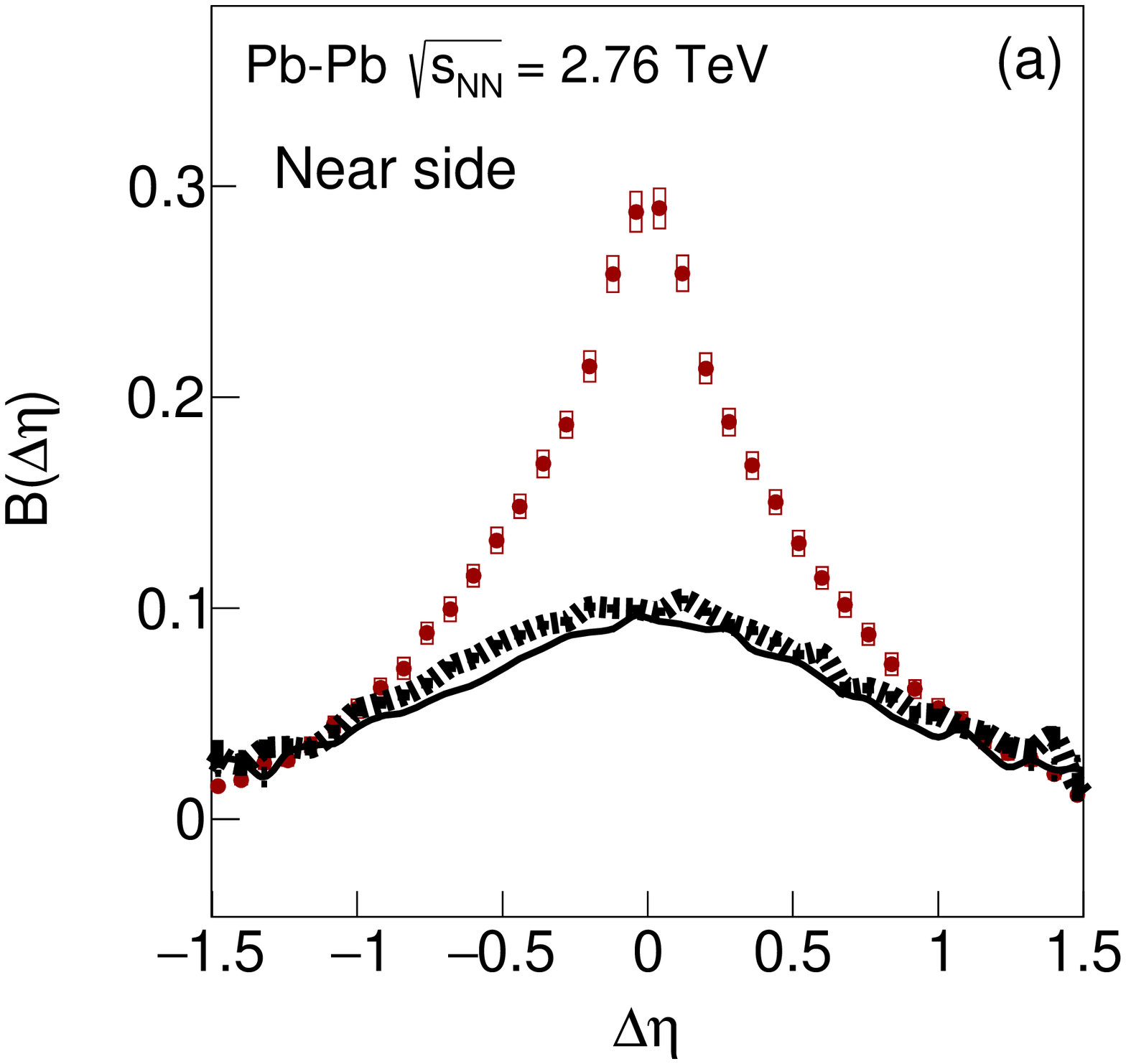}
\includegraphics[width=0.32\textwidth]{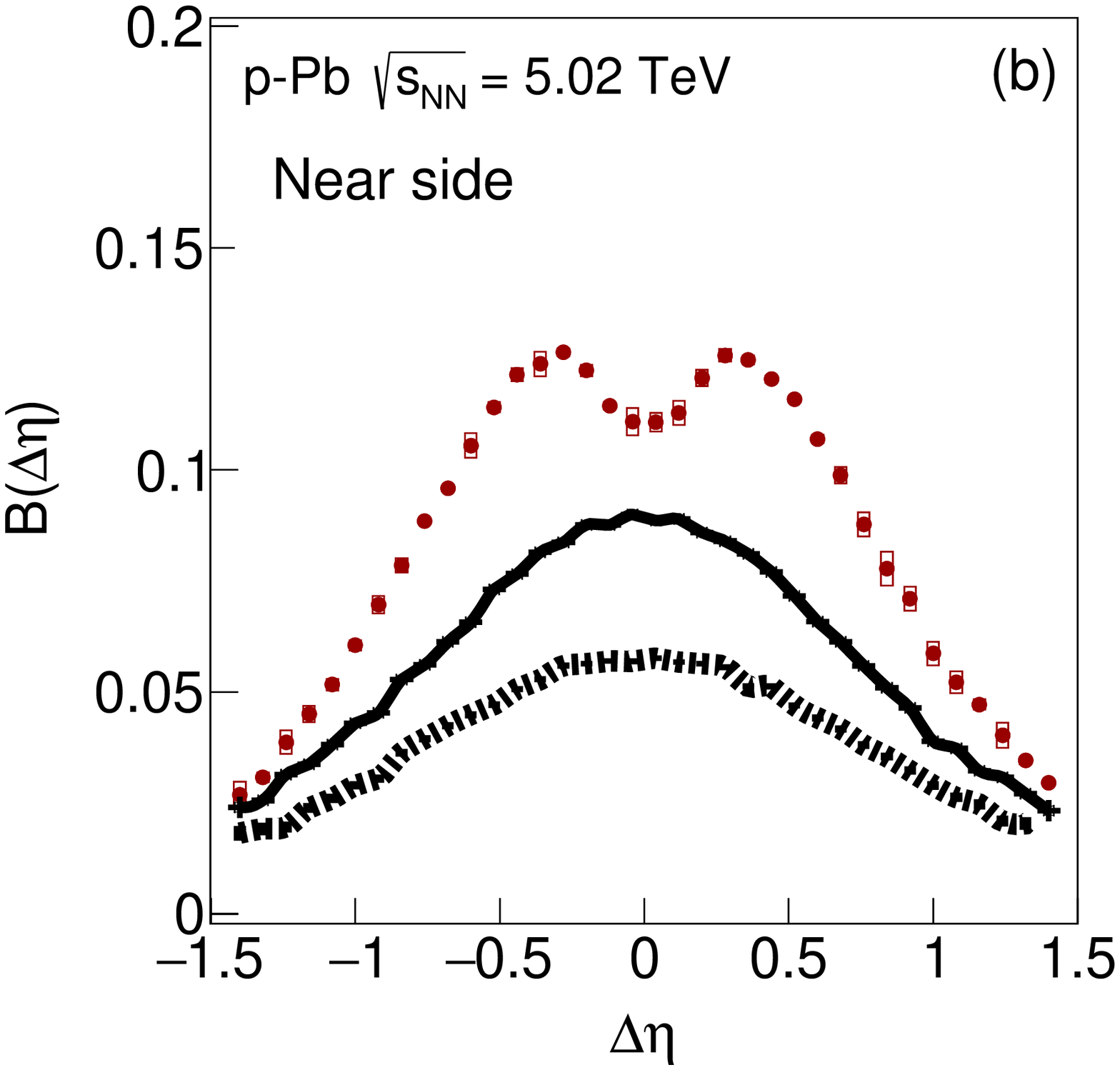}
\includegraphics[width=0.32\textwidth]{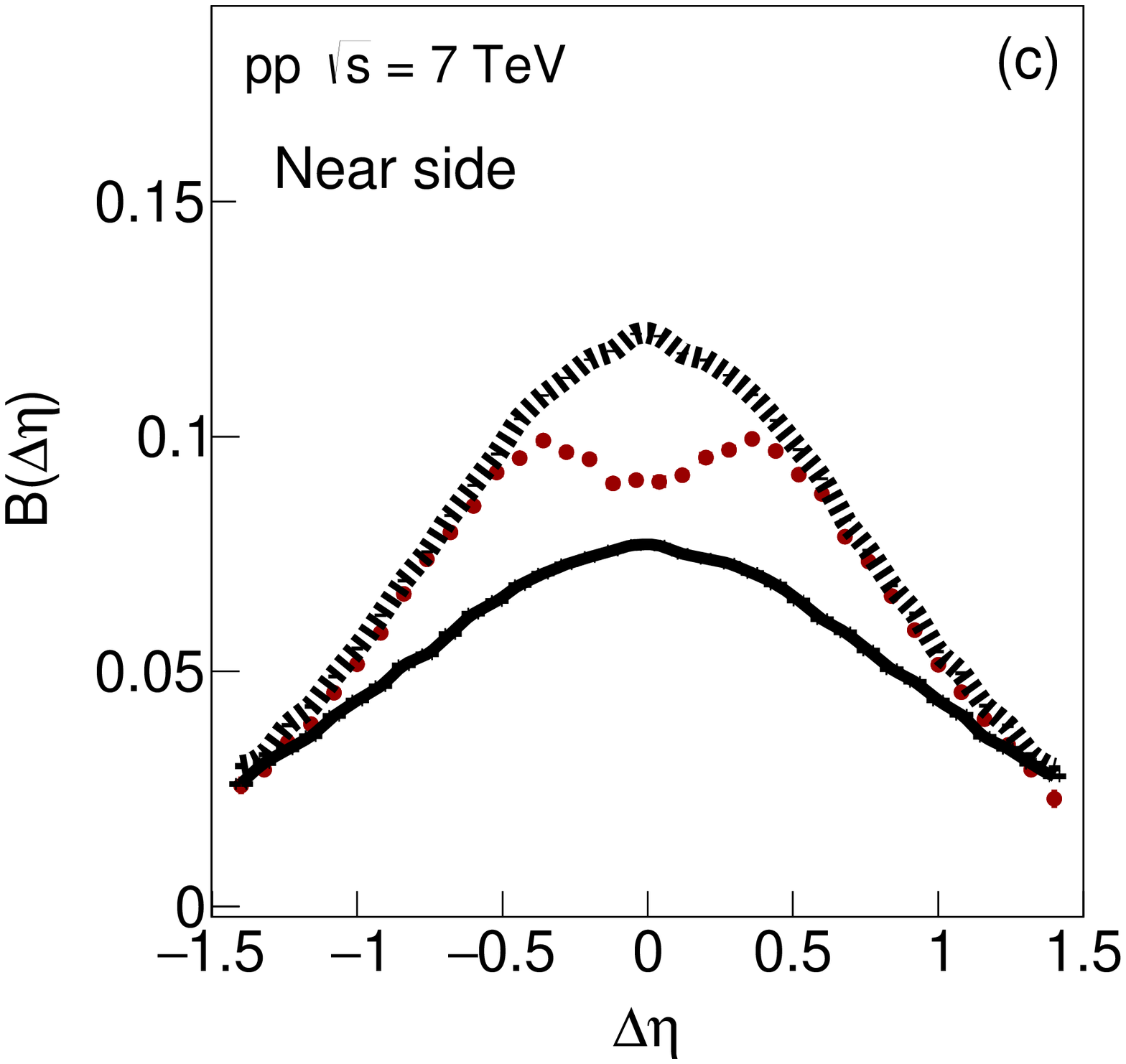}
\includegraphics[width=0.32\textwidth]{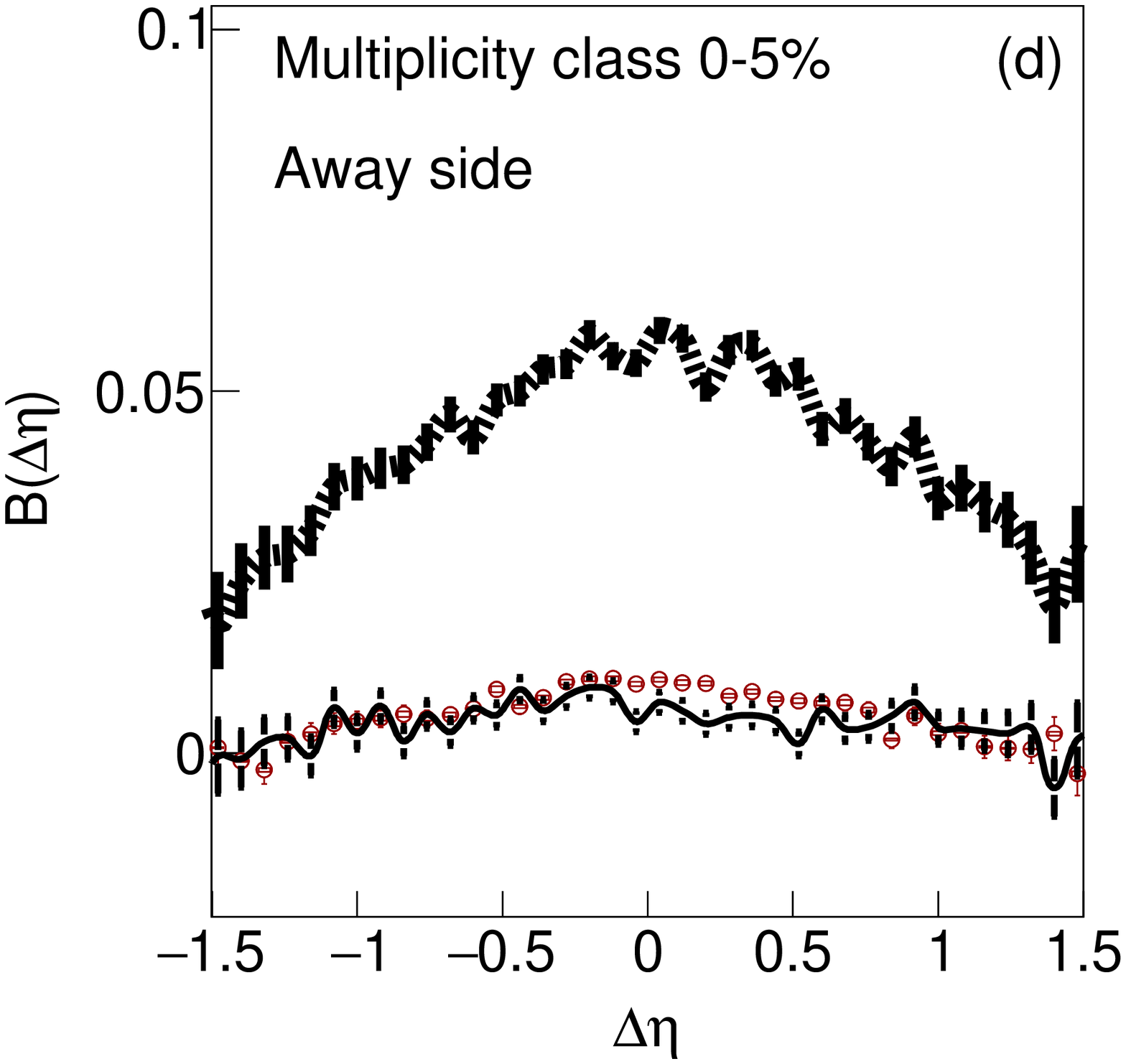}
\includegraphics[width=0.32\textwidth]{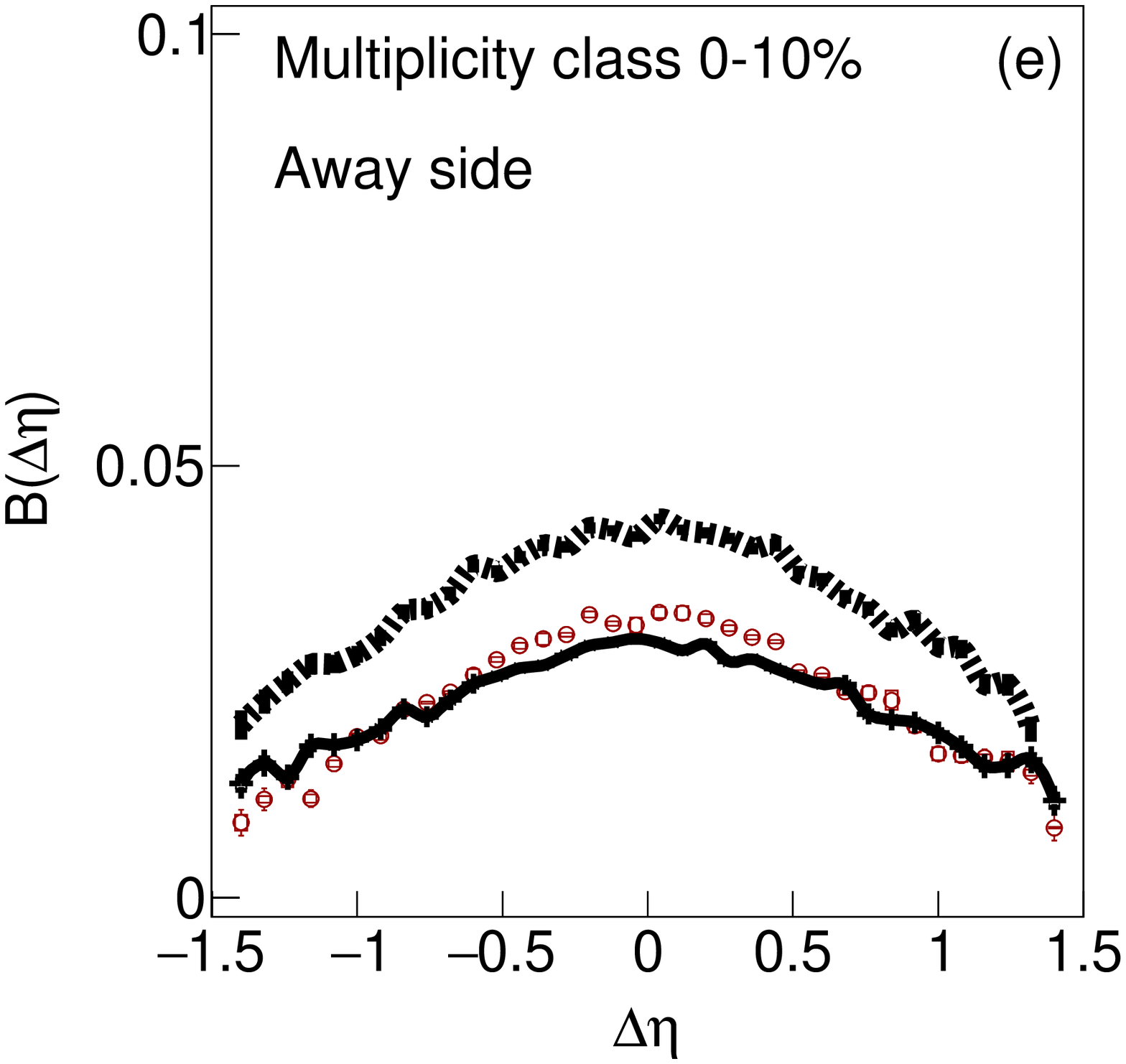}
\includegraphics[width=0.32\textwidth]{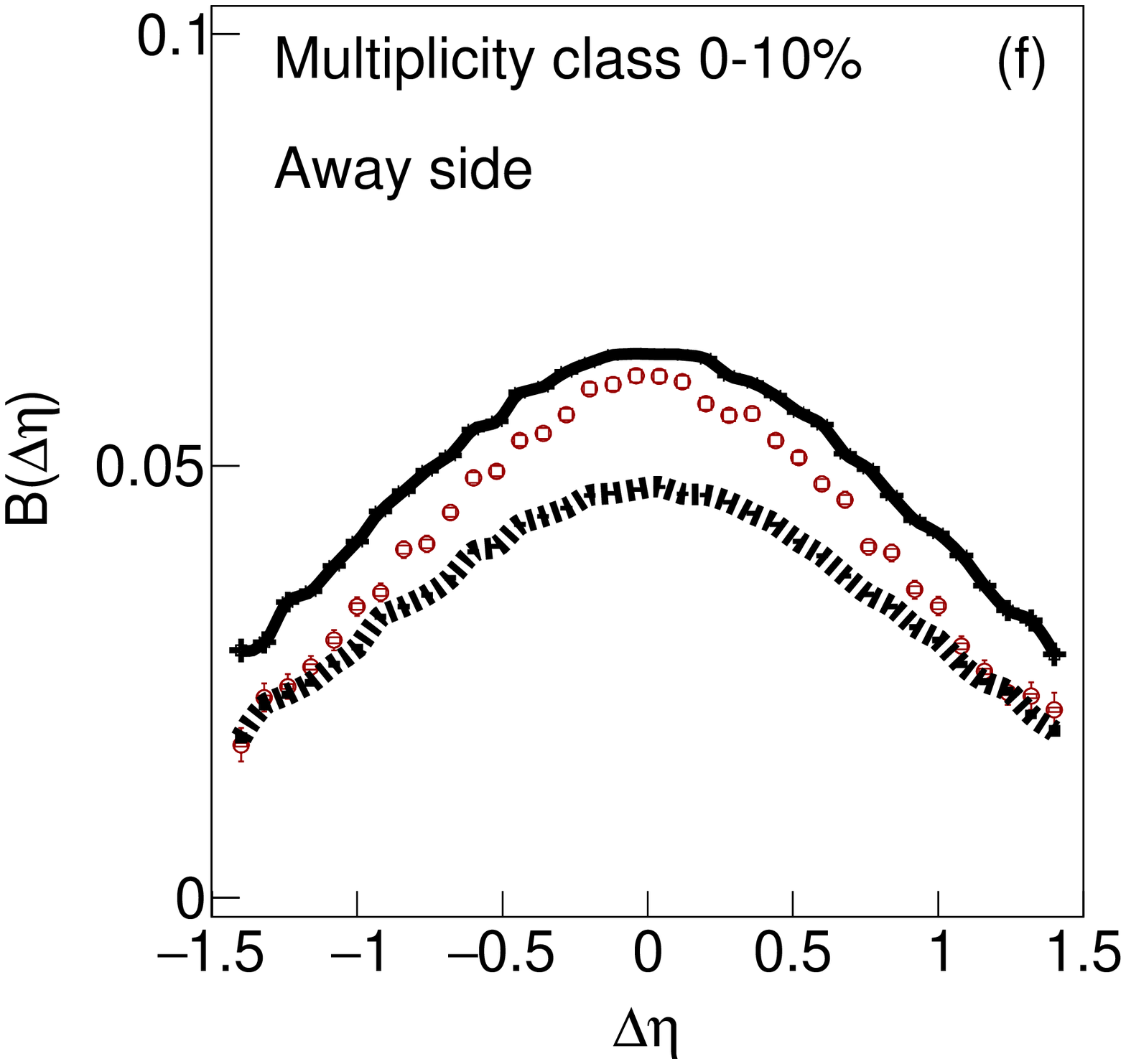}
\includegraphics[width=0.32\textwidth]{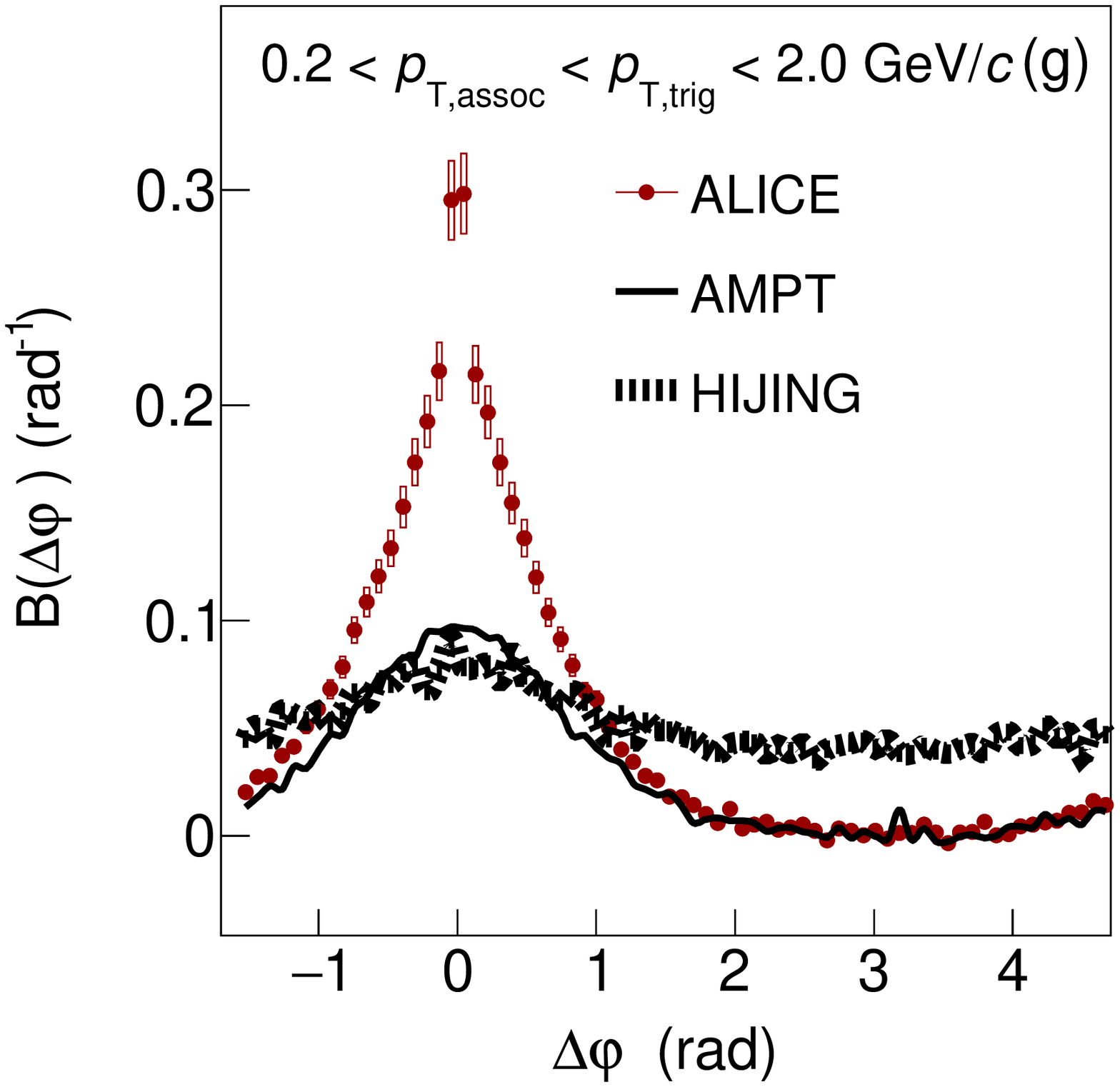}
\includegraphics[width=0.32\textwidth]{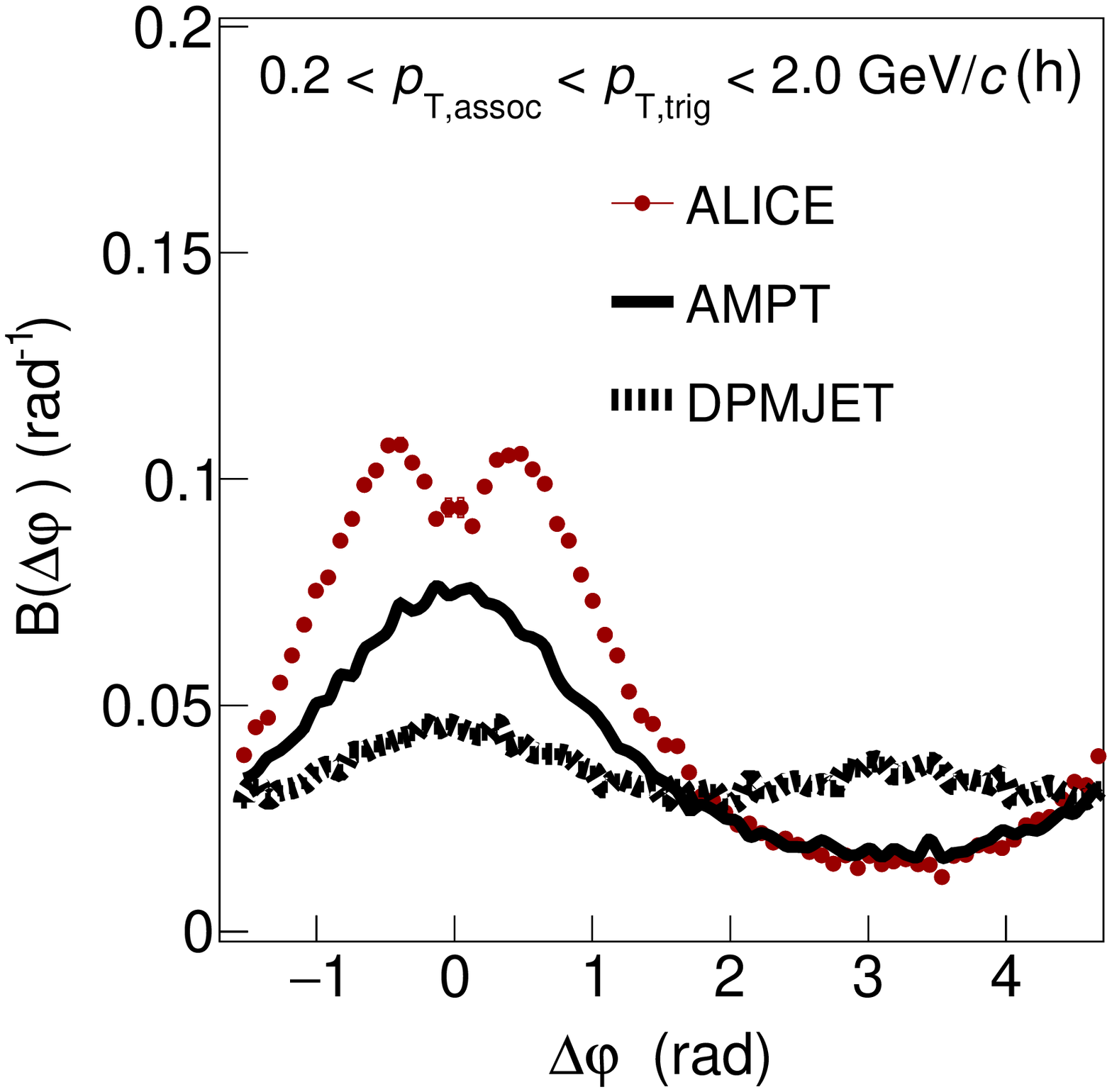}
\includegraphics[width=0.32\textwidth]{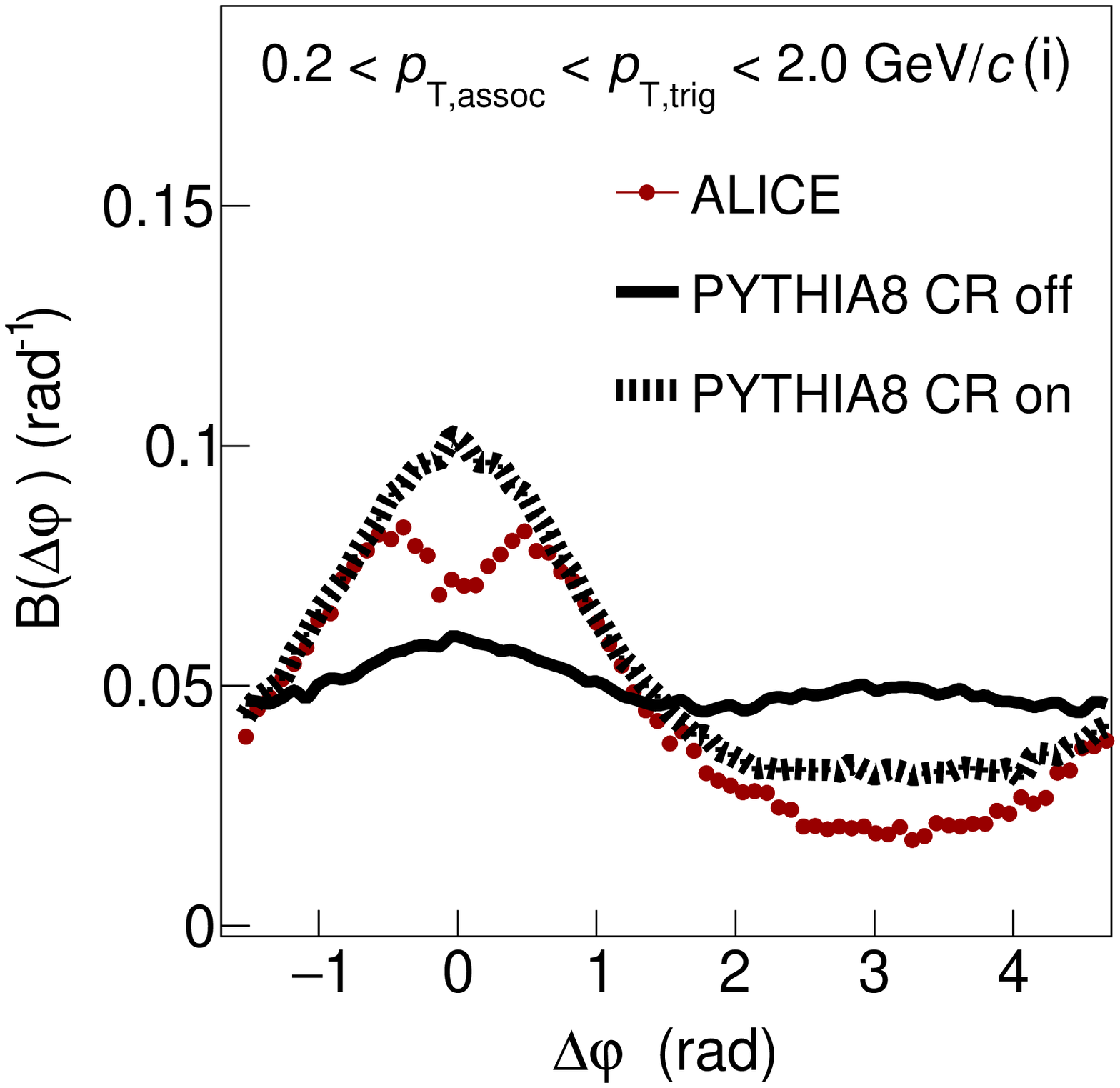}
\caption{The balance function for charged particles with \ptLow~as a function of \deta~on the near--side (upper row) and away--side (middle row) and as a function of \dphi~(lower row) for Pb--Pb (panels (a), (d) and (g)), p--Pb (panels (b), (e) and (h)) and pp collisions (panels (c), (f) and (i)) compared with results from various event generators. Only the highest multiplicity class is shown, i.e. 0-5\% for Pb--Pb and 0-10\% for p--Pb and pp collisions.}
\label{fig:bfIn1dLowpT_models}
\end{figure}

The integral of the balance function over the acceptance is related to measures of charge fluctuations as argued in~\cite{Jeon:2001ue}, and is between 0.25 and 0.35 (i.e. 0.5 and 0.7 in case the $p_{\mathrm{T}}$ requirement between the ``trigger'' and the ``associated'' particles is not imposed) for all systems and multiplicity classes. For each system 
it reveals a mild multiplicity class dependence which, for Pb--Pb, could explain the increase of multiplicity fluctuations for 
central compared to peripheral events reported in~\cite{Abelev:2012pv}.

\subsubsection{\textbf{Balance function projections}}

Figure \ref{fig:bfIn1dLowpT} presents for Pb--Pb, p--Pb, and pp collisions the 
projections of the two-dimensional balance function in \deta~~on the near--side (panels (a), (b), (c) ) and away--side (~in panels (d), (e), (f)), and \dphi~in panels (g), (h), (i), respectively. The statistical uncertainty, usually smaller than the marker size, is represented by the error bar while the systematic uncertainty, calculated as the quadratic sum of the correlated and the uncorrelated part, by the box around each data point. The balance function as a function of the relative pseudorapidity difference \deta~on the near--side exhibits a strong multiplicity 
dependence for all collision systems. In particular, the distribution narrows and the peak value becomes larger 
for high- compared to low-multiplicity events. As a function of the relative azimuthal angle \dphi~on the near--side, 
the balance function exhibits the same qualitative features as for $\mathrm{\Delta}\eta$, i.e. narrower distributions 
with larger magnitude for increasing event multiplicity in all three systems. However, the magnitude of the balance function on the away--side 
exhibits a different trend, with larger values of B(\deta) and B(\dphi) measured for low- compared to high-multiplicity events. 

As already discussed in Section~\ref{Sec:Analysis}, in p--Pb collisions, the nucleon--nucleon centre-of-mass system shifts by a rapidity of -0.465 with respect to the ALICE laboratory system in the direction 
of the proton beam. The influence of this shift was studied with simulations and, although the balance function is not translationally-invariant, the shift does not lead to any significant difference in either the projections of the balance function or the extracted widths.

As indicated previously, starting from mid-central events in Pb--Pb collisions a distinct depletion is observed 
in the two-dimensional distribution around $(\mathrm{\Delta}\eta,\mathrm{\Delta}\varphi)=(0,0)$ that becomes more 
pronounced in events with low multiplicities, and in particular in p--Pb and pp collisions. The fact that the 
aforementioned depletion does not seem to be restricted to a very narrow window in either \deta~or \dphi~(the 
structures extend to $-0.4 < \mathrm{\Delta}\eta < 0.4$ and $-\pi/6 < \mathrm{\Delta}\varphi < \pi/6$) indicates 
that the origin is not due to detector effects, as was confirmed by independent studies involving modification of cuts controlling track splitting and merging. One possible mechanism that could create such a structure is 
the charge-dependent short-range correlations such as Coulomb attraction and repulsion, 
or quantum statistics correlations~\cite{Aamodt:2011kd,Adam:2015pya,Adam:2015vna}. To test this hypothesis, a criterium on the minimum transverse momentum difference 
$\mathrm{\Delta} p_{\mathrm{T}}$ between two particles of a pair was applied. The value was varied from $\mathrm{\Delta} p_{\mathrm{T}}~=~0$~GeV/$c$ to $\mathrm{\mathrm{\Delta}}p_{\mathrm{T}}~=~0.2$~GeV/$c$. 
The choice for the selected values is driven by the fact that the bulk of short-range correlations are expected to have 
$\mathrm{\mathrm{\Delta}}p_{\mathrm{T}}<0.1$~GeV/$c$~\cite{Abelev:2014pja}. The depletion is less pronounced with increasing value of $\mathrm{\mathrm{\Delta}}p_{\mathrm{T}}$ and vanishes for $\mathrm{\mathrm{\Delta}}p_{\mathrm{T}}~=~0.2$~GeV/$c$. The disappearance of the depletion was also achieved by increasing the lower transverse momentum threshold for both the trigger 
and the associated particle to $p_{\mathrm{T}} > 0.5$~GeV/$c$. Both these observations are inline with the hypothesis that the depletion originates from (mainly) quantum statistics correlations and Coulomb effects. The physics conclusion, i.e. narrower distributions with increasing event multiplicity, does not change applying one of these criteria.

\subsubsection{\textbf{Comparison with models}}
In Figs.~\ref{fig:bfIn1dLowpT_models} (a),(d) and (g) the balance function in \deta~on the near-- (a) and away--side (d), and in \dphi~(g) are compared with Monte Carlo calculations using the HIJING \cite{Wang:1991hta} and AMPT \cite{Zhang:1999bd,Lin:2004en} event generators. The figures show the 0--5$\%$ multiplicity class of Pb--Pb collisions. In AMPT simulations, the string melting option was used, with parameters tuned to describe the experimental data on anisotropic 
flow at LHC energies~\cite{Xu:2011fi,Xu:2011jm}. The centrality classes were defined based on the module of the impact parameter. 
It is seen that neither AMPT nor HIJING are able to describe the balance function projections in $\Delta \eta$ on the near--side (see Fig.~\ref{fig:bfIn1dLowpT_models}-a), since they expect not only much broader distributions but they also underestimate the magnitude of the balance function. On the other hand, the projection of the balance function in $\Delta \eta$ on the away--side (Fig.~\ref{fig:bfIn1dLowpT_models}-d) indicates that AMPT is in qualitative agreement with the data points, contrary to HIJING that predicts a significantly larger magnitude of the balance function. Finally, the $\Delta \varphi$ projection of the balance function in Fig.~\ref{fig:bfIn1dLowpT_models} (g) shows that while HIJING is still not able to describe the data points, AMPT predicts narrower distributions on the near--side but with a much smaller magnitude than the one experimentally measured.

The comparison of the experimental results for the 0--10$\%$ multiplicity class in p-Pb collisions with model predictions is presented in 
Figs.~\ref{fig:bfIn1dLowpT_models} (b),(e) and (h). For this comparison, results from Monte Carlo calculations using 
the DPMJET \cite{Ranft:1999qe} and AMPT \cite{Zhang:1999bd,Lin:2004en} event generators were used. DPMJET is a model based 
on independent pp collisions, describing hard processes, hadron--hadron interactions, and hadronic interactions 
involving photons, without any collective effects. This model fails to describe the experimental data points in either of the two projections in $\Delta \eta$, i.e. on the near-- and the away--side in Fig.~\ref{fig:bfIn1dLowpT_models} (b) and (e), respectively, expecting much broader distributions with smaller (larger) magnitude on the near--(away--)side. In addition, for the balance function projection in $\Delta \varphi$ presented in Fig.~\ref{fig:bfIn1dLowpT_models} (h), DPMJET predicts broader distributions with a smaller magnitude compared to the measured data points on the near--side, but also exhibits a correlation peak on the away--side contrary to what is observed experimentally. On the other hand, AMPT, as in the case of the Pb--Pb collisions, seems to describe better the balance function projections in both $\Delta \eta$ and $\Delta \varphi$.

For pp collisions, the experimental results are compared with two variants of calculations using PYTHIA8  
tune 4C \cite{Buckley:2011ms} in Figs.~\ref{fig:bfIn1dLowpT_models} (c),(f) and (i). This tune contains modified multi-parton 
interaction (MPI) parameters that allow it to describe the multiplicity dependence of $\langle p_\mathrm{T} \rangle$ 
\cite{Abelev:2013bla}. The default calculation includes the color reconnection mechanism, which is switched off in the second 
configuration. The version of PYTHIA8 without the inclusion of color reconnection expects a broader balance function near--side projection in $\Delta \eta$ with a smaller magnitude than the one measured as observed in Fig.~\ref{fig:bfIn1dLowpT_models} (c). On the other hand, the same tune predicts larger magnitude than the one measured for the balance function away--side projection in $\Delta \eta$ (see Fig.~\ref{fig:bfIn1dLowpT_models} (f)). Finally, for the projection in $\Delta \varphi$, this tune expects significantly broader distributions on the near--side than the measured ones, with an extra correlation peak developing on the away--side which is not observed experimentally. On the other hand, the tune of PYTHIA8 with the inclusion of color reconnection describes the experimental measurement fairly well in both $\Delta \eta$ and $\Delta \varphi$ projections.

As discussed in the previous paragraphs, there are models that exhibit a correlation peak on the away--side contrary to what is supported by the data. For this reason, the width of the balance function distribution in \deta~and \dphi~will be extracted and compared with models on the near--side only. 

\begin{figure}[tp!]
\centering
\includegraphics[width=0.485\textwidth,height=0.455\textwidth]{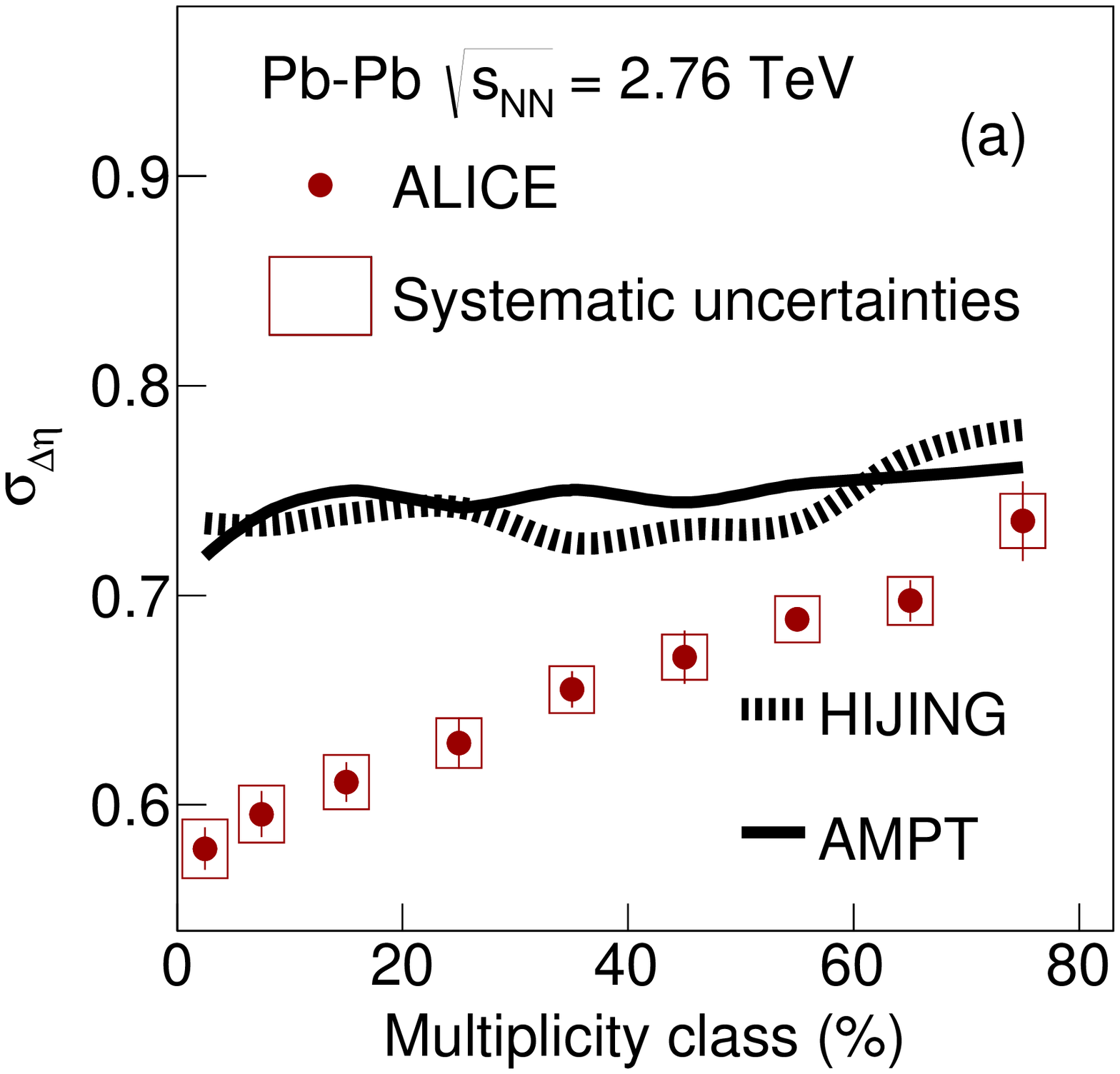}
\includegraphics[width=0.485\textwidth,height=0.455\textwidth]{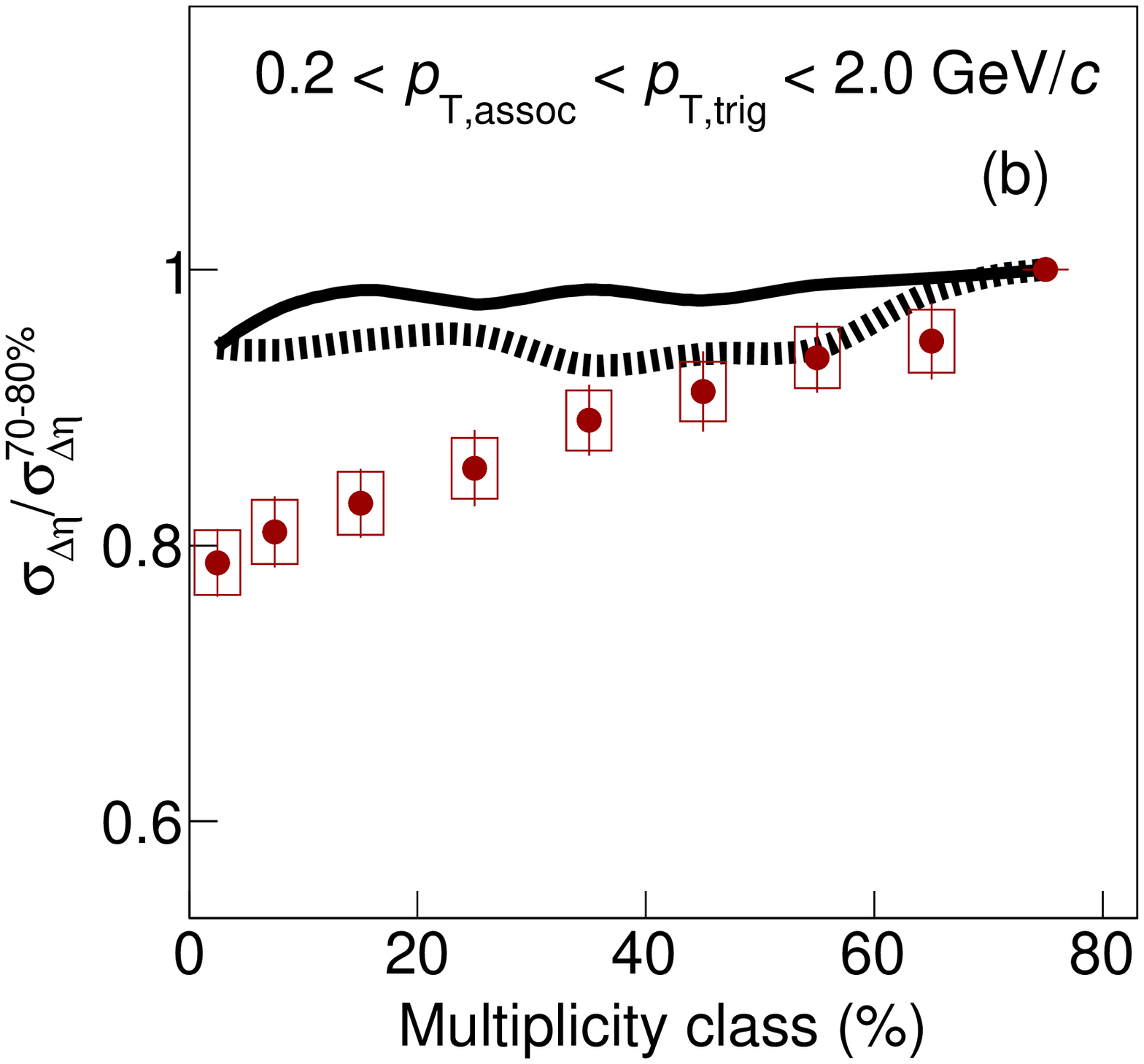}
\includegraphics[width=0.485\textwidth,height=0.455\textwidth]{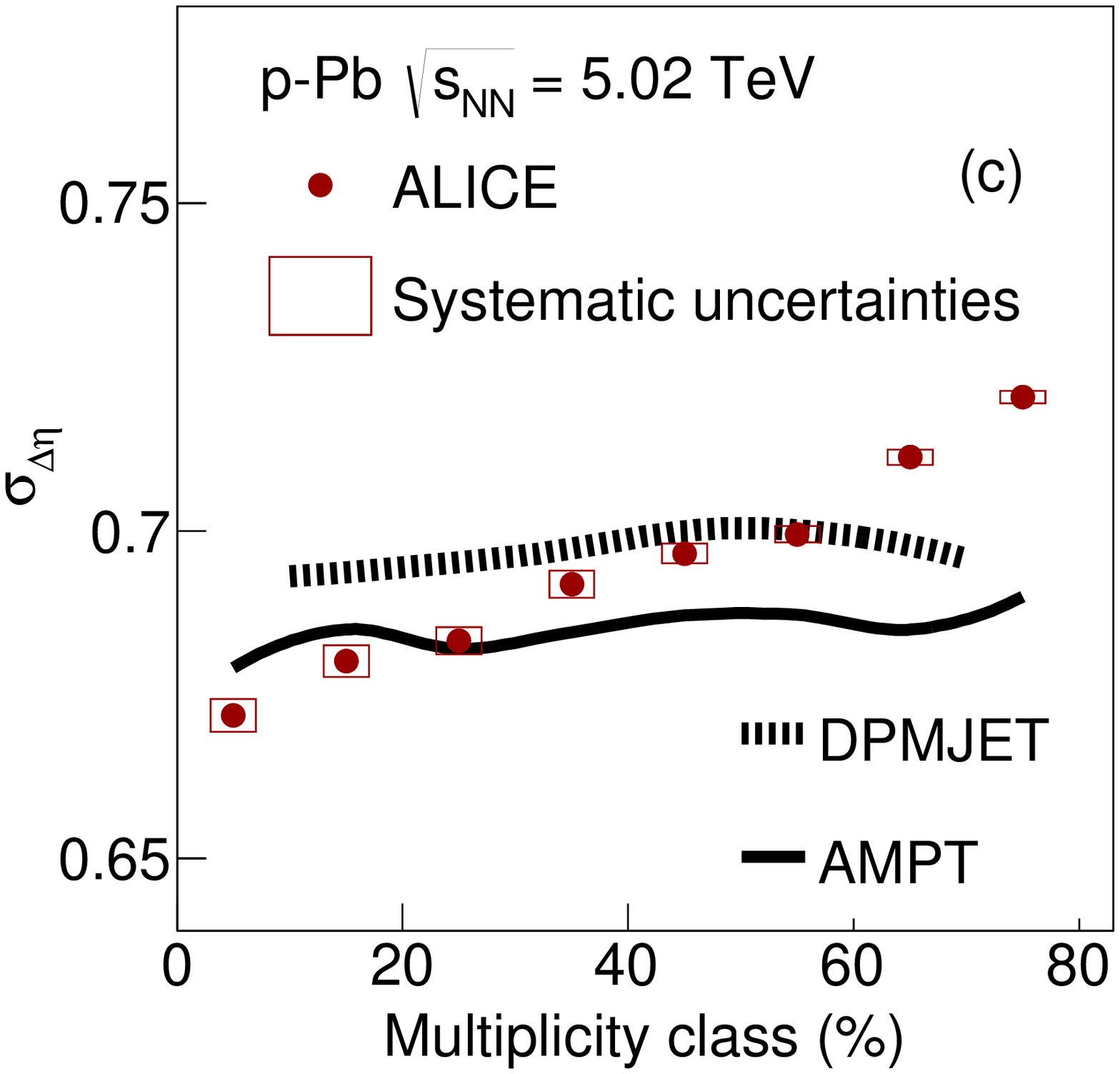}
\includegraphics[width=0.485\textwidth,height=0.455\textwidth]{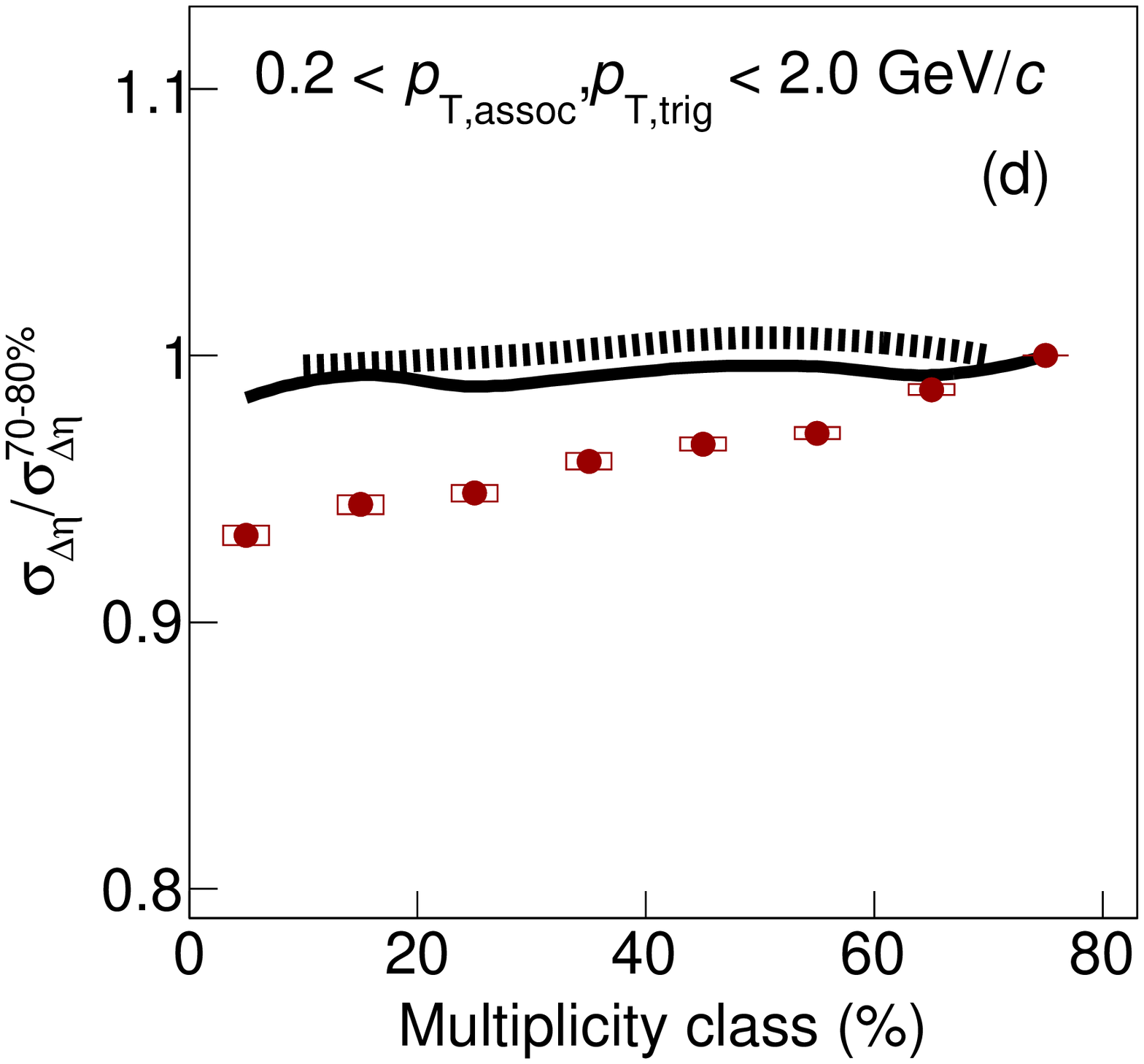}
\includegraphics[width=0.485\textwidth,height=0.455\textwidth]{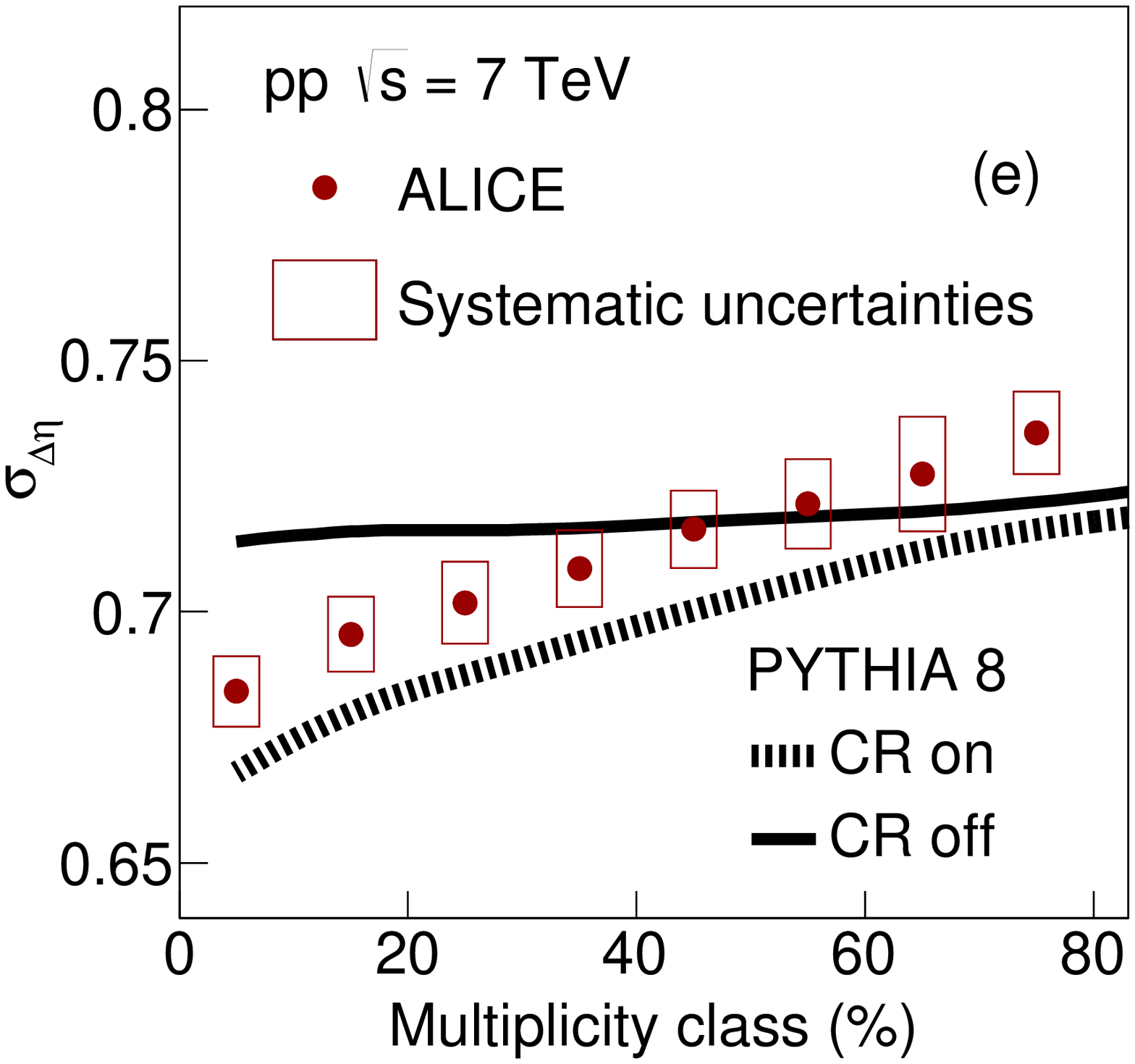}
\includegraphics[width=0.485\textwidth,height=0.455\textwidth]{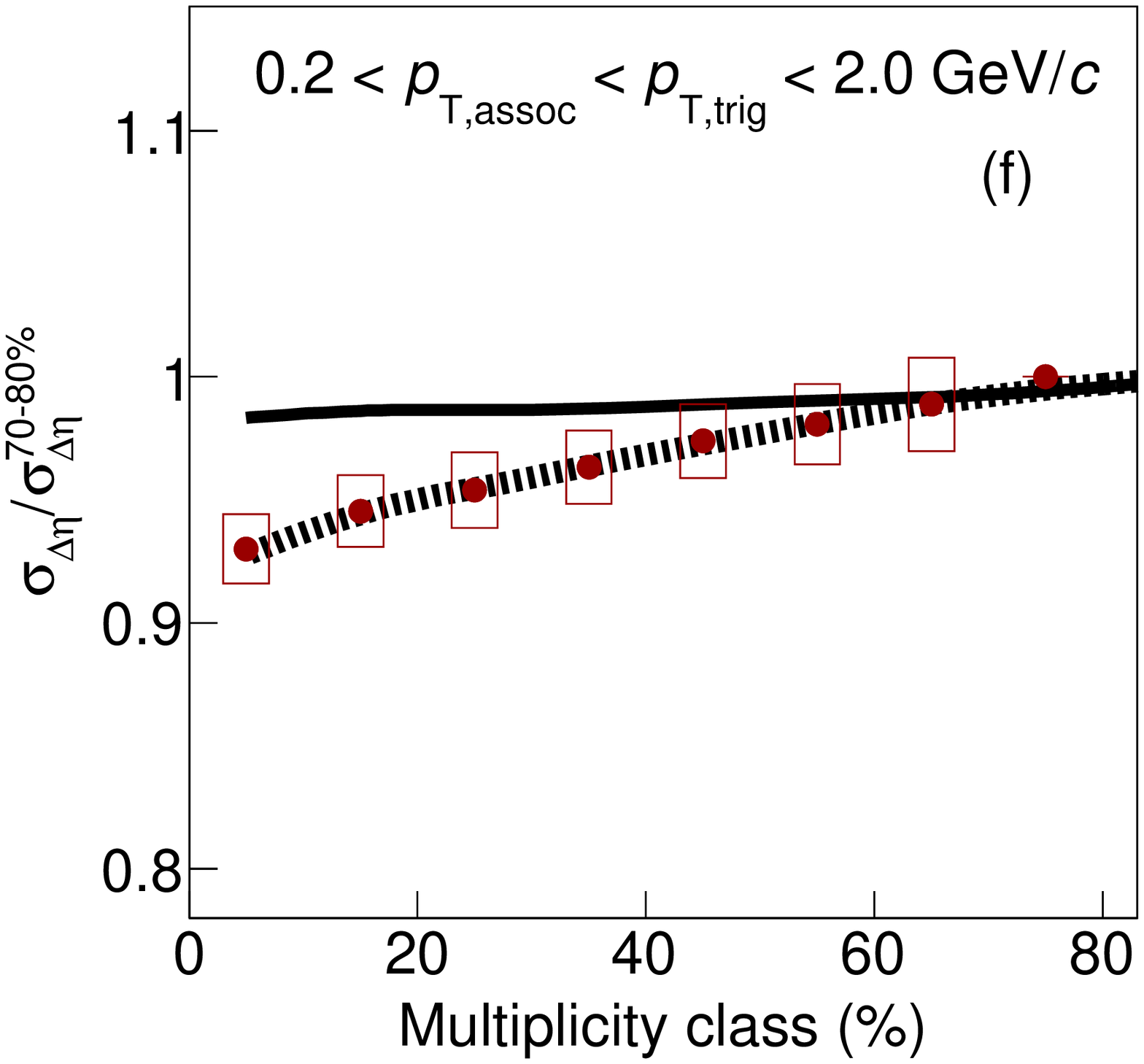}
\caption{The multiplicity-class dependence of \sigmadeta~in Pb--Pb, p--Pb, and pp collisions at $\sqrt{s_{\mathrm{NN}}} = 2.76$, 5.02, and 7~TeV~compared with results from various event generators in panels (a), (c), and (e). Panels (b), (d), and (f) show the relative decrease of \sigmadeta~calculated with respect to $\sigma_{\mathrm{\Delta}\eta}^{70-80\%}$, as a function of the multiplicity class. The transverse momentum values for both the trigger and the associated 
particles satisfy the condition \ptLow.}
\label{fig:widthInDeltaEtaLowpT}
\end{figure}

\subsubsection{\textbf{Balance function width}}
To quantify the narrowing of the balance function width as a function of multiplicity, the standard deviation 
$\sigma$ is calculated as described in Section~\ref{Sec:BalanceFunction}. The panels (a), (c), and (e) of Fig.~\ref{fig:widthInDeltaEtaLowpT}  present the evolution of \sigmadeta~on the near--side with multiplicity class, expressed by the 
multiplicity percentile for Pb--Pb, p--Pb, and pp collisions, respectively. Note that the 
multiplicity decreases from left to right along the horizontal axis. The statistical uncertainties of the data points are 
represented by the error bars and are usually smaller than the marker size. For all collision systems, a 
significant narrowing of the balance function in \deta~with increasing multiplicity is observed.

The panels (b), (d), and (f) of Figs.~\ref{fig:widthInDeltaEtaLowpT} show 
the relative decrease of \sigmadeta, 
expressed by the ratio of \sigmadeta~for each multiplicity class over the value in the lowest 
multiplicity class, i.e. 70-80\% for all collision systems. The narrowing of the balance function with 
increasing multiplicity is most prominent in Pb--Pb collisions where the relative decrease between the largest and lowest multiplicity class is 
$21.2 \pm 2.4 \mathrm{(stat.)} \pm 2.4 \mathrm{(syst.)} \%$. A significant relative decrease is also observed 
for the other two systems with values of $6.7 \pm 0.2 \mathrm{(stat.)} \pm 0.4 \mathrm{(syst.)} \%$ 
and $7.0 \pm 0.3 \mathrm{(stat.)} \pm 1.4 \mathrm{(syst.)} \%$ in p--Pb and pp collisions, respectively. Note though that the multiplicities in these three systems are significantly different (see e.g. Table~\ref{Table:correctedMultiplicity})
	
In Fig.~\ref{fig:widthInDeltaEtaLowpT} (a) the width in \deta for Pb--Pb collisions is compared with the results from HIJING and AMPT. Neither model describes the experimentally observed narrowing of the balance function with increasing multiplicity. This is also reflected in Fig.~\ref{fig:widthInDeltaEtaLowpT} (b) where the relative decrease for both models is around 4$\%$.

\begin{figure}[tp!]
\centering
\includegraphics[width=0.485\textwidth,height=0.455\textwidth]{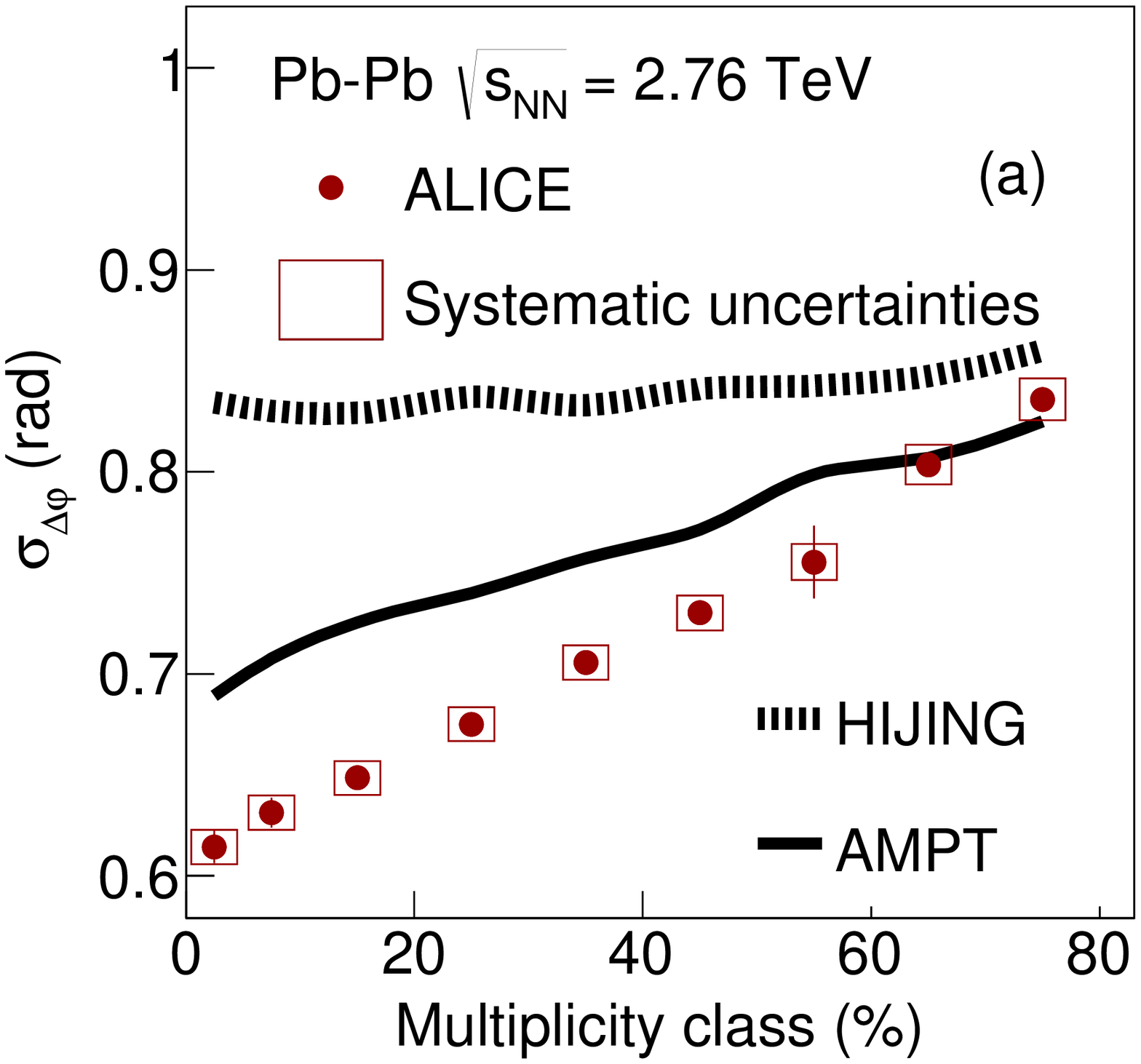}
\includegraphics[width=0.485\textwidth,height=0.455\textwidth]{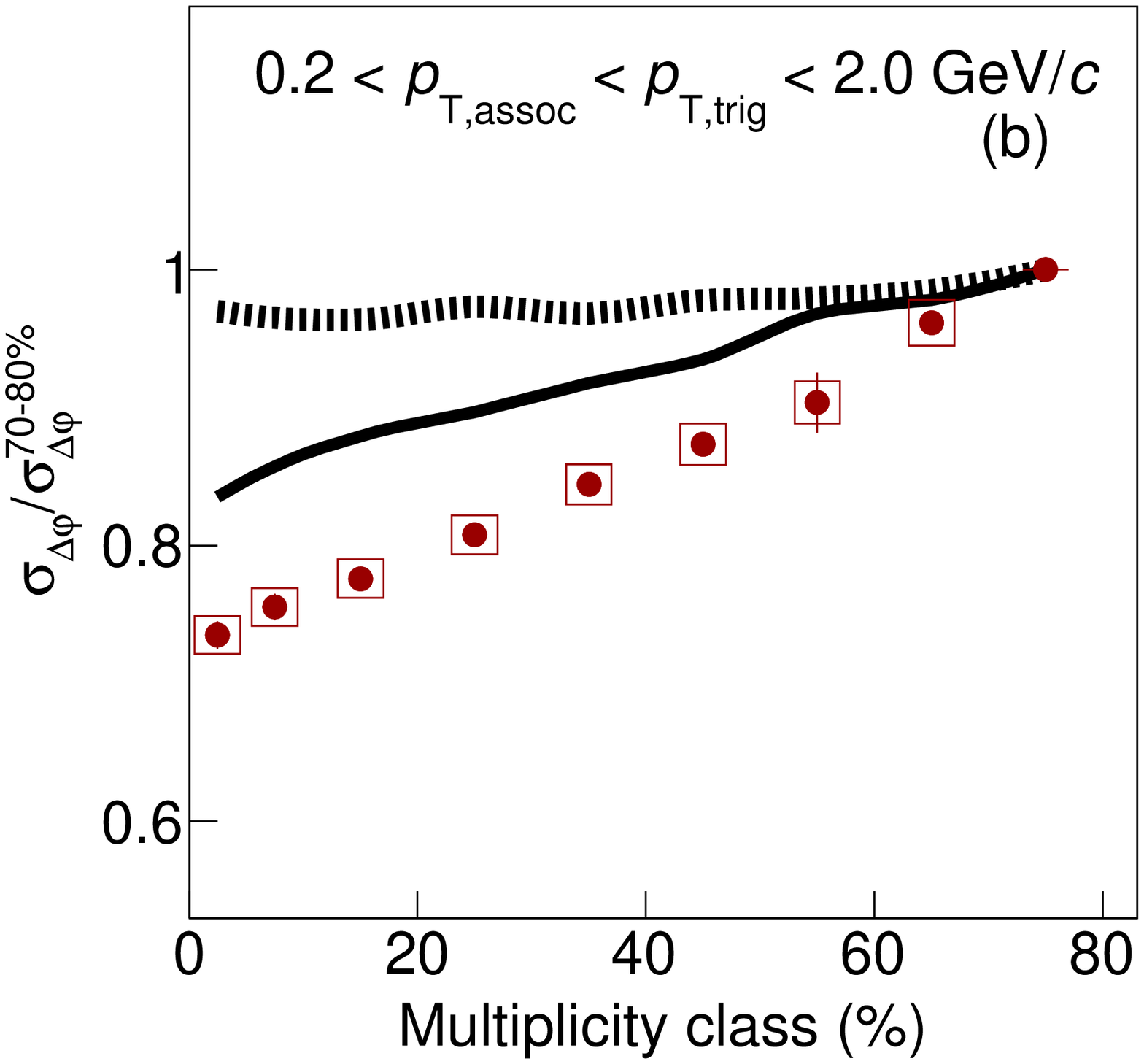}
\includegraphics[width=0.485\textwidth,height=0.455\textwidth]{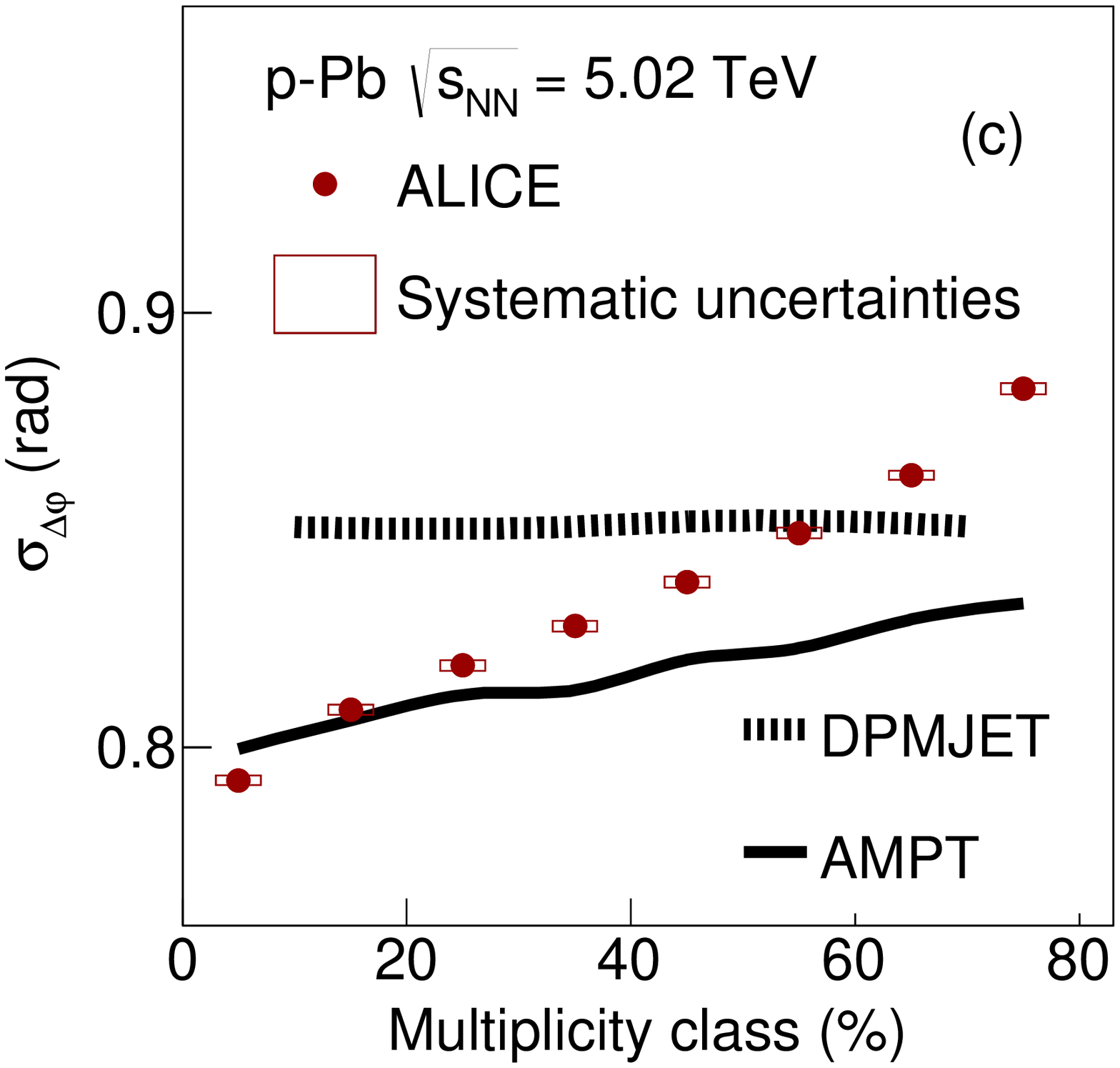}
\includegraphics[width=0.485\textwidth,height=0.455\textwidth]{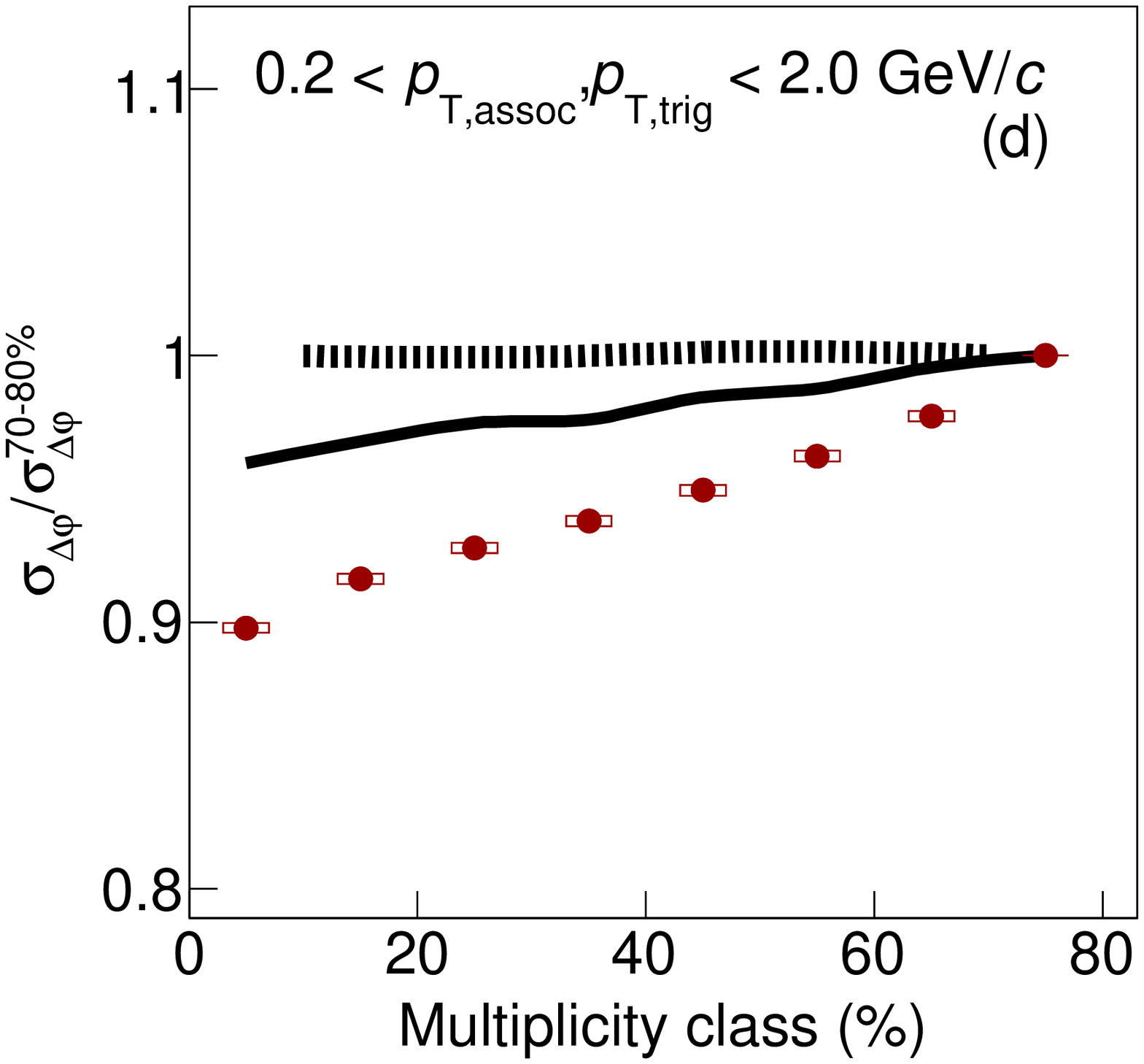}
\includegraphics[width=0.485\textwidth,height=0.455\textwidth]{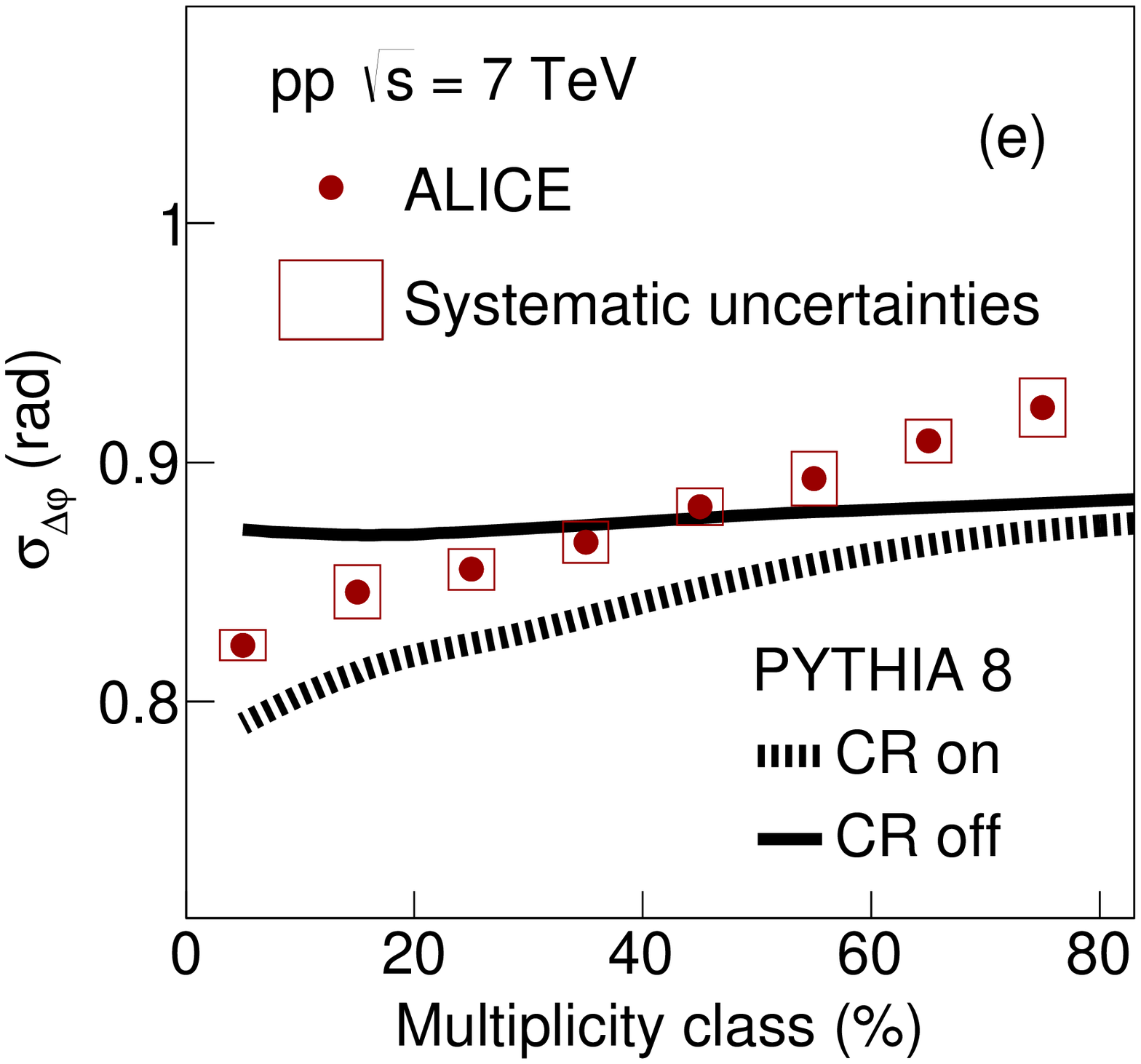}
\includegraphics[width=0.485\textwidth,height=0.455\textwidth]{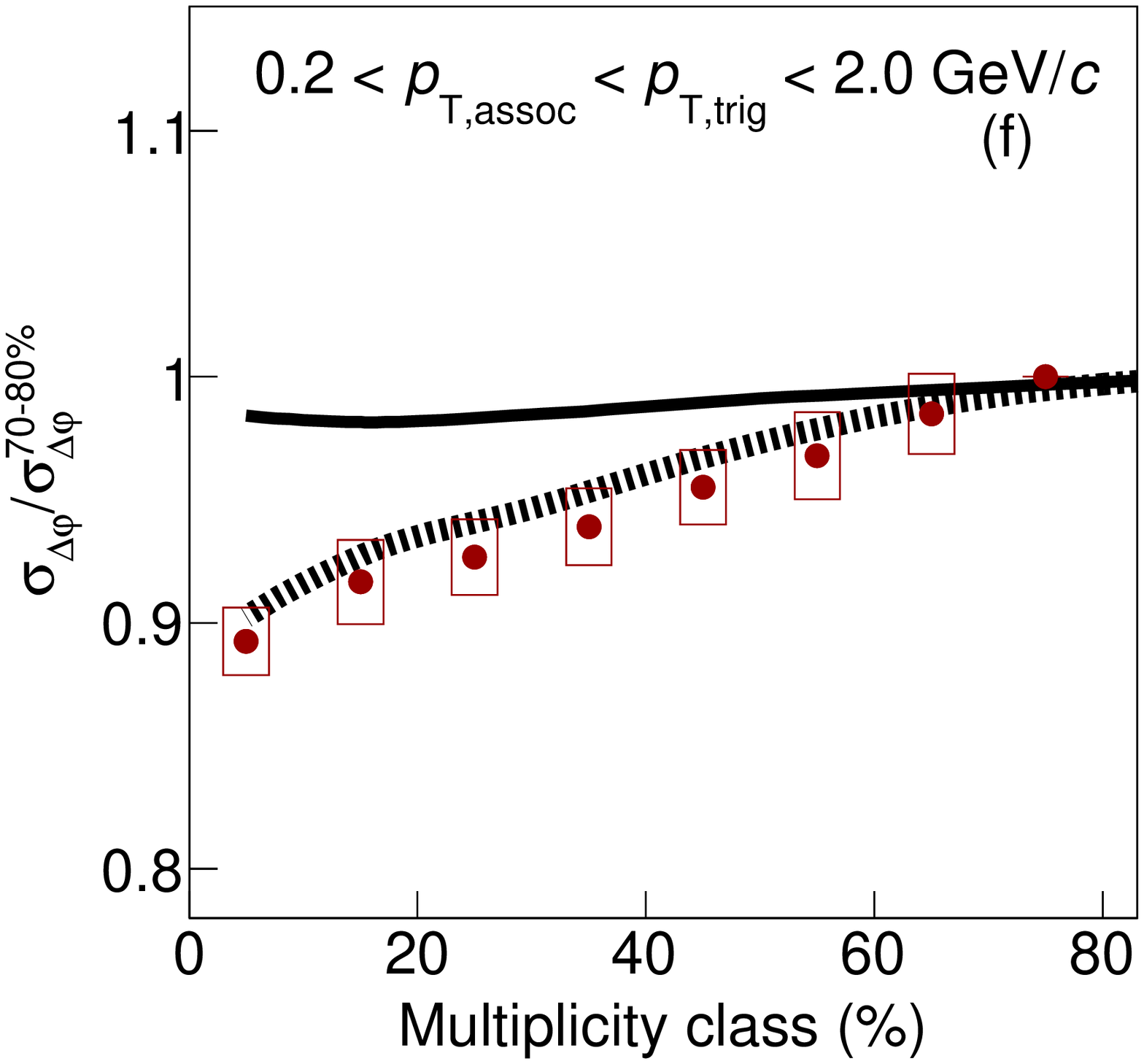}
\caption{The multiplicity-class dependence of \sigmadphi~in Pb--Pb, p--Pb, and pp collisions at $\sqrt{s_{\mathrm{NN}}} = 2.76$, 5.02, and 7~TeV~compared with results from various event generators in panels (a), (c), and (e). Panels (b), (d), and (f) show the relative decrease of \sigmadphi~calculated with respect to $\sigma_{\mathrm{\Delta}\varphi}^{70-80\%}$ as a function of the multiplicity class. The transverse momentum values for both the trigger and the associated 
particles satisfy the condition \ptLow.}
\label{fig:widthInDeltaPhiLowpT}
\end{figure}

Figure \ref{fig:widthInDeltaEtaLowpT} (c) shows the comparison of \sigmadeta~ in p-Pb collisions with model calculations. It is seen that DPMJET results in broader balance function 
distributions compared to AMPT. In addition, both models expect narrower balance function distributions 
compared to experimental measurements for low multiplicity classes (starting from 60$\%$ for DPMJET 
and 40$\%$ for AMPT). However, with increasing multiplicity (i.e. below 60$\%$ for DPMJET and 30$\%$ for 
AMPT) the balance function distributions are significantly narrower in the experiment compared to either of the 
models. Similar to the Pb--Pb case, neither of the models is able to reproduce the significant decrease 
of the width with increasing multiplicity observed in data. This is also 
reflected in Fig.~\ref{fig:widthInDeltaEtaLowpT} (d), where the relative decrease of the width 
between the highest and lowest multiplicity class for DPMJET and AMPT is marginal and not larger than 2$\%$.

The experimental results for pp collisions are compared with model predictions in Fig.~\ref{fig:widthInDeltaEtaLowpT} (e). PYTHIA8 without color reconnection, represented by the solid line, fails to describe the 
significant narrowing of the balance function with increasing multiplicity. The values of \sigmadeta~for 
this calculation are comparable within uncertainties to the ones obtained for the lowest multiplicity class in data. 
On the other hand, the inclusion of color reconnection, see the dashed line in Fig.~\ref{fig:widthInDeltaEtaLowpT} (e), results in a qualitatively 
similar narrowing as the one observed in the measurements. The absolute value of \sigmadeta~is lower 
than the experimental results for almost all multiplicity classes. Quantum statistics correlations are not included in the simulation, which might be the reason for this difference. Figure~\ref{fig:widthInDeltaEtaLowpT} (f) that presents the relative decrease of \sigmadeta~
quantifies the previous observations. It is seen that PYTHIA8 without color reconnection shows a rather weak 
(i.e. around 2$\%$) narrowing of the balance function with increasing multiplicity. This narrowing may result from the increased resonance yield for high- compared to low-multiplicity pp events~\cite{Abelev:2013csa}. 
The version of PYTHIA8 with the inclusion of color reconnection expects a relative reduction of around 7$\%$, 
in quantitative agreement with the measurement.

Figure~\ref{fig:widthInDeltaPhiLowpT}
presents the multiplicity dependence of \sigmadphi~in Pb--Pb, p--Pb, and pp collisions in panels (a), (c), and (e), respectively. 
All three systems exhibit a significant multiplicity-dependent narrowing of the balance function in $\mathrm{\Delta}\varphi$. Panels (b), (d), and (f) quantify this narrowing by presenting the decrease of the width in $\mathrm{\Delta}\varphi$ for 
each multiplicity class relative to the lowest multiplicity class. The data exhibit a narrowing of 
$26.5 \pm 1.0 \mathrm{(stat.)} \pm 1.4 \mathrm{(syst.)} \%$, $10.2 \pm 0.3 \mathrm{(stat.)} \pm 0.2 \mathrm{(syst.)} \%$, 
and $10.8 \pm 0.4 \mathrm{(stat.)} \pm 1.4 \mathrm{(syst.)} \%$ in Pb--Pb, p--Pb, and pp collisions.

The multiplicity dependence of the width in $\mathrm{\Delta}\varphi$ in Pb--Pb collisions is compared with expectations 
from HIJING and AMPT in Fig.~\ref{fig:widthInDeltaPhiLowpT} (a). HIJING fails to describe the 
experimental measurements while AMPT expects a significant decrease of \sigmadphi~with increasing multiplicity. 
The relative decrease in AMPT is about $18\%$, see Fig.~\ref{fig:widthInDeltaPhiLowpT} (b), and can be attributed to a rather strong multiplicity-dependent 
radial flow in the model that acts over the balancing pairs, retaining their initial correlations in $\mathrm{\Delta}\varphi$.

The measurements in p--Pb collisions are compared with the results from DPMJET and 
AMPT in Fig.~\ref{fig:widthInDeltaPhiLowpT} (c). Neither DPMJET, which does not exhibit a significant dependence 
on the event multiplicity, nor AMPT, which exhibits a relative decrease of around $4\%$, can quantitatively describe 
the experimental findings, as demonstrated in Fig.~\ref{fig:widthInDeltaPhiLowpT} (d). 

Finally, the values of \sigmadphi~in pp collisions are compared in Fig.~\ref{fig:widthInDeltaPhiLowpT} (e) with the two 
variants of PYTHIA8 calculations described before. Similarly to the picture that emerged from the comparison of \sigmadeta, 
the variant of PYTHIA8 calculation without the inclusion of color reconnection does not describe the strong multiplicity dependence 
reported in pp collision data. However, the calculation with color reconnection exhibits a qualitatively similar 
decrease of \sigmadphi~with increasing multiplicity. The relative decrease for this model is around 10$\%$, in 
quantitative agreement with the experimental results, as indicated in Fig.~\ref{fig:widthInDeltaPhiLowpT} (f).

The comparison between the data and the corresponding expectations from models like PYTHIA, 
illustrates the potentially significant role of color reconnection on charge-dependent correlations for small systems 
such as pp collisions. The effect of color reconnection in PYTHIA8 is strongly connected to MPIs, whose number increases with increasing multiplicity. In high-multiplicity pp events, MPIs lead to many color strings
that will overlap in physical space. Within PYTHIA8 approach, these strings are given a probability to be reconnected and 
hence hadronize not independently, but rather in a process that resembles collective final-state effects. This 
process results in a transverse boost of the fragments that leads to the development of final-state correlations between 
charged particles in a similar way as a collective radial boost does.

\subsection{\textbf{Balance function at high transverse momentum}}

\begin{figure}[th!]
\centering
\includegraphics[width=\textwidth]{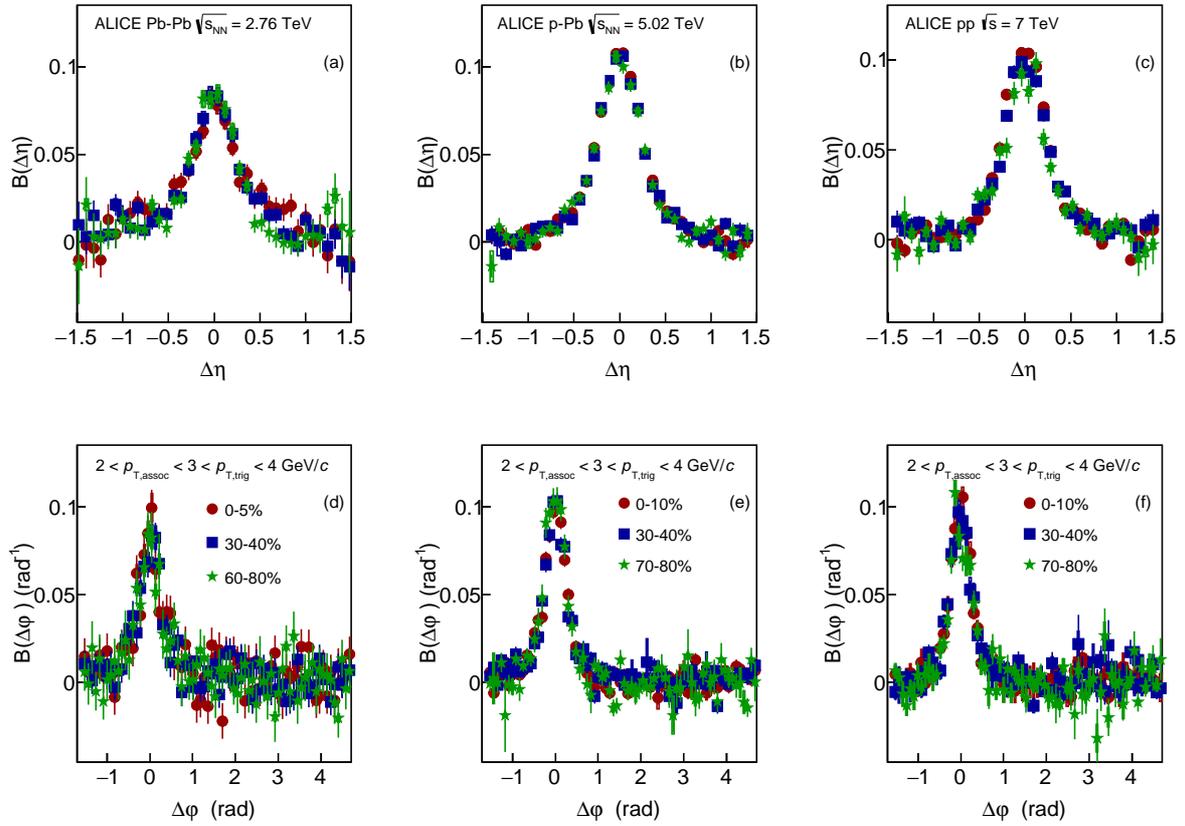}
\caption{The balance function for charged particles with \ptMed~as a function of \deta~(upper row) and \dphi~(lower row) in 
different multiplicity classes of Pb--Pb, in panels (a) and (d), p--Pb, in panels (b) and (e), and pp collisions, in panels (c) and (f), at $\sqrt{s_{\mathrm{NN}}} = 2.76$, 5.02, and 7~TeV, respectively.}
\label{fig:bfIn1dIntermediatepT} 
\end{figure}

\begin{figure}[th!]
\centering
\includegraphics[width=\textwidth]{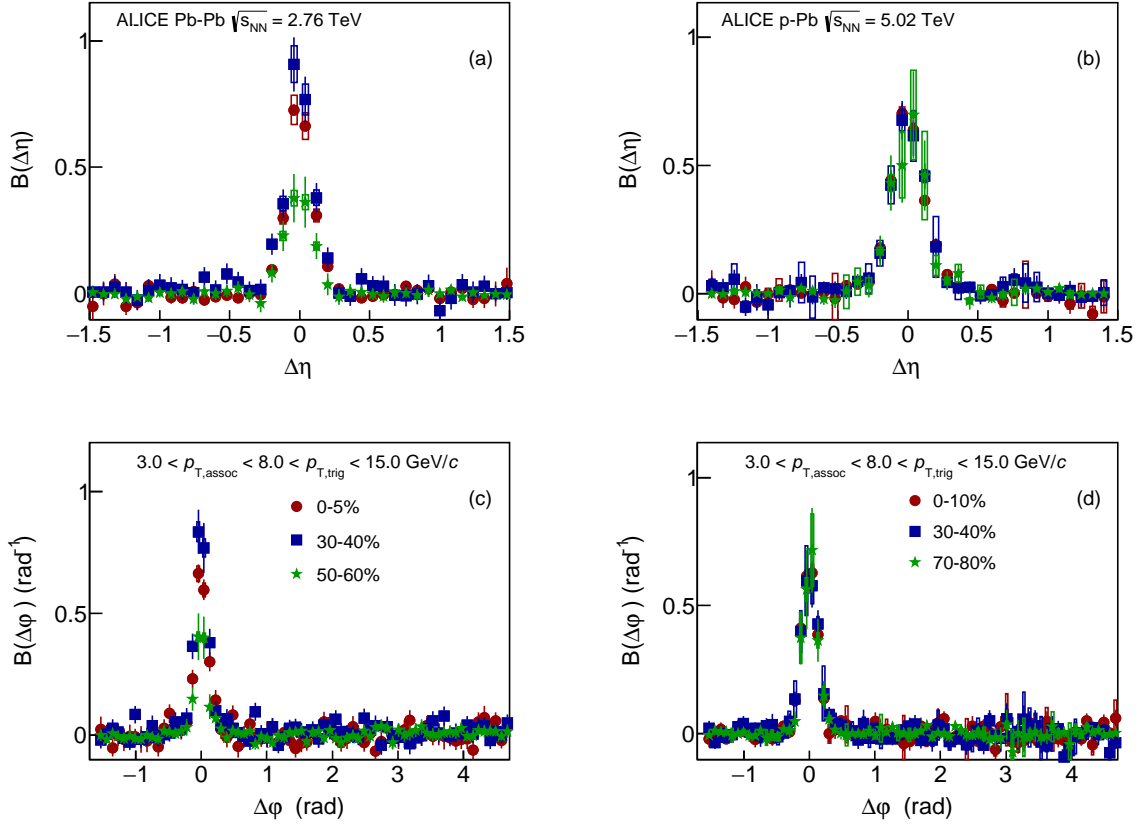}
\caption{The balance function for charged particles with \ptHigh~as a function of \deta~(upper row) and \dphi~(lower row) in 
different multiplicity classes of Pb--Pb in panels (a) and (c) and p--Pb collisions in panels (b) and (d) at $\sqrt{s_{\mathrm{NN}}} = 2.76$ and 5.02~TeV, respectively.}
\label{fig:bfIn1dHighpT} 
\end{figure}

In order to study if the narrowing of the balance function is restricted to the bulk particle production at low \pt~or 
is also connected to hard processes, the balance function was also measured in all collision systems for higher 
values of transverse momentum for both trigger and associated particles. 
Figure~\ref{fig:bfIn1dIntermediatepT} presents the projections of the two-
dimensional balance functions in \deta~in panels (a), (b), (c), and \dphi~in panels (d), (e), (f) for \ptMed~in Pb--Pb, p--Pb, and pp 
collisions, respectively. The analysis of Pb--Pb and p--Pb collisions was also extended to higher transverse momenta, 
\ptHigh~, shown in Fig.~\ref{fig:bfIn1dHighpT}.

The charge-dependent correlations exhibit little if any multiplicity dependence, in contrast to the findings from the 
lower transverse momentum region. In addition, the distributions in the intermediate and high-\pt~range are 
significantly narrower than the corresponding distributions at lower values of \pt~for each multiplicity class.


\begin{figure}[th!]
\centering
\includegraphics[width=0.49\textwidth]{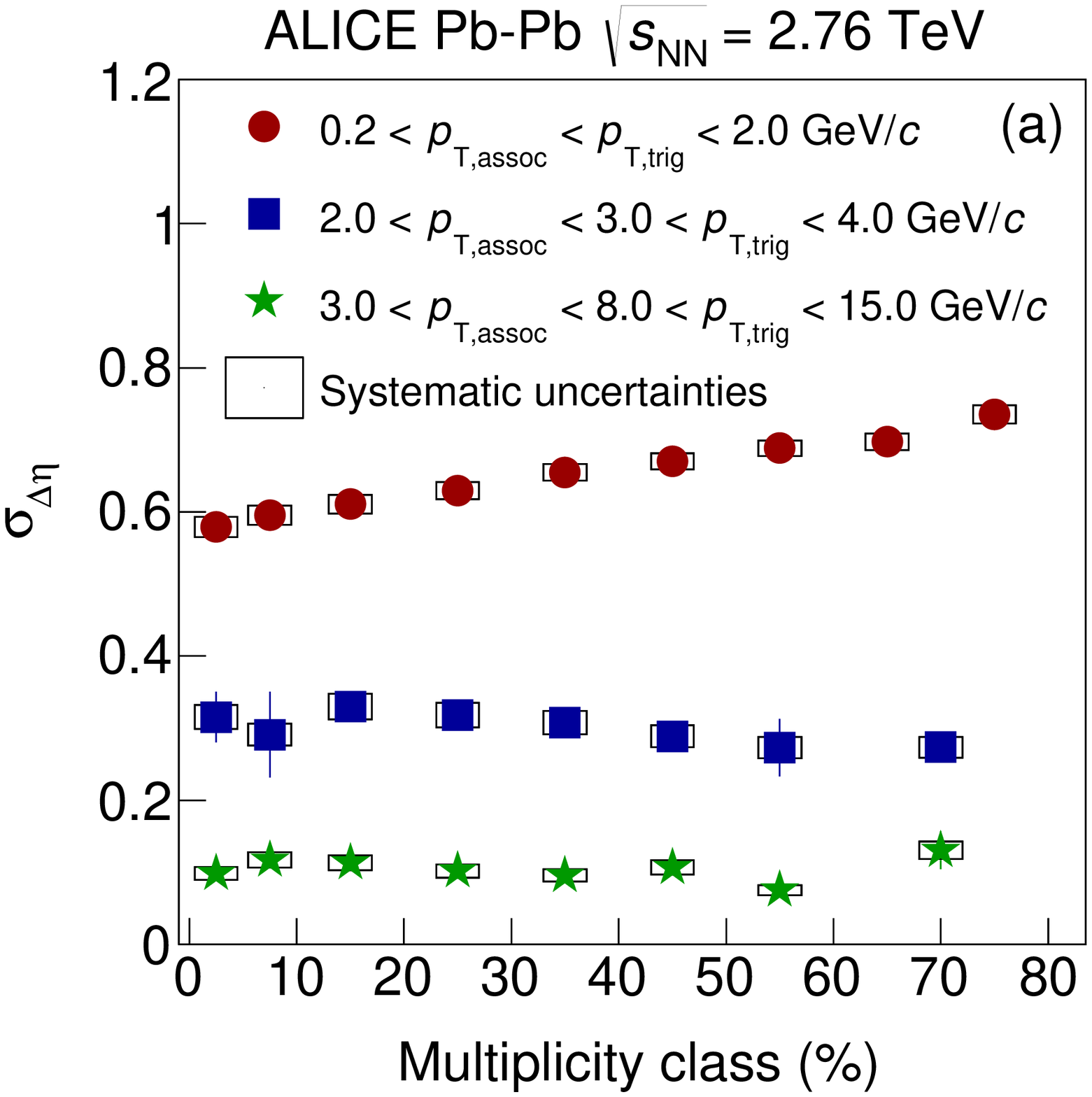}
\includegraphics[width=0.49\textwidth]{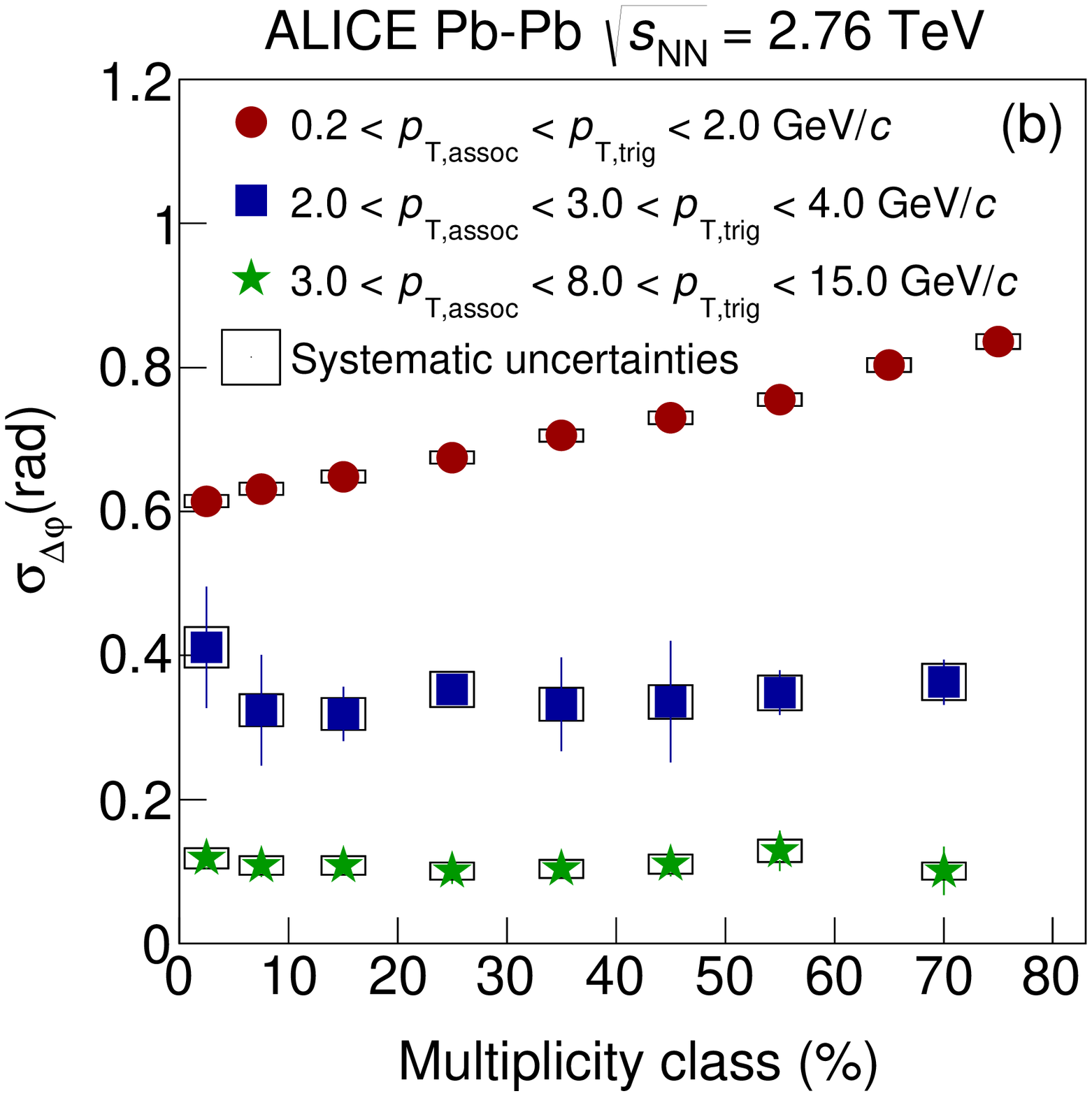}
\caption{The multiplicity-class dependence of \sigmadeta~(a) and \sigmadphi~(b) in Pb--Pb collisions at 
\snnPbPb. The different markers represent the low (i.e. \ptLow~with red circles), intermediate (i.e. 
$2.0 < p_\mathrm{T,assoc} < 3.0 < p_\mathrm{T,trig} < 4.0$~GeV/$c$~with blue squares), and high (i.e. \ptHigh~with green triangles) transverse momentum regions used in this analysis.}
\label{fig:widthPbPbAllpT}
\end{figure}

\begin{figure}[th!]
\centering
\includegraphics[width=0.49\textwidth]{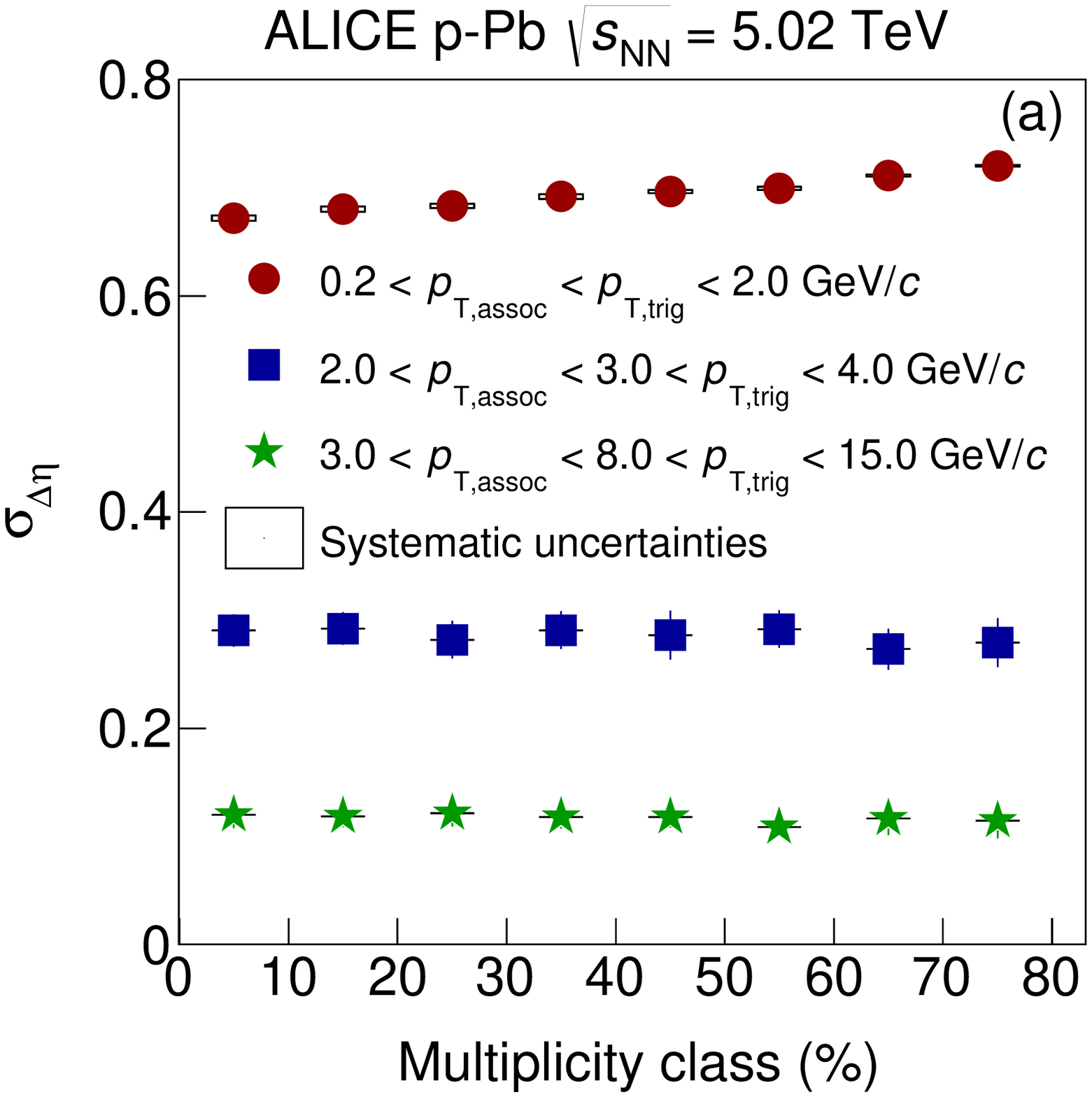}
\includegraphics[width=0.49\textwidth]{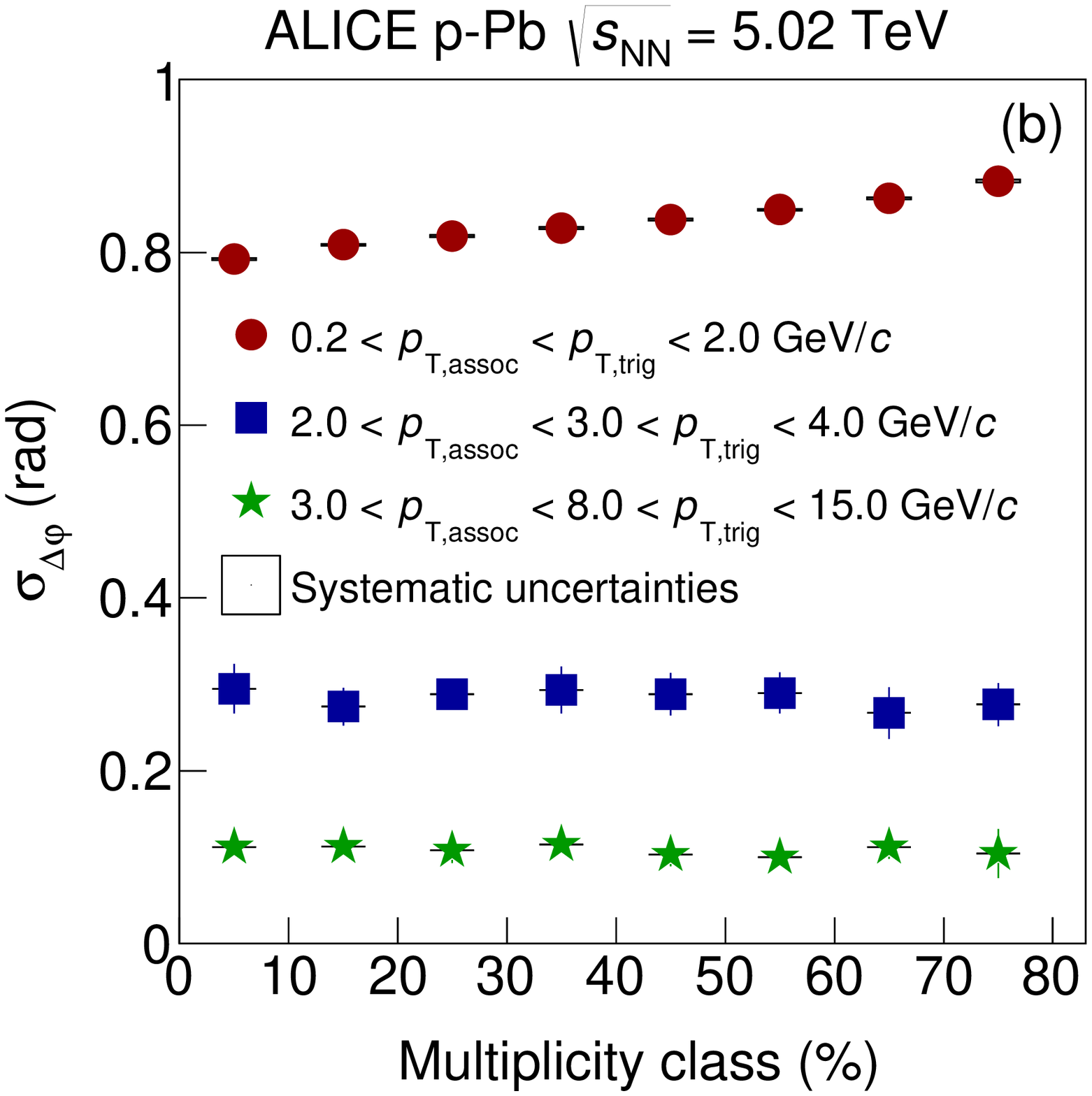}
\caption{\label{fig:widthpPbAllpT}The multiplicity-class dependence of \sigmadeta~(a) and \sigmadphi~(b) in p--Pb collisions at 
\snnpPb. The different markers represent the low (i.e. \ptLow~with red circles), intermediate (i.e. 
$2.0 < p_\mathrm{T,assoc} < 3.0 < p_\mathrm{T,trig} < 4.0$~GeV/$c$~with blue squares), and high (i.e. \ptHigh~with green triangles) transverse momentum regions used in this analysis.}
\end{figure}

\begin{figure}[th!]
\centering
\includegraphics[width=0.49\textwidth]{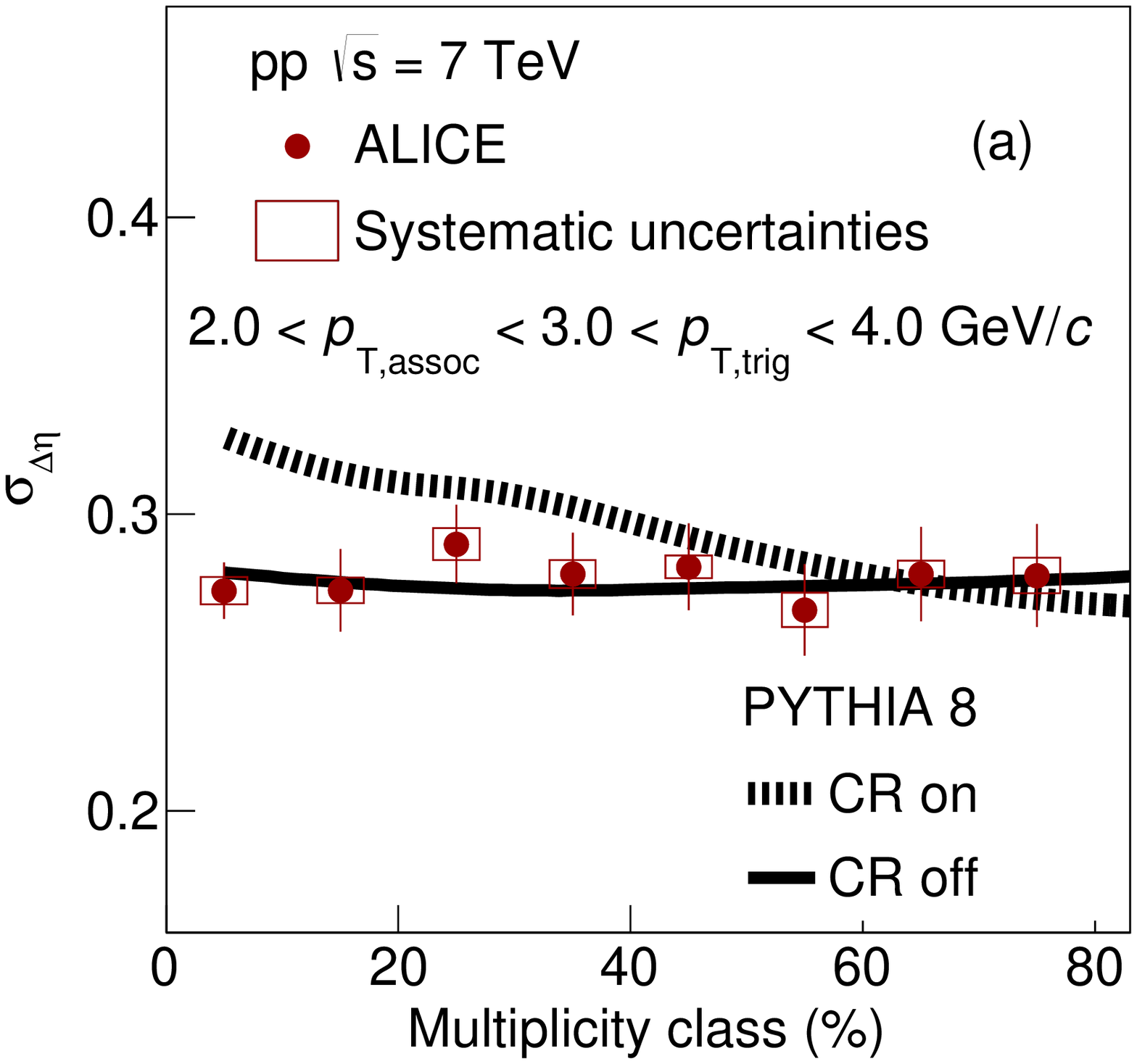}
\includegraphics[width=0.49\textwidth]{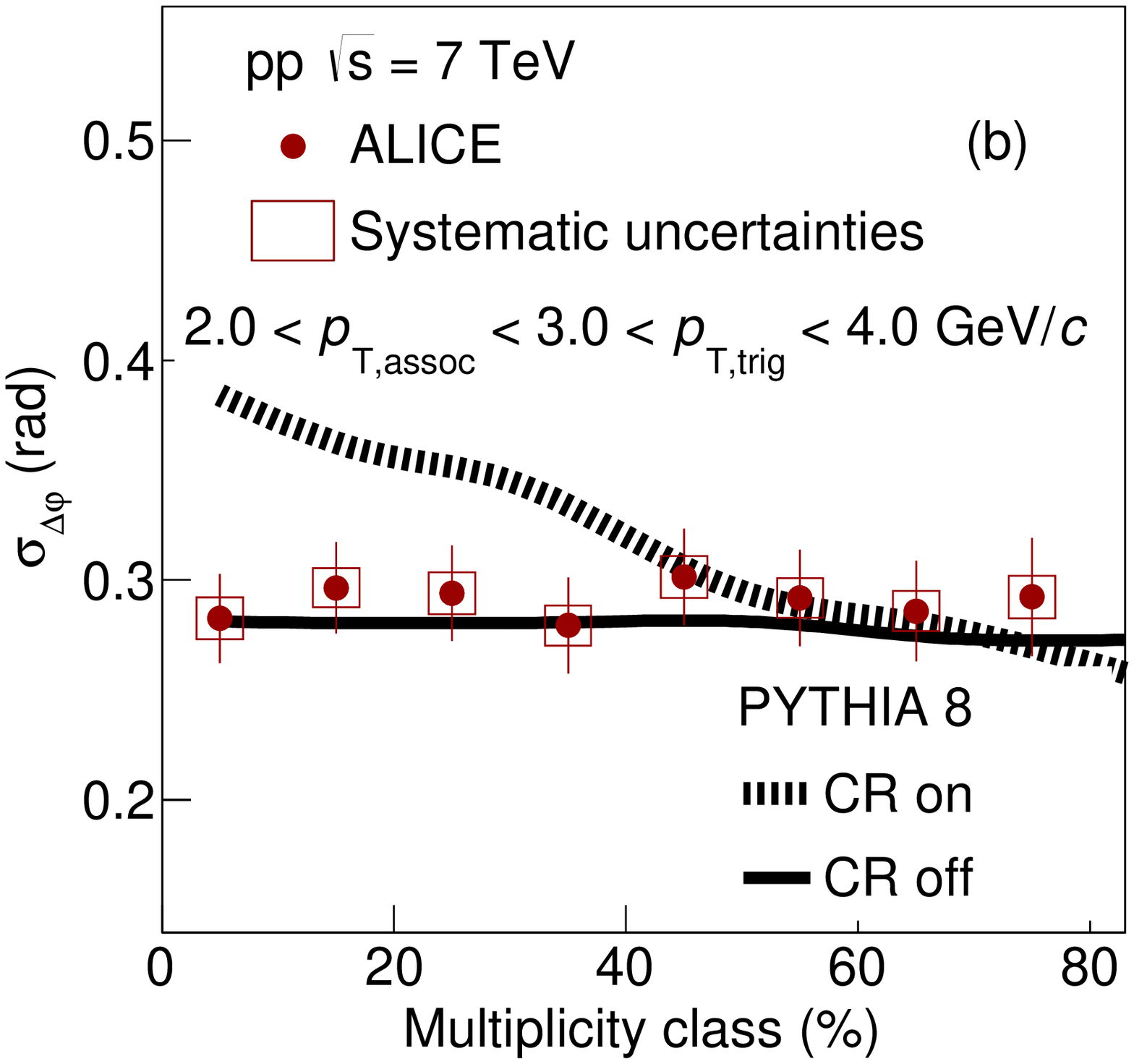}
\caption{The multiplicity-class dependence of the width of the balance function in \deta~(a) and in \dphi~(b) 
in pp collisions at \snnpp. The results correspond to the intermediate transverse momentum region (i.e. 
$2.0 < p_\mathrm{T,assoc} < 3.0 < p_\mathrm{T,trig} < 4.0$~GeV/$c$). The data points are compared with two 
versions of PYTHIA8 calculations.}
\label{fig:widthInppIntermediatepT}
\end{figure}

The widths of the balance function, \sigmadeta~and \sigmadphi~for the different transverse momentum regions, 
are presented in Figs.~\ref{fig:widthPbPbAllpT} - \ref{fig:widthpPbAllpT} as a 
function of the multiplicity class, for Pb--Pb and p--Pb collisions, respectively. The observed narrowing of the balance function with increasing multiplicity 
is restricted to the lower transverse momentum region, i.e. where the bulk of particles are produced. 
For higher transverse momenta, the multiplicity class dependence is significantly reduced, or even vanishes. In 
addition, the values of \sigmadeta~and \sigmadphi~decrease with increasing \pt~for 
a given multiplicity class. This decrease can be attributed to the transition to a region where initial hard-scattering processes and parton fragmentation become the dominant particle production mechanism. The 
emerging hadrons are thus correlated within a cone whose angular size decreases with increasing \pt. 

For pp collisions, the widths of the balance function \sigmadeta~and \sigmadphi~are compared with results from 
PYTHIA in Fig.~\ref{fig:widthInppIntermediatepT}. The tune of PYTHIA8 without the inclusion of color reconnection 
is found to describe the data at a quantitative level, for both \sigmadeta~and \sigmadphi. On the other hand, PYTHIA8 with the inclusion of color 
reconnection shows a broadening of the distributions with increasing multiplicity in both \deta~and \dphi, which 
is not supported by the data.

\subsection{\textbf{Comparison between the three systems}}

\begin{figure}[t!]
\centering
\includegraphics[width=0.49\textwidth]{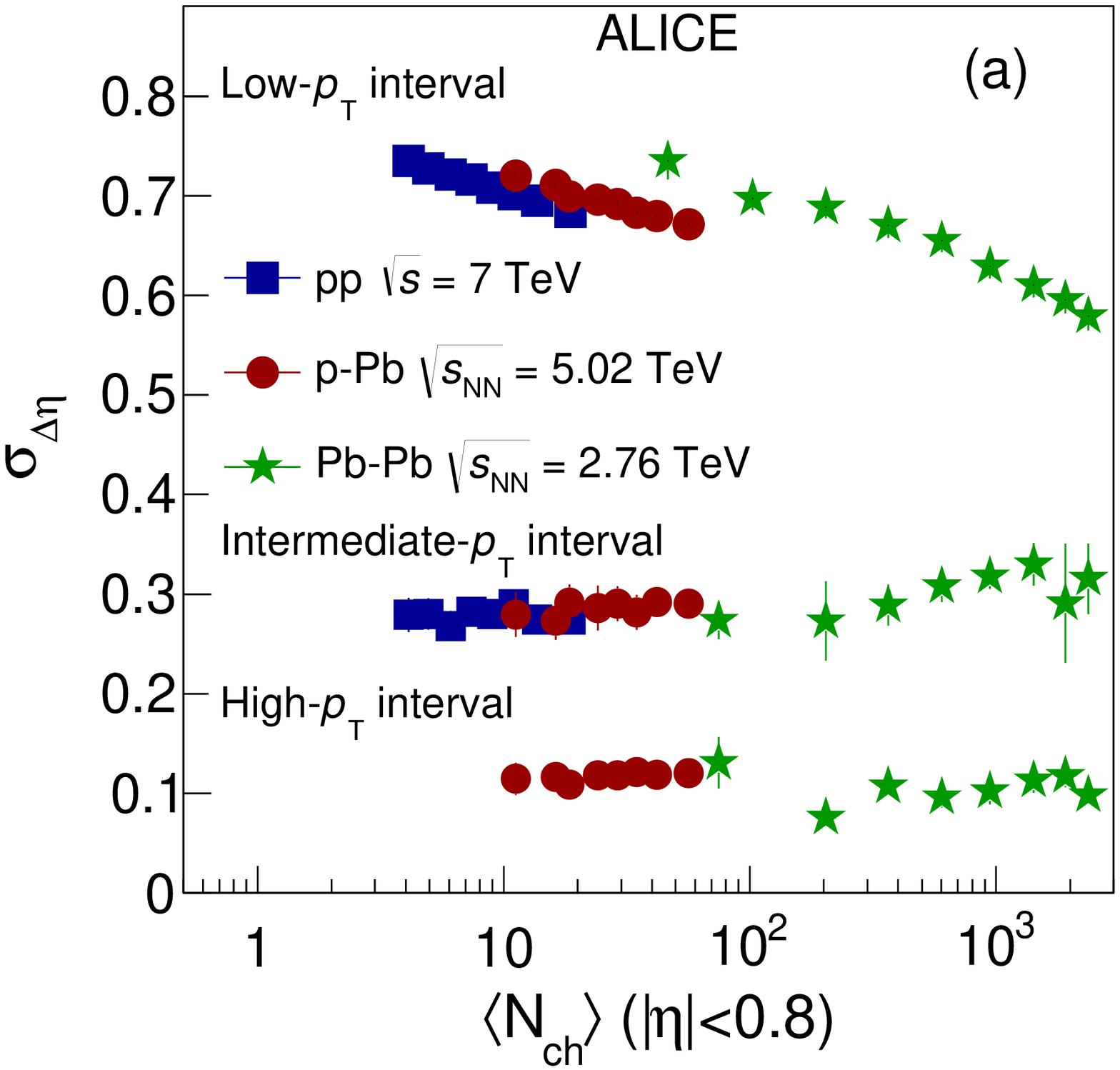}
\includegraphics[width=0.49\textwidth]{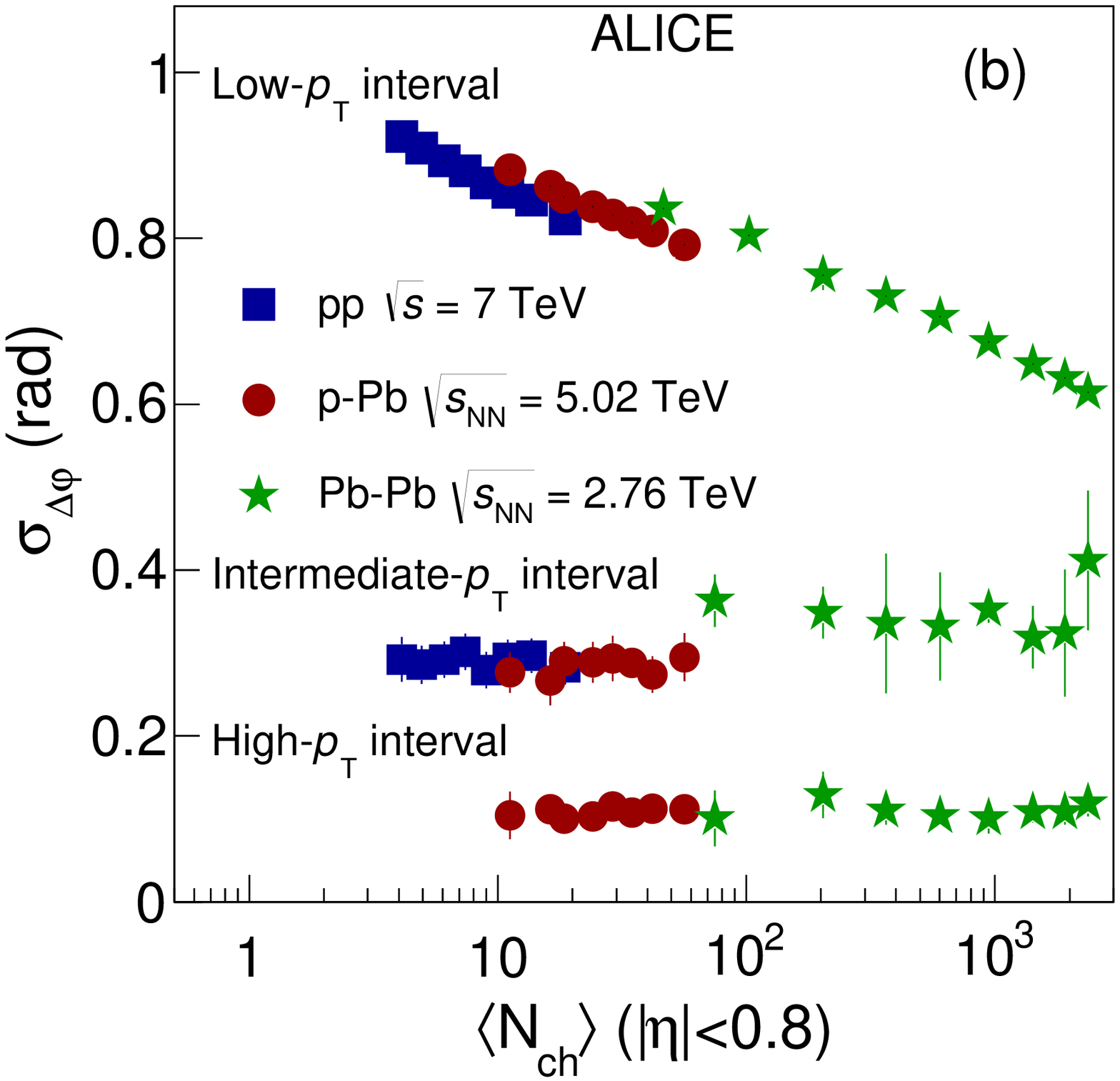}
\caption{The width of the balance function in \deta~(a) and in \dphi~(b) for the three systems 
analyzed (pp, p--Pb, and Pb--Pb), as a function of the charged-particle multiplicity, estimated with the V0A for 
$|\eta| < 0.8$ and $p_{\rm{T}} > 0.2$~GeV/$c$. The low-, intermediate-, and high-\pt~intervals correspond to \ptLow, 
\ptMed,~and \ptHigh, respectively.}
\label{fig:systemComparisonWidthvsMultiplicityLowPt}
\end{figure}

A comparison of the widths of the balance function in pp, p--Pb, and Pb--Pb as a function of particle multiplicity can 
provide direct information about differences and similarities between these systems in e.g. particle production 
mechanisms. It is important to note though, that this is performed for different center-of-mass energies which could complicate the comparison.

Figure~\ref{fig:systemComparisonWidthvsMultiplicityLowPt} presents the charged-particle multiplicity dependence 
of the width of the balance function in \deta~(a) and \dphi~(b) for all three systems. The 
results for the low-, intermediate- and high-\pt~intervals are shown in the 
same plot. Multiplicity is defined as the number of charged particles reconstructed in $|\eta| < 0.8$ and \pt~$> 0.2$~GeV/$c$, as described in Section~\ref{Sec:Analysis}. It is seen that between the pp and the p--Pb systems, and 
for overlapping multiplicities in the low-\pt~region, the width in both $\mathrm{\Delta}\eta$ and $\mathrm{\Delta}\varphi$ has similar values. This could indicate that the charge-dependent correlations have similar origin in these two 
systems. On the other hand, the comparison of the results between p--Pb and Pb--Pb at the overlapping 
multiplicities indicate differences for both \sigmadeta~and (to a smaller extent in) \sigmadphi. The origin of the charge-dependent correlations 
probed with the balance function in Pb--Pb collisions is believed to be related to radial flow and/or to a delayed hadronization scenario. The differences 
observed in the results of the Pb--Pb system compared with the ones in pp and p--Pb collisions at similar multiplicities 
could be explained by a different mechanism that drives the charge-dependent correlations in smaller systems.

With increasing values of transverse momentum, the balance functions become narrower and exhibit no significant 
multiplicity dependence for all systems, as discussed previously. The origin of these correlations at these 
transverse momentum ranges could be connected to initial hard parton scattering and subsequent fragmentation. 
The agreement of the values of both \sigmadeta~and \sigmadphi~for all multiplicities over all three systems clearly 
indicates that the dynamics responsible for the high-\pt~charge-dependent correlations do not change significantly between 
pp, p--Pb, and Pb--Pb.  

\begin{figure}[t!]
\centering
\includegraphics[width=0.49\textwidth]{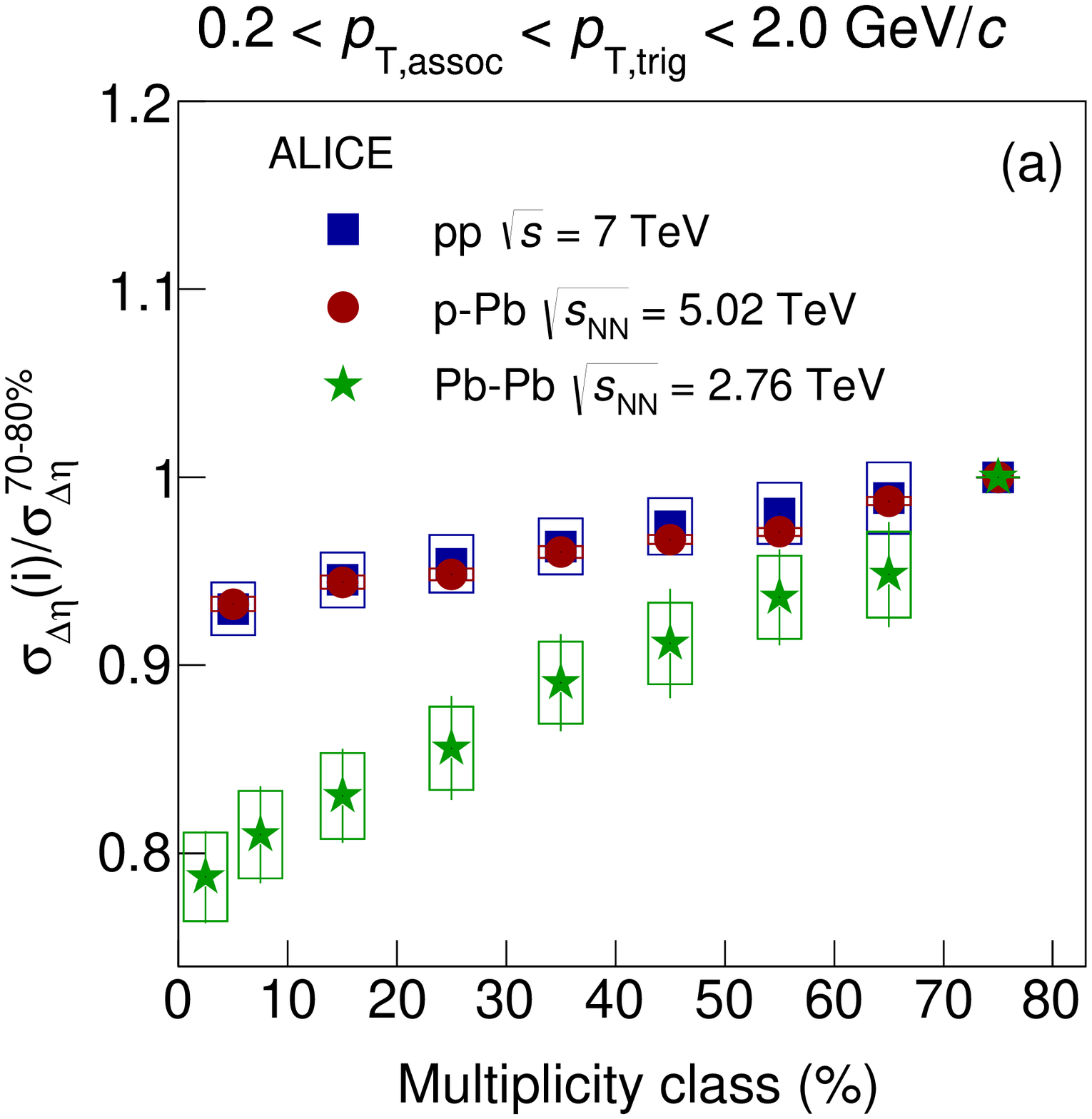}
\includegraphics[width=0.49\textwidth]{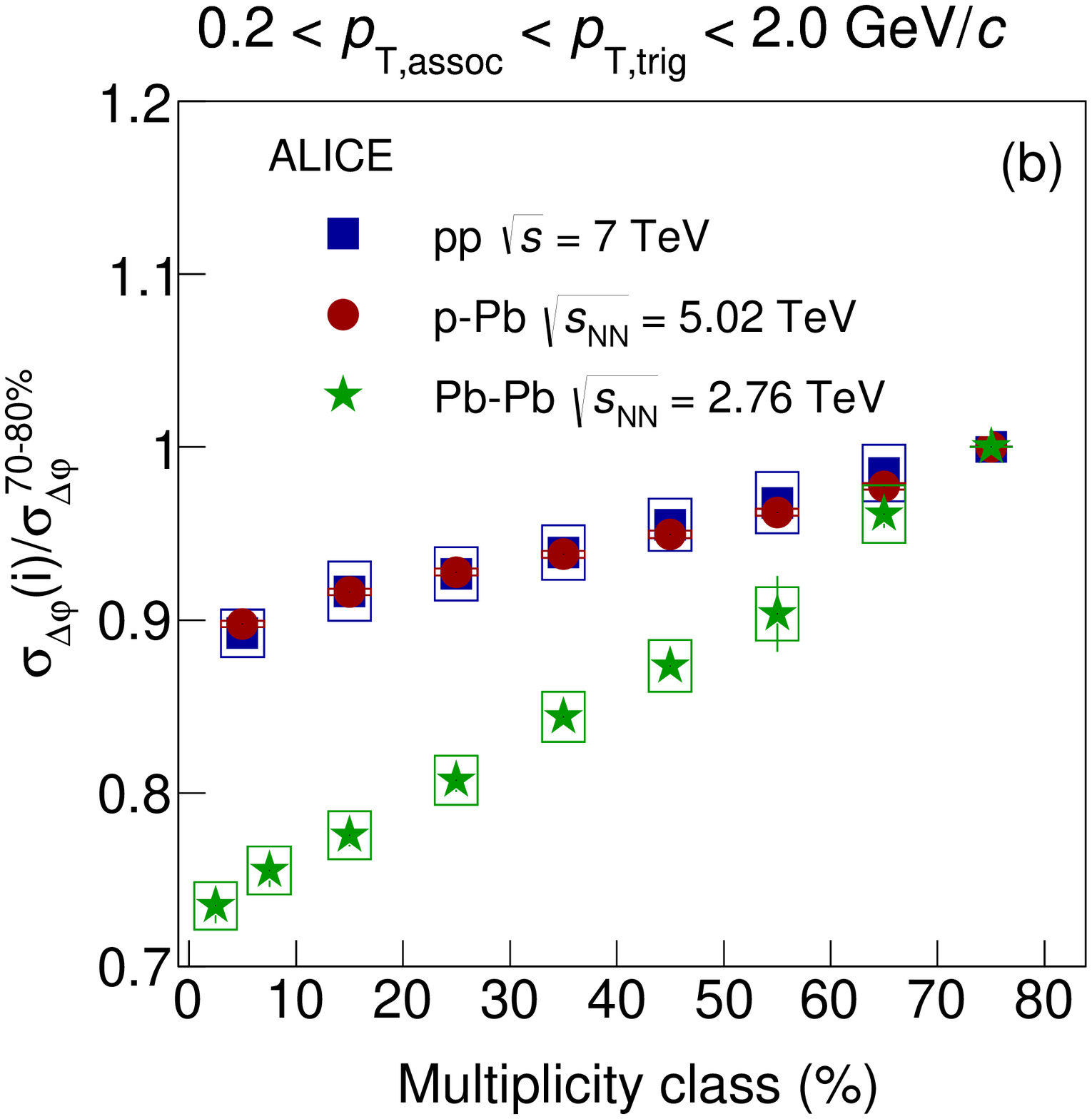}
\caption{The multiplicity-class dependence of the width of the balance function in \deta~(a) and in \dphi~(b) 
for the three systems analyzed (pp, p--Pb, and Pb--Pb) relative to the 70-80$\%$ multiplicity class.}
\label{fig:systemComparisonRelativeWidthLowPt}
\end{figure}

The narrowing of the balance function in both $\mathrm{\Delta}\eta$ and $\mathrm{\Delta}\varphi$ is a 
distinct characteristic of the low transverse momentum region. Figure~\ref{fig:systemComparisonRelativeWidthLowPt} visually illustrates the relative decrease of the different systems in this region. It is interesting that the relative decrease in this representation is similar between the two small systems (around 7$\%$ and 
10.5$\%$ in $\mathrm{\Delta}\eta$ and $\mathrm{\Delta}\varphi$, respectively) while different for Pb-Pb. The results from the analysis of 
Pb--Pb collisions illustrate a significantly larger relative decrease of 21.2 $\%$ in \deta~(26.5$\%$ for \dphi). A 
direct comparison of the width at the same multiplicity class can not be done because, for the same class, the physics conditions are quite different for 
pp, p--Pb, and Pb--Pb collisions. However, the comparison of the relative decrease and the 
agreement of the results in both \deta~and \dphi~between the two small systems could indicate that 
they share a similar mechanism which is responsible for the decrease of the width with increasing multiplicity. 


\section{Summary}
\label{Sec:Summary}

This article reports the first measurements of the balance function for charged particles in pp, p--Pb, and Pb--Pb collisions at the LHC measured with the ALICE detector. For all three systems, the balance function in both relative pseudorapidity (\deta) and relative azimuthal angle (\dphi) was studied for up to 9 multiplicity classes, and different trigger and associated particle transverse momentum. The widths of the balance functions in 
\deta~and \dphi~were found to decrease 
with increasing multiplicity for all systems only in the low-\pt~region (for 
\pt~$< 2.0$~GeV/$c$). For higher values of \pt, the multiplicity-class dependence is significantly reduced, if not 
vanished, and the correlations of balancing partners are stronger with respect to the low-\pt~region. Models incorporating collective effects, such as AMPT, reproduce the narrowing of the experimental points 
qualitatively in \dphi, but fail to reproduce the dependence in \deta. On the other hand, models based on independent 
pp collisions such as DPMJET and HIJING do not show any narrowing in p--Pb and Pb--Pb. The comparison of the results in 
pp collisions with different PYTHIA8 tunes indicates the importance of MPIs and of the color reconnection mechanism, whose inclusion within this model allows for a qualitative description of the experimentally 
measured narrowing with increasing multiplicity at low values of transverse momentum. 


\newenvironment{acknowledgement}{\relax}{\relax}
\begin{acknowledgement}
\section*{Acknowledgements}

The ALICE Collaboration would like to thank all its engineers and technicians for their invaluable contributions to the construction of the experiment and the CERN accelerator teams for the outstanding performance of the LHC complex.
The ALICE Collaboration gratefully acknowledges the resources and support provided by all Grid centres and the Worldwide LHC Computing Grid (WLCG) collaboration.
The ALICE Collaboration acknowledges the following funding agencies for their support in building and
running the ALICE detector:
State Committee of Science,  World Federation of Scientists (WFS)
and Swiss Fonds Kidagan, Armenia;
Conselho Nacional de Desenvolvimento Cient\'{\i}fico e Tecnol\'{o}gico (CNPq), Financiadora de Estudos e Projetos (FINEP),
Funda\c{c}\~{a}o de Amparo \`{a} Pesquisa do Estado de S\~{a}o Paulo (FAPESP);
National Natural Science Foundation of China (NSFC), the Chinese Ministry of Education (CMOE)
and the Ministry of Science and Technology of China (MSTC);
Ministry of Education and Youth of the Czech Republic;
Danish Natural Science Research Council, the Carlsberg Foundation and the Danish National Research Foundation;
The European Research Council under the European Community's Seventh Framework Programme;
Helsinki Institute of Physics and the Academy of Finland;
French CNRS-IN2P3, the `Region Pays de Loire', `Region Alsace', `Region Auvergne' and CEA, France;
German Bundesministerium fur Bildung, Wissenschaft, Forschung und Technologie (BMBF) and the Helmholtz Association;
General Secretariat for Research and Technology, Ministry of Development, Greece;
Hungarian Orszagos Tudomanyos Kutatasi Alappgrammok (OTKA) and National Office for Research and Technology (NKTH);
Department of Atomic Energy and Department of Science and Technology of the Government of India;
Istituto Nazionale di Fisica Nucleare (INFN) and Centro Fermi -
Museo Storico della Fisica e Centro Studi e Ricerche ``Enrico Fermi'', Italy;
MEXT Grant-in-Aid for Specially Promoted Research, Ja\-pan;
Joint Institute for Nuclear Research, Dubna;
National Research Foundation of Korea (NRF);
Consejo Nacional de Cienca y Tecnologia (CONACYT), Direccion General de Asuntos del Personal Academico(DGAPA), M\'{e}xico, Amerique Latine Formation academique - 
European Commission~(ALFA-EC) and the EPLANET Program~(European Particle Physics Latin American Network);
Stichting voor Fundamenteel Onderzoek der Materie (FOM) and the Nederlandse Organisatie voor Wetenschappelijk Onderzoek (NWO), Netherlands;
Research Council of Norway (NFR);
National Science Centre, Poland;
Ministry of National Education/Institute for Atomic Physics and National Council of Scientific Research in Higher Education~(CNCSI-UEFISCDI), Romania;
Ministry of Education and Science of Russian Federation, Russian
Academy of Sciences, Russian Federal Agency of Atomic Energy,
Russian Federal Agency for Science and Innovations and The Russian
Foundation for Basic Research;
Ministry of Education of Slovakia;
Department of Science and Technology, South Africa;
Centro de Investigaciones Energeticas, Medioambientales y Tecnologicas (CIEMAT), E-Infrastructure shared between Europe and Latin America (EELA), 
Ministerio de Econom\'{i}a y Competitividad (MINECO) of Spain, Xunta de Galicia (Conseller\'{\i}a de Educaci\'{o}n),
Centro de Aplicaciones Tecnológicas y Desarrollo Nuclear (CEA\-DEN), Cubaenerg\'{\i}a, Cuba, and IAEA (International Atomic Energy Agency);
Swedish Research Council (VR) and Knut $\&$ Alice Wallenberg
Foundation (KAW);
Ukraine Ministry of Education and Science;
United Kingdom Science and Technology Facilities Council (STFC);
The United States Department of Energy, the United States National
Science Foundation, the State of Texas, and the State of Ohio;
Ministry of Science, Education and Sports of Croatia and  Unity through Knowledge Fund, Croatia;
Council of Scientific and Industrial Research (CSIR), New Delhi, India;
Pontificia Universidad Cat\'{o}lica del Per\'{u}.
\end{acknowledgement}

\bibliographystyle{utphys}   
\bibliography{balanceFunctionCDS}{}

\newpage
\appendix
\section{The ALICE Collaboration}
\label{app:collab}



\begingroup
\small
\begin{flushleft}
J.~Adam\Irefn{org40}\And
D.~Adamov\'{a}\Irefn{org83}\And
M.M.~Aggarwal\Irefn{org87}\And
G.~Aglieri Rinella\Irefn{org36}\And
M.~Agnello\Irefn{org110}\And
N.~Agrawal\Irefn{org48}\And
Z.~Ahammed\Irefn{org132}\And
S.U.~Ahn\Irefn{org68}\And
S.~Aiola\Irefn{org136}\And
A.~Akindinov\Irefn{org58}
% ...
\renewcommand\labelenumi{\textsuperscript{\theenumi}~}

\section*{Affiliation notes}
\renewcommand\theenumi{\roman{enumi}}
\begin{Authlist}
\item \Adef{0}Deceased
\item \Adef{idp1746384}{Also at: Georgia State University, Atlanta, Georgia, United States}
\item \Adef{idp3782640}{Also at: M.V. Lomonosov Moscow State University, D.V. Skobeltsyn Institute of Nuclear, Physics, Moscow, Russia}
\end{Authlist}

\section*{Collaboration Institutes}
\renewcommand\theenumi{\arabic{enumi}~}
\begin{Authlist}

\item \Idef{org1}A.I. Alikhanyan National Science Laboratory (Yerevan Physics Institute) Foundation, Yerevan, Armenia
\item \Idef{org2}Benem\'{e}rita Universidad Aut\'{o}noma de Puebla, Puebla, Mexico
\item \Idef{org3}Bogolyubov Institute for Theoretical Physics, Kiev, Ukraine
\item \Idef{org4}Bose Institute, Department of Physics and Centre for Astroparticle Physics and Space Science (CAPSS), Kolkata, India
\item \Idef{org5}Budker Institute for Nuclear Physics, Novosibirsk, Russia
\item \Idef{org6}California Polytechnic State University, San Luis Obispo, California, United States
\item \Idef{org7}Central China Normal University, Wuhan, China
\item \Idef{org8}Centre de Calcul de l'IN2P3, Villeurbanne, France
\item \Idef{org9}Centro de Aplicaciones Tecnol\'{o}gicas y Desarrollo Nuclear (CEADEN), Havana, Cuba
\item \Idef{org10}Centro de Investigaciones Energ\'{e}ticas Medioambientales y Tecnol\'{o}gicas (CIEMAT), Madrid, Spain
\item \Idef{org11}Centro de Investigaci\'{o}n y de Estudios Avanzados (CINVESTAV), Mexico City and M\'{e}rida, Mexico
\item \Idef{org12}Centro Fermi - Museo Storico della Fisica e Centro Studi e Ricerche ``Enrico Fermi'', Rome, Italy
\item \Idef{org13}Chicago State University, Chicago, Illinois, USA
\item \Idef{org14}China Institute of Atomic Energy, Beijing, China
\item \Idef{org15}Commissariat \`{a} l'Energie Atomique, IRFU, Saclay, France
\item \Idef{org16}COMSATS Institute of Information Technology (CIIT), Islamabad, Pakistan
\item \Idef{org17}Departamento de F\'{\i}sica de Part\'{\i}culas and IGFAE, Universidad de Santiago de Compostela, Santiago de Compostela, Spain
\item \Idef{org18}Department of Physics and Technology, University of Bergen, Bergen, Norway
\item \Idef{org19}Department of Physics, Aligarh Muslim University, Aligarh, India
\item \Idef{org20}Department of Physics, Ohio State University, Columbus, Ohio, United States
\item \Idef{org21}Department of Physics, Sejong University, Seoul, South Korea
\item \Idef{org22}Department of Physics, University of Oslo, Oslo, Norway
\item \Idef{org23}Dipartimento di Elettrotecnica ed Elettronica del Politecnico, Bari, Italy
\item \Idef{org24}Dipartimento di Fisica dell'Universit\`{a} 'La Sapienza' and Sezione INFN Rome, Italy
\item \Idef{org25}Dipartimento di Fisica dell'Universit\`{a} and Sezione INFN, Cagliari, Italy
\item \Idef{org26}Dipartimento di Fisica dell'Universit\`{a} and Sezione INFN, Trieste, Italy
\item \Idef{org27}Dipartimento di Fisica dell'Universit\`{a} and Sezione INFN, Turin, Italy
\item \Idef{org28}Dipartimento di Fisica e Astronomia dell'Universit\`{a} and Sezione INFN, Bologna, Italy
\item \Idef{org29}Dipartimento di Fisica e Astronomia dell'Universit\`{a} and Sezione INFN, Catania, Italy
\item \Idef{org30}Dipartimento di Fisica e Astronomia dell'Universit\`{a} and Sezione INFN, Padova, Italy
\item \Idef{org31}Dipartimento di Fisica `E.R.~Caianiello' dell'Universit\`{a} and Gruppo Collegato INFN, Salerno, Italy
\item \Idef{org32}Dipartimento di Scienze e Innovazione Tecnologica dell'Universit\`{a} del  Piemonte Orientale and Gruppo Collegato INFN, Alessandria, Italy
\item \Idef{org33}Dipartimento Interateneo di Fisica `M.~Merlin' and Sezione INFN, Bari, Italy
\item \Idef{org34}Division of Experimental High Energy Physics, University of Lund, Lund, Sweden
\item \Idef{org35}Eberhard Karls Universit\"{a}t T\"{u}bingen, T\"{u}bingen, Germany
\item \Idef{org36}European Organization for Nuclear Research (CERN), Geneva, Switzerland
\item \Idef{org37}Excellence Cluster Universe, Technische Universit\"{a}t M\"{u}nchen, Munich, Germany
\item \Idef{org38}Faculty of Engineering, Bergen University College, Bergen, Norway
\item \Idef{org39}Faculty of Mathematics, Physics and Informatics, Comenius University, Bratislava, Slovakia
\item \Idef{org40}Faculty of Nuclear Sciences and Physical Engineering, Czech Technical University in Prague, Prague, Czech Republic
\item \Idef{org41}Faculty of Science, P.J.~\v{S}af\'{a}rik University, Ko\v{s}ice, Slovakia
\item \Idef{org42}Faculty of Technology, Buskerud and Vestfold University College, Vestfold, Norway
\item \Idef{org43}Frankfurt Institute for Advanced Studies, Johann Wolfgang Goethe-Universit\"{a}t Frankfurt, Frankfurt, Germany
\item \Idef{org44}Gangneung-Wonju National University, Gangneung, South Korea
\item \Idef{org45}Gauhati University, Department of Physics, Guwahati, India
\item \Idef{org46}Helsinki Institute of Physics (HIP), Helsinki, Finland
\item \Idef{org47}Hiroshima University, Hiroshima, Japan
\item \Idef{org48}Indian Institute of Technology Bombay (IIT), Mumbai, India
\item \Idef{org49}Indian Institute of Technology Indore, Indore (IITI), India
\item \Idef{org50}Inha University, Incheon, South Korea
\item \Idef{org51}Institut de Physique Nucl\'eaire d'Orsay (IPNO), Universit\'e Paris-Sud, CNRS-IN2P3, Orsay, France
\item \Idef{org52}Institut f\"{u}r Informatik, Johann Wolfgang Goethe-Universit\"{a}t Frankfurt, Frankfurt, Germany
\item \Idef{org53}Institut f\"{u}r Kernphysik, Johann Wolfgang Goethe-Universit\"{a}t Frankfurt, Frankfurt, Germany
\item \Idef{org54}Institut f\"{u}r Kernphysik, Westf\"{a}lische Wilhelms-Universit\"{a}t M\"{u}nster, M\"{u}nster, Germany
\item \Idef{org55}Institut Pluridisciplinaire Hubert Curien (IPHC), Universit\'{e} de Strasbourg, CNRS-IN2P3, Strasbourg, France
\item \Idef{org56}Institute for Nuclear Research, Academy of Sciences, Moscow, Russia
\item \Idef{org57}Institute for Subatomic Physics of Utrecht University, Utrecht, Netherlands
\item \Idef{org58}Institute for Theoretical and Experimental Physics, Moscow, Russia
\item \Idef{org59}Institute of Experimental Physics, Slovak Academy of Sciences, Ko\v{s}ice, Slovakia
\item \Idef{org60}Institute of Physics, Academy of Sciences of the Czech Republic, Prague, Czech Republic
\item \Idef{org61}Institute of Physics, Bhubaneswar, India
\item \Idef{org62}Institute of Space Science (ISS), Bucharest, Romania
\item \Idef{org63}Instituto de Ciencias Nucleares, Universidad Nacional Aut\'{o}noma de M\'{e}xico, Mexico City, Mexico
\item \Idef{org64}Instituto de F\'{\i}sica, Universidad Nacional Aut\'{o}noma de M\'{e}xico, Mexico City, Mexico
\item \Idef{org65}iThemba LABS, National Research Foundation, Somerset West, South Africa
\item \Idef{org66}Joint Institute for Nuclear Research (JINR), Dubna, Russia
\item \Idef{org67}Konkuk University, Seoul, South Korea
\item \Idef{org68}Korea Institute of Science and Technology Information, Daejeon, South Korea
\item \Idef{org69}KTO Karatay University, Konya, Turkey
\item \Idef{org70}Laboratoire de Physique Corpusculaire (LPC), Clermont Universit\'{e}, Universit\'{e} Blaise Pascal, CNRS--IN2P3, Clermont-Ferrand, France
\item \Idef{org71}Laboratoire de Physique Subatomique et de Cosmologie, Universit\'{e} Grenoble-Alpes, CNRS-IN2P3, Grenoble, France
\item \Idef{org72}Laboratori Nazionali di Frascati, INFN, Frascati, Italy
\item \Idef{org73}Laboratori Nazionali di Legnaro, INFN, Legnaro, Italy
\item \Idef{org74}Lawrence Berkeley National Laboratory, Berkeley, California, United States
\item \Idef{org75}Moscow Engineering Physics Institute, Moscow, Russia
\item \Idef{org76}Nagasaki Institute of Applied Science, Nagasaki, Japan
\item \Idef{org77}National Centre for Nuclear Studies, Warsaw, Poland
\item \Idef{org78}National Institute for Physics and Nuclear Engineering, Bucharest, Romania
\item \Idef{org79}National Institute of Science Education and Research, Bhubaneswar, India
\item \Idef{org80}Niels Bohr Institute, University of Copenhagen, Copenhagen, Denmark
\item \Idef{org81}Nikhef, Nationaal instituut voor subatomaire fysica, Amsterdam, Netherlands
\item \Idef{org82}Nuclear Physics Group, STFC Daresbury Laboratory, Daresbury, United Kingdom
\item \Idef{org83}Nuclear Physics Institute, Academy of Sciences of the Czech Republic, \v{R}e\v{z} u Prahy, Czech Republic
\item \Idef{org84}Oak Ridge National Laboratory, Oak Ridge, Tennessee, United States
\item \Idef{org85}Petersburg Nuclear Physics Institute, Gatchina, Russia
\item \Idef{org86}Physics Department, Creighton University, Omaha, Nebraska, United States
\item \Idef{org87}Physics Department, Panjab University, Chandigarh, India
\item \Idef{org88}Physics Department, University of Athens, Athens, Greece
\item \Idef{org89}Physics Department, University of Cape Town, Cape Town, South Africa
\item \Idef{org90}Physics Department, University of Jammu, Jammu, India
\item \Idef{org91}Physics Department, University of Rajasthan, Jaipur, India
\item \Idef{org92}Physik Department, Technische Universit\"{a}t M\"{u}nchen, Munich, Germany
\item \Idef{org93}Physikalisches Institut, Ruprecht-Karls-Universit\"{a}t Heidelberg, Heidelberg, Germany
\item \Idef{org94}Purdue University, West Lafayette, Indiana, United States
\item \Idef{org95}Pusan National University, Pusan, South Korea
\item \Idef{org96}Research Division and ExtreMe Matter Institute EMMI, GSI Helmholtzzentrum f\"ur Schwerionenforschung, Darmstadt, Germany
\item \Idef{org97}Rudjer Bo\v{s}kovi\'{c} Institute, Zagreb, Croatia
\item \Idef{org98}Russian Federal Nuclear Center (VNIIEF), Sarov, Russia
\item \Idef{org99}Russian Research Centre Kurchatov Institute, Moscow, Russia
\item \Idef{org100}Saha Institute of Nuclear Physics, Kolkata, India
\item \Idef{org101}School of Physics and Astronomy, University of Birmingham, Birmingham, United Kingdom
\item \Idef{org102}Secci\'{o}n F\'{\i}sica, Departamento de Ciencias, Pontificia Universidad Cat\'{o}lica del Per\'{u}, Lima, Peru
\item \Idef{org103}Sezione INFN, Bari, Italy
\item \Idef{org104}Sezione INFN, Bologna, Italy
\item \Idef{org105}Sezione INFN, Cagliari, Italy
\item \Idef{org106}Sezione INFN, Catania, Italy
\item \Idef{org107}Sezione INFN, Padova, Italy
\item \Idef{org108}Sezione INFN, Rome, Italy
\item \Idef{org109}Sezione INFN, Trieste, Italy
\item \Idef{org110}Sezione INFN, Turin, Italy
\item \Idef{org111}SSC IHEP of NRC Kurchatov institute, Protvino, Russia
\item \Idef{org112}Stefan Meyer Institut f\"{u}r Subatomare Physik (SMI), Vienna, Austria
\item \Idef{org113}SUBATECH, Ecole des Mines de Nantes, Universit\'{e} de Nantes, CNRS-IN2P3, Nantes, France
\item \Idef{org114}Suranaree University of Technology, Nakhon Ratchasima, Thailand
\item \Idef{org115}Technical University of Ko\v{s}ice, Ko\v{s}ice, Slovakia
\item \Idef{org116}Technical University of Split FESB, Split, Croatia
\item \Idef{org117}The Henryk Niewodniczanski Institute of Nuclear Physics, Polish Academy of Sciences, Cracow, Poland
\item \Idef{org118}The University of Texas at Austin, Physics Department, Austin, Texas, USA
\item \Idef{org119}Universidad Aut\'{o}noma de Sinaloa, Culiac\'{a}n, Mexico
\item \Idef{org120}Universidade de S\~{a}o Paulo (USP), S\~{a}o Paulo, Brazil
\item \Idef{org121}Universidade Estadual de Campinas (UNICAMP), Campinas, Brazil
\item \Idef{org122}University of Houston, Houston, Texas, United States
\item \Idef{org123}University of Jyv\"{a}skyl\"{a}, Jyv\"{a}skyl\"{a}, Finland
\item \Idef{org124}University of Liverpool, Liverpool, United Kingdom
\item \Idef{org125}University of Tennessee, Knoxville, Tennessee, United States
\item \Idef{org126}University of the Witwatersrand, Johannesburg, South Africa
\item \Idef{org127}University of Tokyo, Tokyo, Japan
\item \Idef{org128}University of Tsukuba, Tsukuba, Japan
\item \Idef{org129}University of Zagreb, Zagreb, Croatia
\item \Idef{org130}Universit\'{e} de Lyon, Universit\'{e} Lyon 1, CNRS/IN2P3, IPN-Lyon, Villeurbanne, France
\item \Idef{org131}V.~Fock Institute for Physics, St. Petersburg State University, St. Petersburg, Russia
\item \Idef{org132}Variable Energy Cyclotron Centre, Kolkata, India
\item \Idef{org133}Warsaw University of Technology, Warsaw, Poland
\item \Idef{org134}Wayne State University, Detroit, Michigan, United States
\item \Idef{org135}Wigner Research Centre for Physics, Hungarian Academy of Sciences, Budapest, Hungary
\item \Idef{org136}Yale University, New Haven, Connecticut, United States
\item \Idef{org137}Yonsei University, Seoul, South Korea
\item \Idef{org138}Zentrum f\"{u}r Technologietransfer und Telekommunikation (ZTT), Fachhochschule Worms, Worms, Germany
\end{Authlist}
\endgroup

\end{document}